\documentclass[slac_one]{revtex4} 
\usepackage{graphicx} 
\usepackage{fancyhdr} 
\usepackage{epsfig} 
\newcommand{\bmp}{\mathbf p}

\def\beq{\begin{equation}}
\def\eeq#1{\label{#1}\end{equation}}
\def\eeqn{\end{equation}}
\def\beqa{\begin{eqnarray}}
\def\eeqa#1{\label{#1}\end{eqnarray}}
\def\eeqan{\end{eqnarray}}

\def\d{\hbox{d}}
\def\e{\epsilon}

\begin{document}

\hfill$\vcenter{
\hbox{\bf FERMILAB-CONF-06-006-T}
\hbox{\bf LTH-688}
\hbox{\bf MADPH-06-1252}
\hbox{\bf SLAC-PUB-11620}}
$

\title{{\small{2005 ALCPG \& ILC Workshops - Snowmass,
U.S.A.}}\\ 
\vspace{12pt}
Report of the 2005 Snowmass Top/QCD Working Group} 

\author{
A. Juste$^a$,
Y. Kiyo$^b$,
F. Petriello$^c$,
T. Teubner$^d$,
K. Agashe$^e$,
P. Batra$^f$,
U. Baur$^g$,
C.F. Berger$^h$,
J.A.R. Cembranos$^i$,
A. Gehrmann-De Ridder$^j$,
T. Gehrmann$^k$,
E.W.N. Glover$^l$,
S. Godfrey$^m$,
A. Hoang$^n$,
M. Perelstein$^o$,
Z. Sullivan$^f$,
T. Tait$^f$,
S. Zhu$^p$
}

\begin{abstract}

This report discusses several topics in both top quark physics and QCD at an International Linear Collider (ILC).  
Issues such as measurements at the $t\bar{t}$ threshold, including both theoretical and machine requirements, and the determination of electroweak 
top quark couplings are reviewed.  New results concerning the potential of a 500 GeV $e^+e^-$ collider for 
measuring $Wtb$ couplings and the top quark Yukawa coupling are presented.  The status of higher order QCD corrections 
to jet production cross sections, heavy quark form factors, and longitudinal gauge boson scattering, needed for percent-level 
studies at the ILC, are reviewed.  A new study of the measurement of the hadronic structure of the photon at a 
$\gamma\gamma$ collider is presented.  The effects on top 
quark properties from several models of new physics, including composite models, Little Higgs theories, and CPT violation, 
are studied.

\end{abstract}

\maketitle

\begin{center}

$^a$Fermi National Accelerator Laboratory, P.O. Box 500, MS 357, Batavia, IL 60510, USA\\
$^b$Institut f\"{u}r Theoretische Physik E, RWTH Aachen, D-52056 Aachen, Germany\\
$^c$University of Wisconsin, Madison, WI 53706, USA {\it and} Fermi National Accelerator Laboratory, P.O. Box 500, MS 106, Batavia, IL 60510, USA\\
$^d$Department of Mathematical Sciences, The University of Liverpool, Liverpool L69 3BX, England, U.K.\\
$^e$Johns Hopkins University, Baltimore, MD 21218, USA {\it and} Institute for Advanced Studies, Princeton, NJ 08540, USA 
   {\it and} Syracuse University, Syracuse, NY 13244, USA\\
$^f$High Energy Physics Division, Argonne National Laboratory, Argonne, IL 60439, USA\\
$^g$State University of New York at Buffalo, Buffalo, NY 14260, USA\\
$^h$Stanford Linear Accelerator Center, Stanford University, Stanford, CA 94309, USA\\
$^i$University of California, Irvine, CA 92697, USA\\
$^j$Institut f\"{u}r Theoretische Physik, ETH, CH-8093 Z\"{u}rich, Switzerland\\
$^k$Institut f\"{u}r Theoretische Physik, Universit\"{a}t Z\"{u}rich, CH-8057 Z\"{u}rich, Switzerland\\
$^l$Institute for Particle Physics Phenomenology, University of Durham, Durham DH1 3LE, UK\\
$^m$Ottawa Carleton Institute for Physics, Carleton University, Ottawa K1S 5B6, Canada\\
$^n$Max-Planck-Institute for Physics, Munich, Germany\\
$^o$CIHEP, Cornell University, Ithaca, NY 14853, USA\\
$^p$Institute of Theoretical Physics, School of Physics, Peking University, Bejing 100871, China

\end{center}

\thispagestyle{fancy}


\section{INTRODUCTION}

The precision study of both new and already discovered particles will be a major component of the 
experimental programs at both the Large Hadron Collider (LHC) and a future International Linear Collider (ILC).  
The past two decades in particle physics established the importance of this precision physics program.  The outstanding 
success of the $Z$-pole program at LEP and SLC elevated the global fit to the precision electroweak data 
into the primary experimental constraint on extensions of the Standard Model (SM).  When combined with 
input from the Tevatron, it probes energy scales far beyond the kinematic limits of current colliders.  

The enormous production rates for top quarks at future colliders, reaching $10^7$ $t\bar{t}$ pairs in a $10 \, {\rm fb}^{-1}$ year at 
the LHC and $10^5$ in a $100 \, {\rm fb}^{-1}$ year at a 500 GeV $e^+e^-$ collider, will allow a similar 
program studying the top quark to be pursued.  Rare decays of the top quark, deviations from its chirality 
structure in the SM, and its electroweak couplings will be studied.  A high precision measurement of the top 
quark mass will greatly reduce the uncertainties in important electroweak parameters.  A variety of work is 
needed for this program to be successful, including the precision calculation of top quark properties in the 
SM, the determination of experimental capabilities for performing measurements, and finding the most 
likely deviations predicted by models of new physics.  The calculation of higher-order QCD corrections to 
top quark cross sections can have an important effect on top quark physics; for example, the threshold 
corrections to the process $e^+e^- \rightarrow t\bar{t}H$ increase its rate by a factor of two, drastically 
increasing the sensitivity of a 500 GeV ILC to the top quark Yukawa coupling (see Section~2.5 of 
this report).  The precision needed for the $t\bar{t}$ threshold scan imposes strong requirements on the monitoring 
of the luminosity spectrum, which are discussed in Section~2.7.  
Detailed analyses can reveal surprising experimental possibilities, such as the measurement of 
the $Wtb$ coupling below the $t\bar{t}$ threshold (see section~2.1 of this report).  The study of the 
predictions coming from models of new physics show that the top quark can be a powerful discriminator 
between different theories; for example, composite theories such as the Randall-Sundrum model predict shifts 
in the coupling of right-handed top quarks to the $Z$, while Little Higgs models generically predict shifts 
in the left-handed top couplings (see Sections~4.1 and~4.2 of this report).      

In addition to the study of the top quark, precision programs studying the Higgs boson or Higgs mechanism, the 
$W$ and $Z$ bosons, and other new particles discovered will be possible.  To fully utilize the percent-level 
experimental precisions for each of these programs, higher order electroweak and QCD corrections in the SM must be included.  For example, 
the two-loop QCD corrections to the $g-2$ of the $b$-quark reach the $2-3\%$ level, and must be included when 
studying $b$ production during a Giga-$Z$ phase at the ILC (see Section~2.3 of this report).  Benchmark 
processes for studying a strongly-interacting Higgs sector such as $V_L V_L \rightarrow t\bar{t}$ receive QCD 
corrections reaching $10-20\%$, which can mask the effects of new physics if not taken into account (see Section~3.2 
of this report).  Finally, fundamental properties of QCD such as the running of the strong coupling constant and the 
hadronic structure of the photon can be studied with unprecedented precision (see Sections~3.1 and~3.3 of 
this report, respectively).

In this report we discuss several issues in top quark physics and precision QCD, with a focus on the physics program at the ILC.  
Section~2 discusses precision studies of the top quark at the ILC.  Important issues such as the 
measurement of the $t\bar{t}$ threshold cross section are reviewed, and new results such as the study of the $Wtb$ 
coupling below the $t\bar{t}$ threshold and the measurement of the top quark Yukawa at $\sqrt{s}=500$ GeV are presented.   A 
preliminary study on the precise determination of the average beam energy and luminosity spectrum at the ILC, 
required for the $t\bar{t}$ threshold scan, is presented in section~2.7.
Section~3 discusses important QCD physics that can be performed at the ILC.  Higher order QCD corrections needed 
for several important measurements are discussed, and new results for the determination of the hadronic structure 
of the photon are presented.  Section~4 studies modifications of top quark properties in several models of 
new physics, and analyzes the potential of the ILC to measure these shifts and use them to discriminate between 
different extensions of the SM.

\section{PRECISION STUDIES OF THE TOP QUARK AT THE ILC}
\label{sec_ptop}


\subsection{Measuring the \mbox{\boldmath$Wtb$} Coupling Below the \mbox{\boldmath$t\bar{t}$} Threshold \\ \small{{\it P. Batra, T. Tait}}}
\label{sbat}

A crucial test of the top quark's electroweak interactions is the strength of the left-handed charged 
current interaction $W$-$t$-$b$.  In the Standard Model, unitarity of the CKM
matrix implies that $g_{Wtb} \sim g V_{tb} \sim g$, 
but in extended models, including
the simple extension by a fourth generation of fermions, this interaction can
differ significantly from the SM expectation.  Currently, it is known
that the $W$-$t$-$b$ vertex is sufficiently strong that the dominant top decay
is $t \rightarrow W b$, but even an imprecise measurement is lacking.  Single
top production at the Tevatron and LHC will help fill this gap in our
knowledge, and is expected to lead to a measurement at the $10\%$ level,
dominated by systematics \cite{Beneke:2000hk}. 

Unlike other top measurements, a direct test of the $W$-$t$-$b$
coupling is challenging at a 500 GeV $e^+ e^-$ collider.  A scan over
the $t \bar{t}$ threshold region is expected to yield precise
measurements of many top parameters in the SM, including the top mass,
width and Yukawa coupling (see \cite{Martinez:2002st} and this report
for projections), while above-threshold measurements may constrain
anomalous, non-SM Lorentz structures \cite{Boos:1999ca}.
Nevertheless, only an indirect measurement of the left-handed
$W$-$t$-$b$ coupling is offered from the $t \bar{t}$ threshold region,
by inferring its value from the SM relation and a precise value of the
top width. If, for example, there is a small non-standard decay mode
of top, it will alter the width and distort the inferred coupling.  It
would be more desirable to have a direct measurement of $W$-$t$-$b$,
by making use of a process which is directly proportional to it. Close
to the $t \bar{t}$ threshold, sensitivity to the coupling is quite
weak, because the rate is essentially the $t \bar{t}$ production cross
section times the branching ratios for $t \rightarrow W b$.  Since we
expect that the BR is very close to one, it does not in fact depend
strongly on the magnitude of the $W$-$t$-$b$ interaction. Meanwhile, single-top production above threshold, which is sensitive to the $W$-$t$-$b$
coupling, is swamped by the $t \bar{t}$ background unless a $\gamma
e$ collision mode is present \cite{Boos:2001sj}.

\begin{figure}[t]  
\centerline{\includegraphics[width=0.65\textwidth]{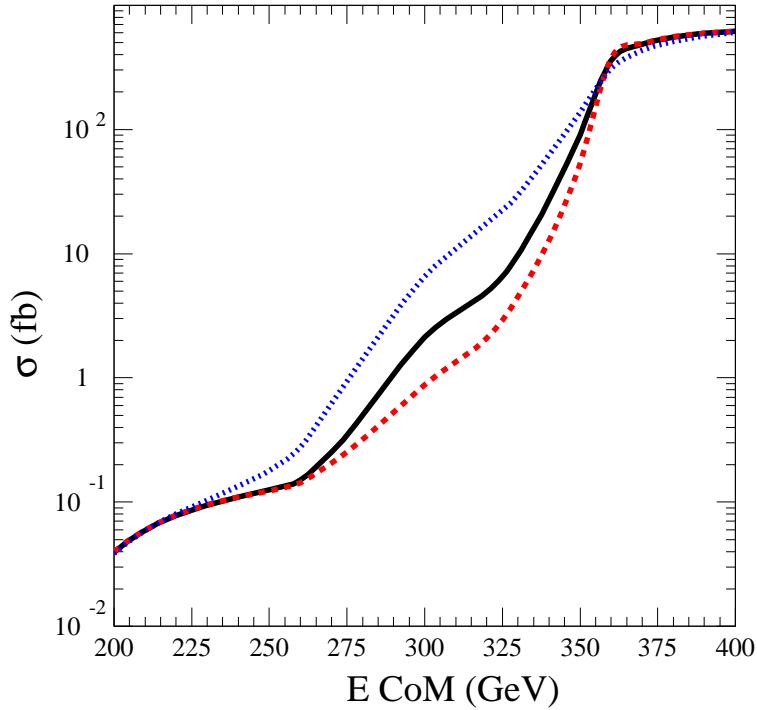}}
\caption{Inclusive rates 
for $e^+ e^- \rightarrow W^+ b W^- \bar{b}$ as a function
of the center-of-mass energy for $g_{Wtb} = g_{SM}$ (black solid), 
$g_{Wtb} = 2 g_{SM}$ (blue dashed), and $g_{Wtb} = g_{SM} /2 $ (red dotted).}
\label{fig:wtb1}  
\end{figure}  

Just below the $t \bar{t}$ threshold, the reaction
$e^+ e^- \rightarrow W^+ b W^- \bar{b}$ still occurs, through a mixture of
non-resonant Feynman diagrams as well as through off-shell top quarks.
At center-of-mass energies far enough below $2 m_t$ but still above $m_t$,
the rate is dominated by contributions from the virtual $t \bar t$ diagrams
in a kinematic configuration where one top is on-shell and the other is 
off-shell.  The rate becomes very sensitive to the $W$-$t$-$b$ interaction,
by virtue of the off-shell top \cite{Wtbinprogress}.  This is illustrated in
Figure~1, which plots the cross section as a function of energy
for several values of $g_{Wtb}$, assuming a $175$ GeV
top mass and a $115$ GeV Higgs mass. 
All analysis was performed using the
MadEvent package \cite{Maltoni:2002qb} at tree level. 
The lines asymptote to the same value at both ends of the
energy spectrum, as on-shell $t \bar{t}$ production dominates close to
threshold and graphs not involving top dominate far below
threshold. Both of these extremes are independent of the $W$-$t$-$b$
coupling.  Thus, energies in between these two extremes are suitable to measure
$g_{Wtb}$.  We avoid the region very close to $2 m_t$ (despite its large rate),
because the details of the transition from off-shell to on-shell do depend
sensitively on the top width, which could obscure $g_{Wtb}$ if there are
non-standard decay modes of the top.  Instead, we focus on the energy
$\sqrt{s}= 340$ GeV, where good leverage on this coupling appears to
be attained with small dependence on the width.  We will explore the interplay
between $g_{Wtb}$ and $\Gamma_t$ below.

Here we restrict ourselves to simple cuts to model the acceptance.
To that end, we require the jets to have $p_T > 10$~GeV and rapidities
$|y| < 2$.  We assume $W$ bosons can be reconstructed with little background
and for simplicity assume perfect $b$-tagging efficiency and no mis-tags.
We improve the purity by requiring that one of the $b$ quarks and one of the
reconstructed $W$'s reconstruct an invariant mass within $m_t \pm 10$ GeV,
though we do not assume the charge of either the $b$ or the $W$ can be
determined.
The dominant background that is independent of the $W$-$t$-$b$ coupling 
comes from diagrams with an intermediate Higgs, which can be eliminated
by subtracting events with $b \bar{b}$ that have an invariant mass
close to the Higgs mass, once the mass is known.  However, we do not
impose such a cut in order to retain the most statistics possible.

The number of events will depend strongly on the top mass, the Higgs mass,
the top width and $g_{Wtb}$.  It is expected that the ILC will determine
the top and Higgs masses to order 100 MeV or  better, 
which is enough to render the
uncertainty in the rates from these parameters much smaller than the
expected statistical uncertainties.  The remaining dependence on the width
and $g_{Wtb}$ allows us to determine a combination of both these quantities.
To illustrate the results, we assume 100~fb$^{-1}$ collected at 
$\sqrt{s} = 340$~GeV.  In Figure~2 we present the contours of
constant event numbers in the plane of $g_{Wtb}$ and $\Gamma_t$ which
reproduce the expected SM event rate of $\sim 1500$ events.
Also shown are the contours corresponding to 1$\sigma$ 
and 2$\sigma$ deviations from such a measurement (assuming that the
SM rate is observed and considering purely statistical uncertainties).  
The result is the expected bound one would obtain on $g_{Wtb}$ 
and $\Gamma_t$, which can be combined with the $\Gamma_t$ from the 
above-threshold scan to extract $g_{Wtb}$ itself (or alternately, one can
go to lower energies where the sensitivity to $\Gamma_t$ is less, though at
the price of the loss of some statistics).  From Figure~2, we 
see that assuming the width is measured with an uncertainty of 100 MeV, 
$g_{Wtb}$ can be measured to the $2\%$ level, 
which would represent better than a factor of 5 improvement compared to the
LHC, and a major improvement in our understanding of the $W$-$t$-$b$ 
interaction.

\begin{figure}[t]  
\centerline{\includegraphics[width=0.65\textwidth]{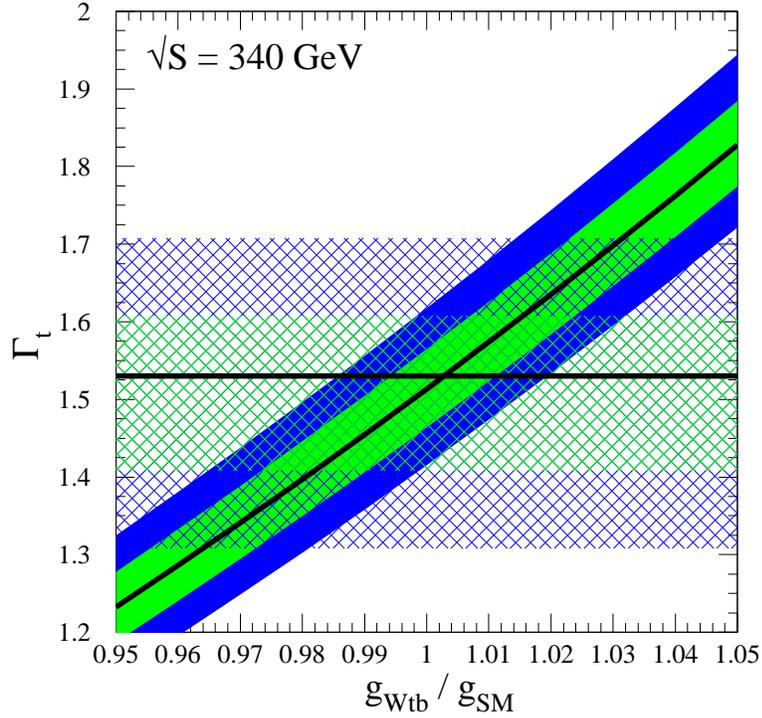}}
\caption{Curve corresponding to the SM rate and its 1$\sigma$ and 2$\sigma$
deviations in the plane of $g_{Wtb}$ and $\Gamma_t$.  Also overlaid is an
expected measurement of $\Gamma_t$ from the on-shell threshold scan with
an uncertainty of $100$ MeV.} 
\label{fig:wtb2}  
\end{figure}  

Many improvements on these rough estimates are possible.  Certainly a more
detailed and exhaustive study of the background would be interesting, as well
as more sophisticated study of the signal, including higher order corrections
and theory uncertainties, and detailed modelling of the $W$ decays and
the observability of the $b$ quarks.  In particular, higher order QCD and
EW corrections to the signal will be essential to include in a realistic 
analysis in order to obtain the desired accuracy in $g_{Wtb}$, but are not
likely to strongly change our conclusions as to how accurately one will be
able to measure the coupling.  
Finally, since the cross section is strongly energy
dependent, it could also be beneficial to consider the utility of a number of
smaller data sets at several different energies below threshold.  We leave such
refinements for future work.


\subsection{Probing Electroweak Top Quark Couplings at the ILC and the LHC\\ \small{{\it U. Baur}}}
\label{sec_Baur}

Although the top quark was discovered almost ten years
ago~\cite{topcdf,topd0}, many of its properties are still only poorly
known~\cite{Chakraborty:2003iw}.  In particular, the couplings of the
top quark to the electroweak (EW) gauge bosons have not yet been
directly measured.  
Current data provide only weak constraints on the couplings of the top
quark with the EW gauge bosons, except for the $ttZ$ vector and axial
vector couplings which are rather tightly but indirectly constrained
by LEP data; and the right-handed $tbW$
coupling, which is severely bound by the observed $b\to s\gamma$
rate~\cite{Larios:1999au}.  

At an $e^+e^-$ linear collider with $\sqrt{s}=500$~GeV and an
integrated luminosity of $100-200$~fb$^{-1}$ one can hope to measure
the $ttV$ ($V=\gamma,\,Z$) couplings in top pair production with a few-percent
precision~\cite{Abe:2001nq}.  However, the process
$e^+e^-\to\gamma^*/Z\to t\bar{t}$ is sensitive to both $tt\gamma$
and $ttZ$ couplings and significant cancellations between the various
couplings can occur.  At hadron colliders, $t\bar{t}$ production is so
dominated by the QCD processes $q\bar{q}\to g^*\to t\bar{t}$ and
$gg\to t\bar{t}$ that a measurement of the $tt\gamma$ and $ttZ$
couplings via $q\bar{q}\to\gamma^*/Z^*\to t\bar{t}$ is hopeless.
Instead, the $ttV$ couplings can be measured in QCD $t\bar{t}\gamma$
production, radiative top quark decays in $t\bar{t}$ events
($t\bar{t}\to\gamma W^+W^- b\bar{b}$), and QCD $t\bar{t}Z$
production~\cite{Baur:2004uw}. $t\bar{t}\gamma$ production and radiative
top quark decays 
are sensitive only to the $tt\gamma$ couplings, whereas $t\bar{t}Z$
production gives information only on the structure of the $ttZ$
vertex.  This obviates having to disentangle potential cancellations
between the different couplings.
In this section we briefly review the measurement of the $ttV$
couplings at the LHC and compare the expected sensitivities with the
bounds which one hopes to achieve at an $e^+e^-$ linear collider. 

The most general Lorentz-invariant vertex function describing the
interaction of a neutral vector boson $V$ with two top quarks can be
written in terms of ten form factors~\cite{Hollik:1998vz}, which are
functions of the kinematic invariants.  In the low energy limit,
these correspond to couplings which multiply dimension-four or -five 
operators in an effective Lagrangian, and may be complex.  If $V$ is 
on-shell, or if $V$ couples to effectively massless fermions, the 
number of independent form factors is reduced to eight.  If, in 
addition, both top quarks are on-shell, the number is further reduced 
to four.  In this case, the $ttV$ vertex can be written in the form
\begin{equation}\label{eq:anomvertex}
\Gamma_\mu^{ttV}(k^2,\,q,\,\bar{q}) = -ie \left\{
  \gamma_\mu \, \left( F_{1V}^V(k^2) + \gamma_5F_{1A}^V(k^2) \right)
+ \frac{\sigma_{\mu\nu}}{2m_t}~(q+\bar{q})^\nu 
   \left( iF_{2V}^V(k^2) + \gamma_5F_{2A}^V(k^2) \right)
\right\} \, ,
\end{equation}
where $e$ is the proton charge, 
$m_t$ is the top quark mass, $q~(\bar{q})$ is the outgoing top
(anti-top) quark four-momentum, and $k^2=(q+\bar{q})^2$.  The terms
$F_{1V}^V(0)$ and $F_{1A}^V(0)$ in the low energy limit are the $ttV$ 
vector and axial vector form factors.  The coefficients 
$F_{2V}^\gamma(0)$ and $F_{2A}^\gamma(0)$ are related to the magnetic 
and ($CP$-violating) electric dipole form factors.

In $t\bar{t}V$ production, one of the top quarks coupling to $V$ is
off-shell.  The most general vertex function relevant for $t\bar{t}V$
production thus contains additional couplings, not included in
Eq.~(\ref{eq:anomvertex}). These additional couplings are irrelevant
in $e^+e^-\to t\bar{t}$, where both top quarks are on-shell. 

In $e^+e^-\to t\bar{t}$ one often uses the following parameterization for
the $ttV$ vertex:
\begin{equation}\label{eq:gordon}
\Gamma_\mu^{ttV}(k^2,\,q,\,\bar{q}) = ie \left\{
  \gamma_\mu \, \left(  \widetilde{F}_{1V}^V(k^2)
                      + \gamma_5\widetilde{F}_{1A}^V(k^2) \right)
+ \frac{(q-\bar{q})_\mu}{2m_t}
    \left(  \widetilde{F}_{2V}^V(k^2)
          + \gamma_5\widetilde F_{2A}^V(k^2) \right)
\right\} .
\end{equation}
Using the Gordon decomposition, it is easy to show that
Eqs.~(\ref{eq:anomvertex}) and~(\ref{eq:gordon}) are equivalent for
on-shell top quarks and that the form
factors $\widetilde F^V_{iV,A}$ and $F^V_{iV,A}$ ($i=1,\,2$) are
related by
\begin{equation}
\label{eq:rel1}
\widetilde F^V_{1V} = -\left( F^V_{1V}+F^V_{2V} \right) \, , \qquad
\widetilde F^V_{2V}  =  F^V_{2V} \, , \qquad
\widetilde F^V_{1A} = -F^V_{1A} \, , \qquad
\widetilde F^V_{2A} =  -iF^V_{2A} \, .
\label{eq:rel4}
\end{equation}

The most promising channel for measuring the $tt\gamma$ couplings at the
LHC is $pp\to \gamma\ell\nu_\ell b\bar{b}jj$ which receives
contributions from $t\bar t\gamma$ production and ordinary $t\bar t$
production where one of the top quarks decays radiatively, $t\to
Wb\gamma$. In order to reduce the background, it is advantageous to
require that both $b$-quarks are tagged. We assume a combined efficiency
of $\epsilon_b^2=40\%$ for tagging both $b$-quarks.

The non-resonant $pp \to W(\to\ell\nu)\gamma b\bar{b}jj$ background and
the single-top backgrounds, $(t\bar{b}\gamma + \bar{t}b\gamma)+X$, can
be suppressed by imposing invariant and transverse mass cuts which
require that the event is consistent either 
with $t\bar{t}\gamma$ production, or with $t\bar{t}$ production with
radiative top decay~\cite{Baur:2004uw}. Imposing a large
separation cut of $\Delta R(\gamma,b)>1$ reduces photon
radiation from the $b$ quarks.  Photon emission from $W$ decay
products can essentially be eliminated by requiring that
$m(jj\gamma) > 90~{\rm GeV}$ and $
m_T(\ell\gamma;p\llap/_T) > 90~{\rm GeV,}$
where $m(jj\gamma)$ is the invariant mass of the $jj\gamma$ system, and
$m_T(\ell\gamma;p\llap/_T)$ is the $\ell\gamma p\llap/_T$ cluster
transverse mass, which peaks sharply at $m_W$. 
After imposing the cuts described above, the irreducible backgrounds are
one to two orders of magnitude smaller than the signal.

The potentially most dangerous reducible background is $t\bar{t}j$ 
production where one of the jets in the final state fakes a photon.
\begin{figure}
\begin{center}
\begin{tabular}{cc}
\includegraphics[width=8.7cm]{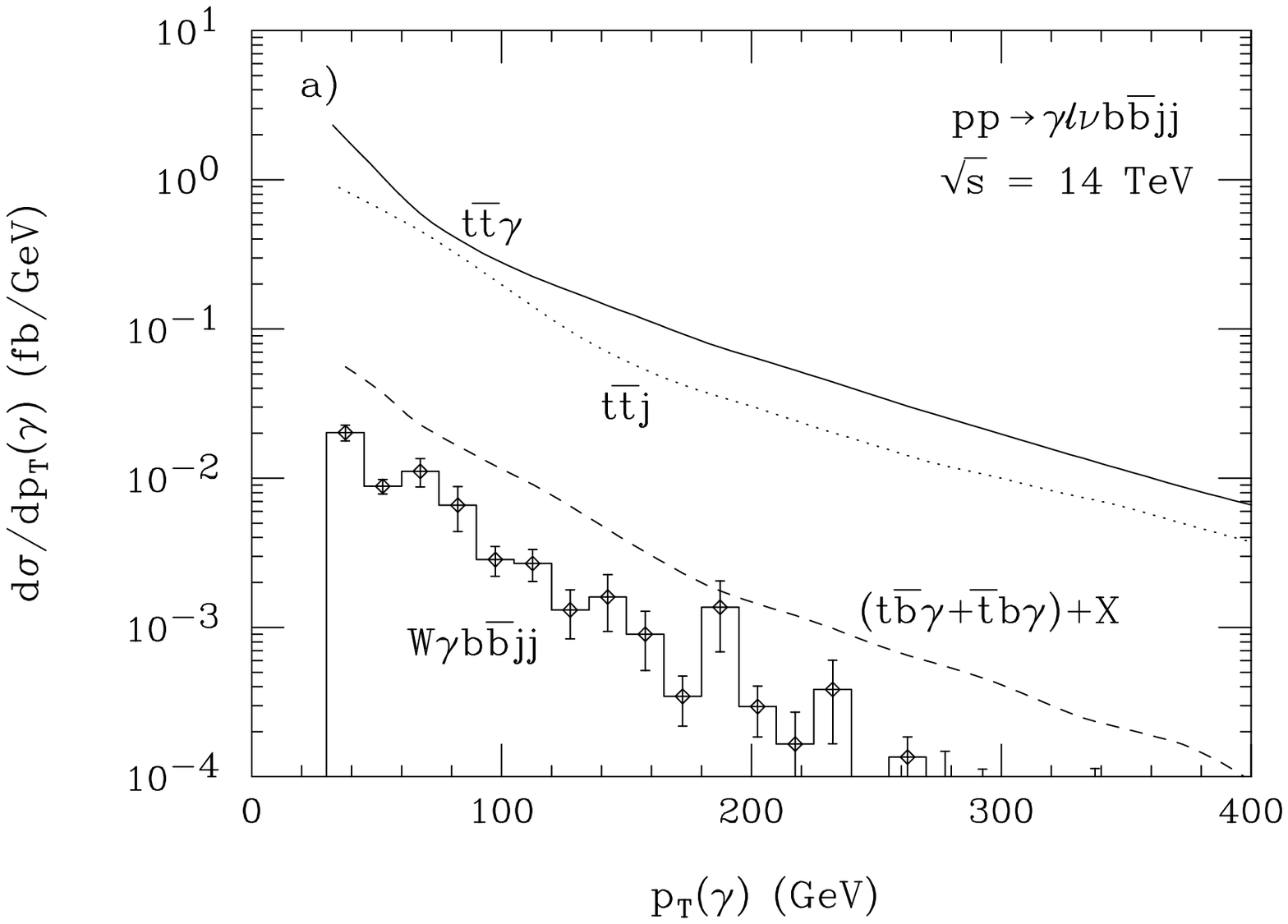} &
\includegraphics[width=8.7cm]{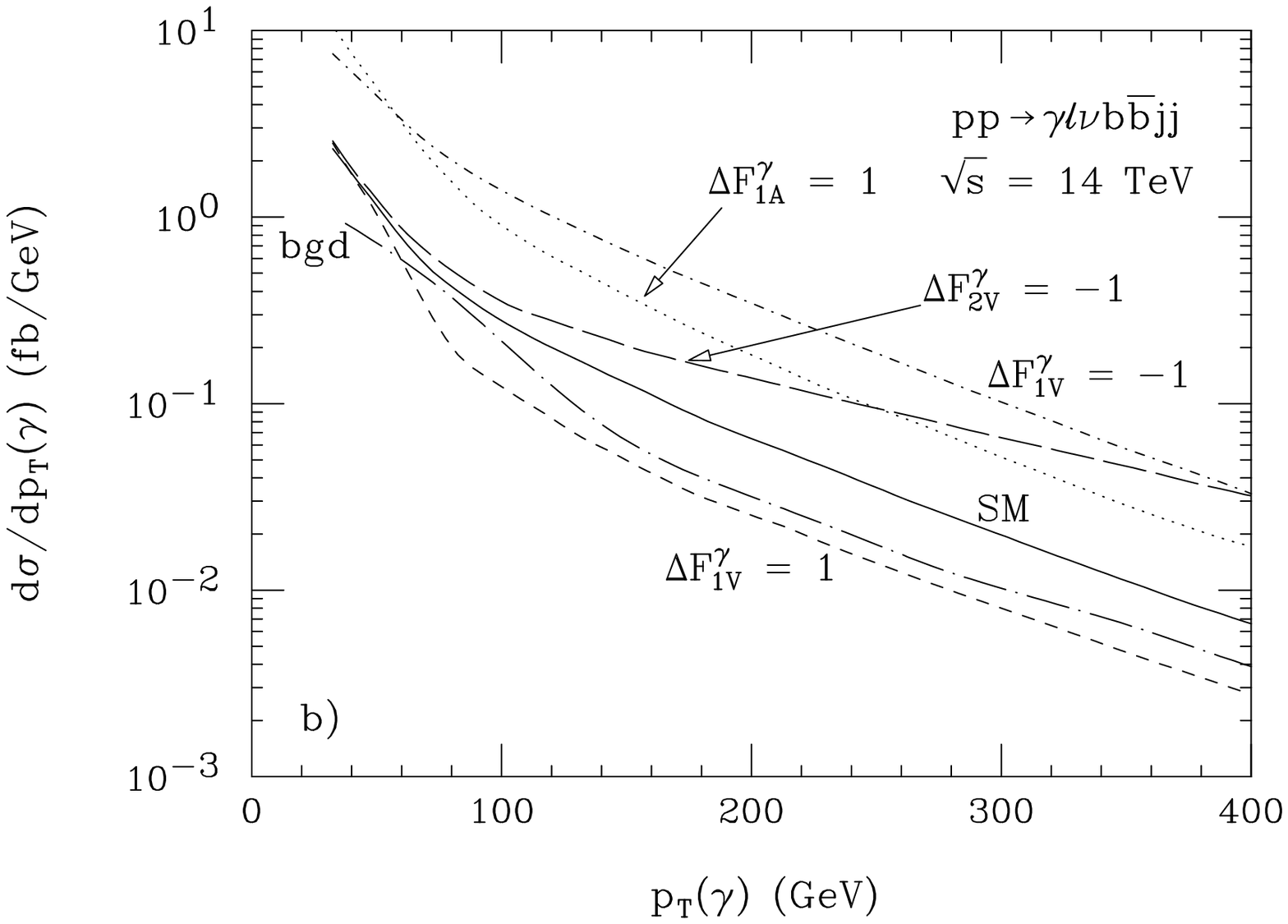}
\end{tabular}
\end{center}
\caption{\label{fig:fig1}{The differential cross sections as a function
of the photon  transverse momentum for $\gamma\ell\nu_\ell b\bar{b}jj$
production at the LHC. Part a) shows the SM signal and the various
contributions to the background. Part b) shows the SM signal and
background, and the signal for various anomalous $tt\gamma$ couplings.}}
\end{figure}
In Fig.~\ref{fig:fig1}a we show the photon transverse momentum
distributions of the $t\bar{t}\gamma$ signal and the backgrounds
discussed above. The $t\bar{t}j$ background is seen to
be a factor~2 to~3 smaller than the $t\bar{t}\gamma$ signal for the
jet-photon misidentification probability
($P_{j\to\gamma}=1/1600$~\cite{atlas_tdr}) used.  

The photon transverse momentum distributions in the SM and for various
anomalous $tt\gamma$ couplings, 
together with the $p_T(\gamma)$ distribution of the
background, are shown in
Fig.~\ref{fig:fig1}b. Only one coupling at a time is allowed to deviate
from its SM prediction. 

The process $pp \to t\bar{t}Z$ leads
to either ${\ell'}^+{\ell'}^-\ell\nu b\bar{b}jj$ or
${\ell'}^+{\ell'}^- b\bar{b}+4j$ final states if the $Z$-boson
decays leptonically and one or both of the $W$ bosons decay
hadronically. If the $Z$ boson decays into neutrinos and both $W$ bosons
decay hadronically, the final state consists of
$p\llap/_Tb\bar{b}+4j$. Since there is essentially no phase space for
$t\to WZb$ decays ($BR(t\to WZb)\approx 3\cdot
10^{-6}$~\cite{Mahlon:1994us}), these final states
arise only from $t\bar tZ$ production. 

In order to identify leptons, $b$ quarks, light jets and the missing
transverse momentum in dilepton and trilepton events, the same
cuts as for $t\bar t\gamma$ production are imposed. One also requires that
there is a same-flavor, opposite-sign lepton pair with 
invariant mass near the $Z$ resonance, $m_Z - 10~{\rm GeV} < m(\ell\ell)
< m_Z + 10~{\rm GeV}$. 

The main backgrounds contributing to the trilepton final state are
singly-resonant $(t\bar{b}Z+\bar{t}bZ)+X$ ($t\bar{b}Zjj$,
$\bar{t}bZjj$, $t\bar{b}Z\ell\nu$ and $\bar{t}bZ\ell\nu$) and
non-resonant $WZb\bar{b}jj$ production. In the dilepton case, the main
background arises from 
$Zb\bar{b}+4j$ production.  To adequately suppress it, one
additionally requires that events have at least one combination of jets
and $b$ quarks which is consistent with the $b\bar b+4j$ system
originating from a $t\bar t$ system. Once these cuts have been imposed,
the $Zb\bar{b}+4j$ background is important only for $p_T(Z)<100$~GeV. 

\begin{figure}
\begin{center}
\begin{tabular}{cc}
\includegraphics[width=8.7cm]{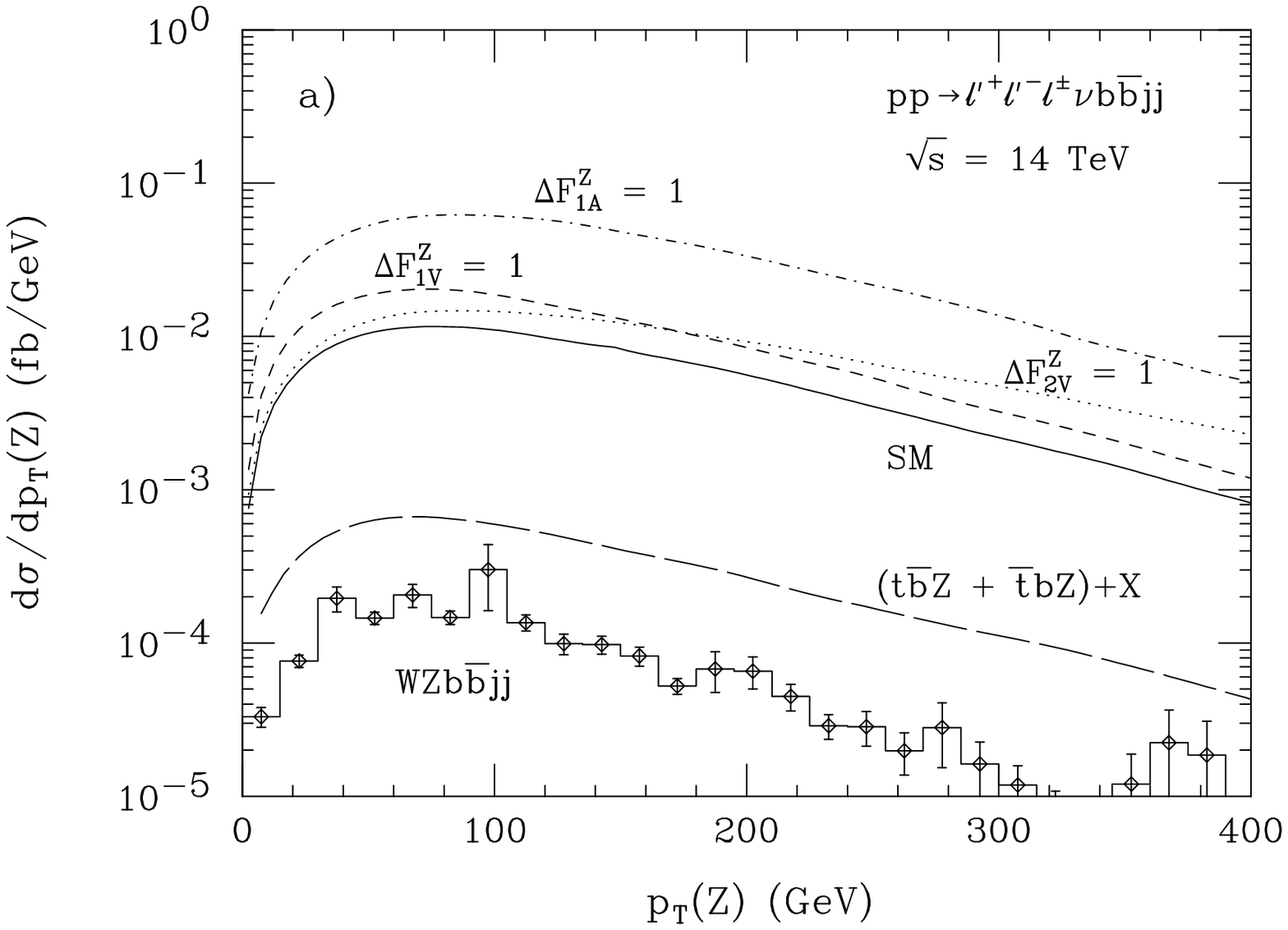} &
\includegraphics[width=8.7cm]{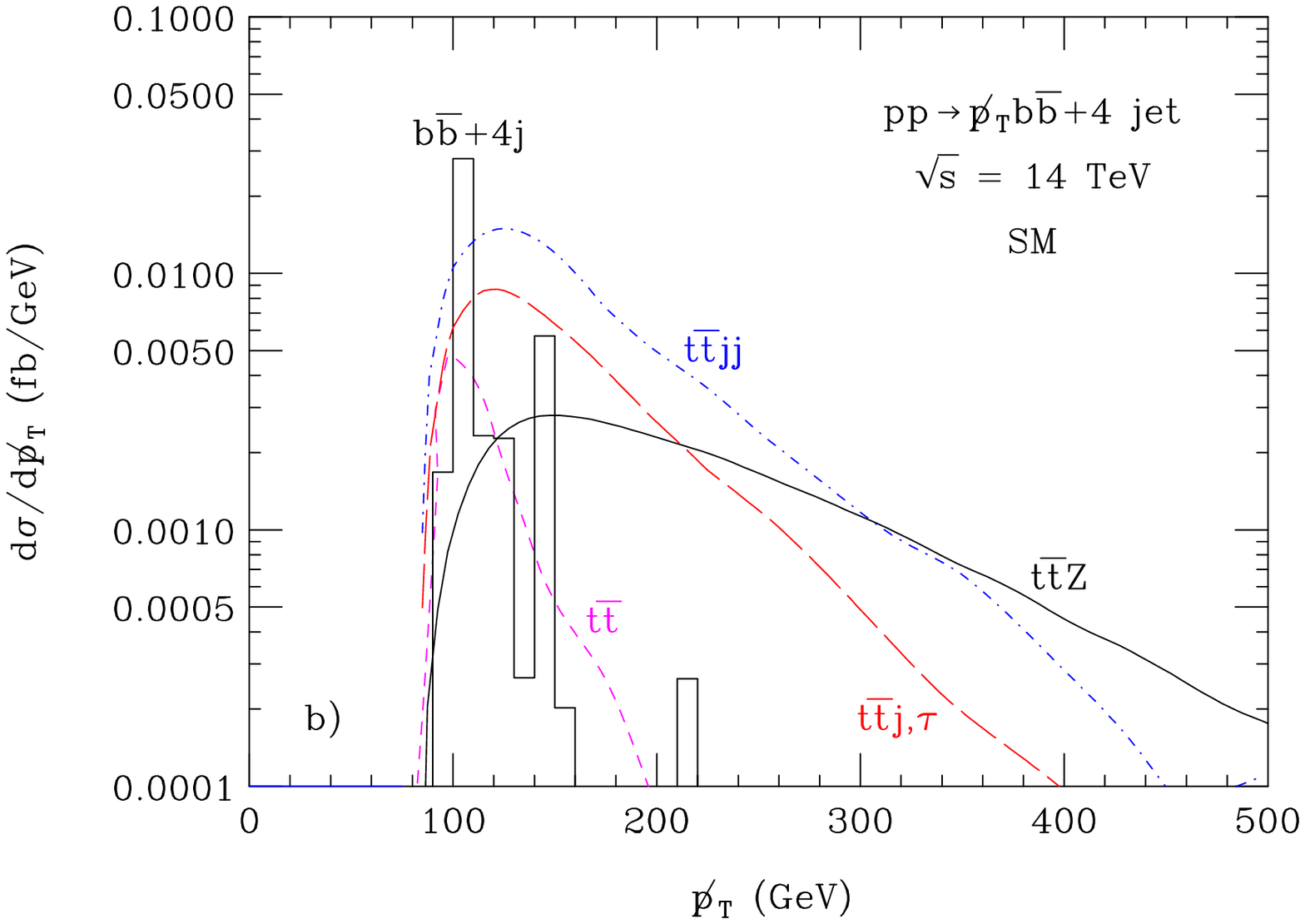}
\end{tabular}
\end{center}
\caption{\label{fig:fig2}{a) The differential cross sections at the LHC
as a function 
of $p_T(Z)$ for ${\ell'}^+{\ell'}^-\ell\nu b\bar{b}jj$ final states.  
Shown are the SM predictions for $t\bar{t}Z$ production, for several
non-standard $ttZ$ couplings, and for various
backgrounds.  Only one coupling at a time is allowed to deviate from 
its SM value. 
b) The differential cross sections as a function of the missing
transverse momentum for $p\llap/_Tb\bar b+4$j production at the
LHC.  Shown are the SM predictions for
$t\bar{t}Z$ production and for various backgrounds. }}
\end{figure}
The $Z$ boson transverse momentum distribution for the trilepton final
state is shown in
Fig.~\ref{fig:fig2}a for the SM signal and backgrounds, as well as for 
the signal with several non-standard $ttZ$ couplings. Only one coupling
at a time is allowed to deviate from its SM 
prediction.  The backgrounds are each more than one order of magnitude
smaller than the SM signal. Note that varying
$F^Z_{1V,A}$ leads mostly to a cross section 
normalization change, hardly affecting the shape of the $p_T(Z)$
distribution.  

For the $p\llap/_Tb\bar{b}+4j$~\cite{new} final state at least 3~jets 
with $p_T>50$~GeV and $p\llap/_T>5~{\rm GeV}^{1/2}\sqrt{\sum p_T}$ are
required. The 
largest backgrounds for this final state come from $t\bar t$ and $b\bar
b+4j$ production where one or several jets are badly mismeasured, from
$pp\to t\bar tjj$ with $t\bar 
t\to\ell^\pm\nu_\ell b\bar bjj$ and the charged lepton
being missed, and from $t\bar tj$ production, where one top decays
hadronically, $t\to Wb\to bjj$, and the other via $t\to Wb\to\tau\nu_\tau b$
with the $\tau$-lepton decaying hadronically, $\tau\to h\nu_\tau$. 

In Fig.~\ref{fig:fig2}b we show the missing transverse momentum
distributions of the SM $t\bar tZ\to p\llap/_Tb\bar{b}+4j$ signal (solid
curve) and various 
backgrounds. The most important backgrounds are $t\bar tjj$
and $t\bar tj$ production. However, the missing transverse momentum
distribution from these processes falls considerably faster than that of
the signal, and for $p\llap/_T>300$~GeV, the SM signal dominates. 

The shape and normalization changes of the photon or $Z$-boson
transverse momentum distribution can be used to derive
quantitative sensitivity bounds on the anomalous $tt\gamma$ and $ttZ$
couplings. For $t\bar{t}Z$ production with $Z\to{\ell'}^+{\ell'}^-$, the
$\Delta\Phi({\ell'}{\ell'})$ 
distribution provides additional information~\cite{Baur:2004uw}. In the
following we assume a normalization uncertainty of the SM cross
section of $\Delta{\cal N}=30\%$. 

Even for a modest integrated luminosity of 30~fb$^{-1}$, it will be 
possible to measure the
$tt\gamma$ vector and axial vector couplings, and the dipole form
factors, with a precision of typically $20\%$ and $35\%$,
respectively. For 300~fb$^{-1}$, the limits improve to $4-7\%$ for
$F^\gamma_{1V,A}$ and to about $20\%$ for $F^\gamma_{2V,A}$. At the
SLHC, assuming an integrated luminosity of 3000~fb$^{-1}$, one can
hope to achieve a $2-3\%$ measurement of the vector and axial vector
couplings, and a $10\%$ measurement of $F^\gamma_{2V,A}$, provided
that particle identification efficiencies are not substantially
smaller, and the reducible backgrounds not much larger, than what we
have assumed.

To extract bounds on the $ttZ$ couplings, we perform a simultaneous
fit to the $p_T(Z)$ and the $\Delta\Phi({\ell'}{\ell'})$
distributions for the trilepton and dilepton final states, and to the
$p\llap/_T$ distribution for the $p\llap/_Tb\bar b+4j$ final state. We
calculate sensitivity bounds for 
300~fb$^{-1}$ and 3000~fb$^{-1}$ at the LHC; for 30~fb$^{-1}$ the
number of events expected is too small to yield meaningful results.
For an integrated luminosity of 300~fb$^{-1}$, it will be possible to
measure the $ttZ$ axial vector coupling with a precision of $10-12\%$,
and $F^Z_{2V,A}$ with a precision of $40\%$.  At the SLHC, these
bounds can be improved by factors of about~1.6 ($F^Z_{2V,A}$)
and~3 ($F^Z_{1A}$).  The bounds which can be achieved for
$F^Z_{1V}$ are much weaker than those projected for $F^Z_{1A}$.  As
mentioned in Sec.~4, the $p_T(Z)$ distributions for the
SM and for $F^Z_{1V,A}=-F^{Z,SM}_{1V,A}$ are almost degenerate.
This is also the case for the $\Delta\Phi({\ell'}{\ell'})$
distribution.  In a fit to these two distributions, therefore, an area
centered at $\Delta F^Z_{1V,A}=-2F^{Z,SM}_{1V,A}$ remains which cannot
be excluded, even at the SLHC.  For $F^Z_{1V}$, the two regions merge,
resulting in rather poor limits.  

\begin{table}[t]
\begin{center}
\caption{Sensitivities achievable at $68.3\%$ CL for the anomalous 
$ttV$ ($V=\gamma,\,Z$) couplings $\widetilde F^V_{1V,A}$ and
$\widetilde F^V_{2V,A}$ of Eq.~(\ref{eq:gordon}) at the LHC for
integrated luminosities of 300~fb$^{-1}$, and the ILC with
$\sqrt{s}=500$~GeV (taken from 
Ref.~\protect\cite{Abe:2001nq}).  Only one coupling at a time is
allowed to deviate from its SM value.\\}
\begin{tabular}{ccc|ccc}
\hline 
 coupling & LHC, 300~fb$^{-1}$ & $e^+e^-$~\protect\cite{Abe:2001nq} &
coupling & LHC, 300~fb$^{-1}$ & $e^+e^-$~\protect\cite{Abe:2001nq} \\
\hline
$\Delta\widetilde F^\gamma_{1V}$ & $\begin{matrix}{ +0.043 \\[-4pt]
-0.041}\end{matrix}$ & $\begin{matrix}{ +0.047 \\[-4pt]
-0.047}\end{matrix}$ , 200~fb$^{-1}$ & $\Delta\widetilde F^Z_{1V}$ &
$\begin{matrix} {+0.24 \\[-4pt] 
-0.62}\end{matrix}$ & $\begin{matrix} {+0.012 \\[-4pt]
-0.012}\end{matrix}$ , 200~fb$^{-1}$ 
\\
$\Delta\widetilde F^\gamma_{1A}$ & $\begin{matrix} {+0.051 \\[-4pt]
-0.048}\end{matrix}$ & $\begin{matrix} {+0.011 \\[-4pt]
-0.011}\end{matrix}$ , 100~fb$^{-1}$  & $\Delta\widetilde F^Z_{1A}$ &
$\begin{matrix} {+0.052 \\[-4pt] 
-0.060}\end{matrix}$ & $\begin{matrix} {+0.013 \\[-4pt]
-0.013}\end{matrix}$ , 100~fb$^{-1}$ 
\\
$\Delta\widetilde F^\gamma_{2V}$ & $\begin{matrix} {+0.038 \\[-4pt]
-0.035}\end{matrix}$ & $\begin{matrix}{ +0.038 \\[-4pt]
-0.038}\end{matrix}$ , 200~fb$^{-1}$  & $\Delta\widetilde F^Z_{2V}$ &
$\begin{matrix} {+0.27 \\[-4pt] 
-0.19}\end{matrix}$ & $\begin{matrix} {+0.009 \\[-4pt]
-0.009}\end{matrix}$ , 200~fb$^{-1}$ 
\\
$\Delta\widetilde F^\gamma_{2A}$ & $\begin{matrix} {+0.16 \\[-4pt]
-0.17}\end{matrix}$ & $\begin{matrix} {+0.014 \\[-4pt]
-0.014}\end{matrix}$ , 100~fb$^{-1}$   & $\Delta\widetilde F^Z_{2A}$ &
$\begin{matrix} {+0.28 \\[-4pt] 
-0.27}\end{matrix}$ & $\begin{matrix} {+0.052 \\[-4pt]
-0.052}\end{matrix}$ , 100~fb$^{-1}$  
\\     
\hline
\end{tabular}
\label{tab:tab1}
\end{center}
\end{table}
It is instructive to compare the bounds for anomalous $ttV$ couplings
achievable at the LHC with those projected for the ILC. The most
complete study of $t\bar{t}$ production at the ILC for general $ttV$
($V=\gamma,\,Z$) couplings so far is 
that of Ref.~\cite{Abe:2001nq}.  It uses the parameterization of
Eq.~(\ref{eq:gordon}) for the $ttV$ vertex function.  In order to
compare the bounds of Ref.~\cite{Abe:2001nq} with those anticipated
at the LHC, the limits on $F^V_{1V,A}$ and $F^V_{2V,A}$ have to be
converted into bounds on $\widetilde 
F^V_{1V,A}$ and $\widetilde F^V_{2V,A}$. Table~\ref{tab:tab1} compares
the bounds we obtain for $\widetilde F^V_{1V,A}$ and $\widetilde
F^V_{2V,A}$ with those reported for the ILC in
Ref.~\cite{Abe:2001nq}. Note that only one coupling at a time is allowed
to deviate from its SM value~\cite{Abe:2001nq}.
We show LHC limits only for an integrated luminosity of
300~fb$^{-1}$. For the SLHC, with 3000~fb$^{-1}$, we obtain bounds which
are a factor $1.3-3$ more stringent than those shown in
Table~\ref{tab:tab1}. Thus, even if the SLHC operates first, and the
$p\llap/_Tb\bar b+4j$ final state is taken into account, a linear
collider will still be able to significantly improve the $ttZ$ anomalous
coupling 
limits, with the possible exception of $\widetilde F^Z_{1A}$. The ILC
will also be able to considerably strengthen the bounds on $\widetilde
F^\gamma_{1A}$ and $\widetilde F^\gamma_{2A}$. It should
be noted, however, that this 
picture could change once correlations between different non-standard
$ttZ$ couplings, and between $tt\gamma$ and $ttZ$ couplings, are taken
into account. Unfortunately, so far no realistic studies for $e^+e^-\to
t\bar t$ which include these correlations have been performed\footnote{However, see
Ref.~\cite{Rindani:2003av} for limits on the $CP$-violating couplings.}.

The LHC will be able to perform first tests of the $ttV$ couplings. 
Already with an
integrated luminosity of 30~fb$^{-1}$, one can probe the $tt\gamma$ couplings
with a precision of about $10-35\%$ per experiment.  With higher
integrated luminosities one will be able to reach the few percent
region. With the exception of $F^Z_{1A}$, the $ttZ$ couplings can only be
measured with a precision of $15-50\%$, even at the SLHC. 
The ILC will
be able to further improve our knowledge of the $ttV$ couplings, in
particular in the $ttZ$ case.


\subsection{Two-loop Corrections to Heavy Quark Form Factors \\ \small{{\it T. Gehrmann}}}
\label{sg2}

The international linear collider will produce large numbers of 
top-antitop quark pairs, thus allowing for precision studies of the 
properties of the top quark. These experimental 
precision studies require equally precise theoretical predictions, i.e.\
higher order corrections in perturbation theory. Up to now, 
the theoretical effort was focused on a precise description of top
quark production at threshold (see~\cite{hoangteubner} for a review),
where QCD corrections to next-to-next-to-leading order (NNLO) in
perturbation theory are known, while observables other than the 
total production cross section in the continuum 
are known only to next-to-leading order (NLO) accuracy. 
In this section, we present results on the virtual two-loop corrections to 
vertex functions involving heavy quarks, which are an important 
ingredient to the NNLO 
corrections to top quark observables in the continuum. 

The vertex function coupling an on-shell heavy quark antiquark pair to an 
external current can be decomposed into so-called form factors, whose 
coefficients follow from Lorentz invariance and symmetry properties of the 
current. For the vector and axial vector current (electroweak
gauge boson),
the vertex function contains, within QCD, four form factors ($F_{1,2}$, 
$G_{1,2}$):
\begin{eqnarray*}
\parbox{1.5cm}{\includegraphics[width=1.4cm]{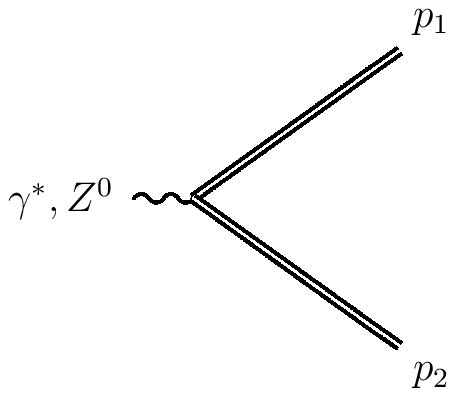}}
&{ = }&{ (-i)  \left(v_Q { F_1(s,m^2)}\gamma^\mu +
  v_Q\frac{1}{2m}{ F_2(s,m^2)} i \sigma^{\mu \nu} (p_1+p_2)_\nu \right. } \nonumber \\
&& \phantom{(-i)}{ \left. + a_Q { G_1(s,m^2)} 
\gamma^\mu \gamma_5 + a_Q
  \frac{1}{2m} { G_2(s,m^2)} \gamma_5 (p_1+p_2)^\mu \right) \;.
}
\end{eqnarray*}
The coupling of heavy quarks to Higgs bosons of positive and 
negative parity contains the 
scalar and pseudoscalar form factors: 
\begin{displaymath}
\parbox{1.5cm}{\includegraphics[width=1.4cm]{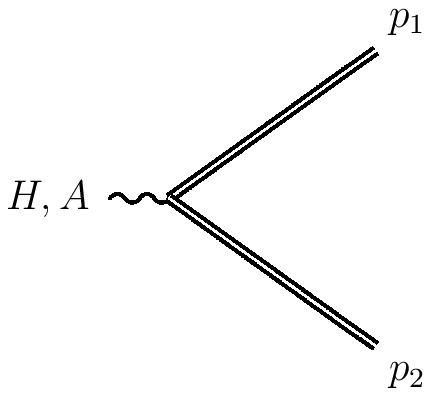}}
{ =   - i \,
\frac{m}{v} \,  
\left[ S_Q { F_{S}(s,m^2)} + i P_Q 
  { F_{P}(s,m^2)} \gamma_5 \right]}.
\end{displaymath}
Here $s$ is the invariant momentum squared of the external current and 
$m$ the heavy quark mass. 

\begin{figure*}[t]
\centering
\includegraphics[width=120mm]{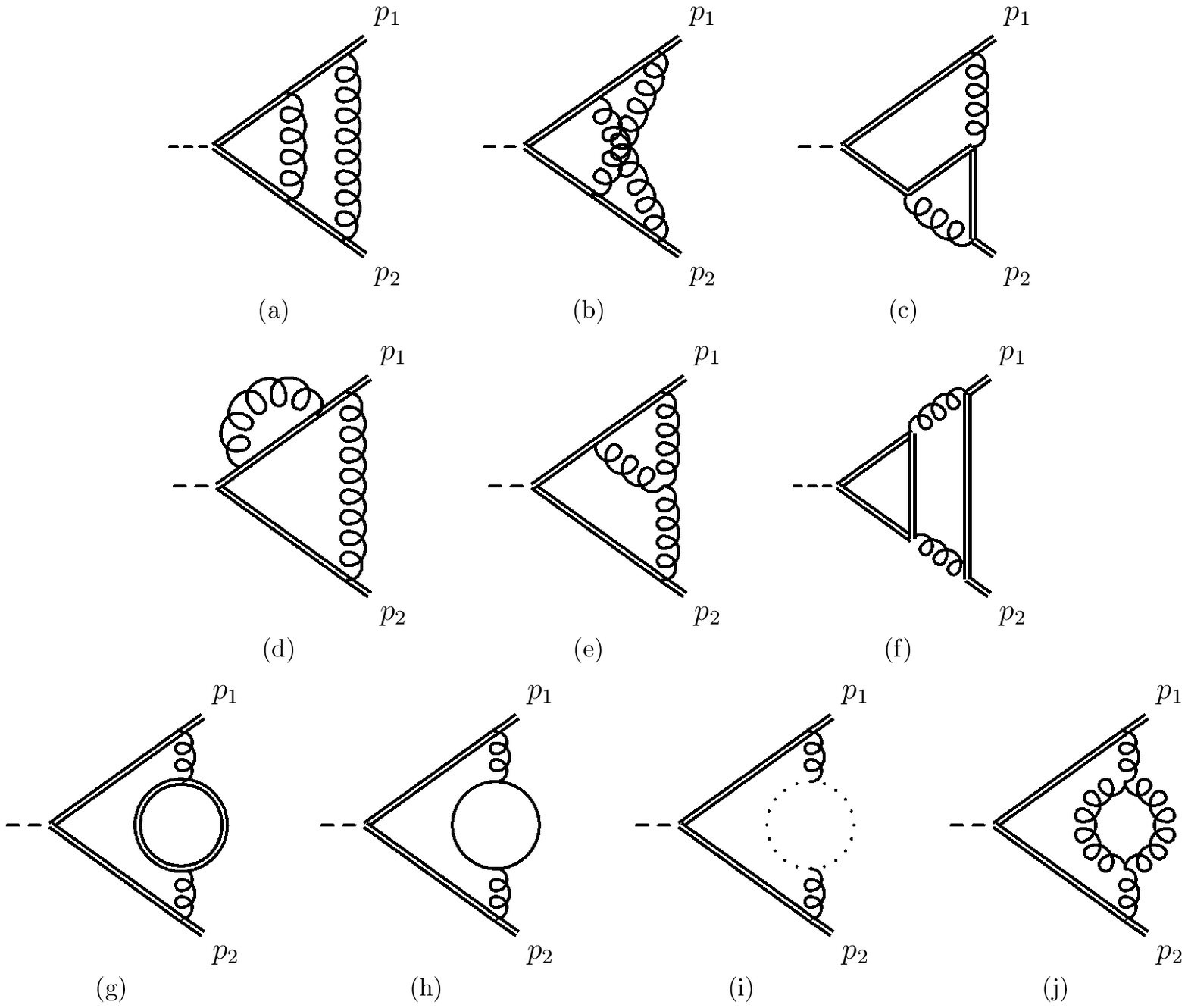}
\caption{Feynman diagrams contributing to 
the two-loop QCD corrections to heavy quark form factors} \label{fd2l}
\end{figure*}
The two-loop corrections to the form factors are obtained 
by applying appropriate projections to the Feynman diagrams 
displayed in Figure~\ref{fd2l}. As a result, the form factors are 
expressed in terms of
hundreds of different scalar integrals. These integrals are reduced to
a small set of master integrals by means of the so-called Laporta
algorithm \cite{Lap} with the help of integration-by-parts identities
\cite{Chet} and Lorentz-invariance identities \cite{Rem3}. 
The master integrals themselves were
evaluated with the method of differential equations
\cite{Rem3,Kot,Rem1,Rem2} in \cite{RoPieRem1,RoPieRem3}. The master
integrals, and thus the form factors are represented as series in the
regularization parameter $\epsilon$ and expressed in terms of
1-dimensional harmonic polylogarithms up to weight 4
\cite{Polylog,Polylog3}.

We obtained the complete two-loop corrections
 for the renormalized vector~\cite{Bernreuther:2004ih},
axial-vector \cite{Bernreuther:2004th,Bernreuther:2005rw}, 
scalar and pseudoscalar~\cite{scalar}
form factors. Agreement was found with 
earlier partial results and with expansions around special kinematical points.

An immediate physical application of the QCD corrections to the 
heavy quark form factors for arbitrary momentum transfer are predictions 
for form factors at zero recoil, the so-called static form factors.

Measurements of these static form factors for heavy quarks could allow indirect 
constraints on new physics scenarios. 
Recently, there has been considerable effort to determine the 
feasibility of such experimental measurements. 
Specifically, the couplings to photons and
$Z$ bosons have been studied
in detail -- both for heavy quark production at hadron colliders
\cite{Baur:2004uw,new,Baur2}
and at a future high-luminosity high-energetic linear electron-positron
collider \cite{tesla,Abe:2001nq,new,Baur3}. At this workshop, this 
issue was revisited in great depth
especially in view of discriminating new physics scenarios
(see sections 4.1 and 4.2).

The most prominent static 
form factor is the electromagnetic spin-flipping form factor: the 
anomalous magnetic moment, which 
is finite in 
the zero-recoil limit, and can be 
obtained from the results of the  previous section~\cite{static}. It reads
\begin{displaymath}
F_{2,Q}(s=0) =
\frac{\alpha_s}{2\pi}\, C_F
+
\left(\frac{\alpha_s}{2\pi}\right)^2
F_{2,Q}^{(2l)} \, ,
\end{displaymath}
with
\begin{eqnarray*}
 F_{2,Q}^{(2l)}  &=&
 \phantom{+}C_F^2\left(
  -\frac{31}{4}+2\,\zeta_2\,(5-6\,\ln(2))+3\,\zeta_3\right)
 +C_F\,C_A\left( \frac{317}{36}
  +3\,\zeta_2\,(-1+2\,\ln(2))-\frac{3}{2}\,\zeta_3\right)\nonumber
\\&&
+C_F\,T_F\left(\frac{119}{9}-8\,\zeta_2\right) 
- \frac{25}{9} C_F\,T_F\,N_l + C_F\,\beta_0\,\ln({\mu^2}/{m_Q^2}),
\end{eqnarray*}
from which we can derive the  static  magnetic and weak magnetic form
factor of a quark $Q$. We consider
\begin{displaymath}
\left(\frac{g-2}{2}\right)^{\gamma,Z}_Q \equiv F_{2,Q}^{\gamma,Z}
\left(0 \right) =
{v}_Q^{\gamma,Z}\,F_{2,Q}\left(0 \right) \, ,
\end{displaymath}
which correspond to the anomalous magnetic (MDM) and weak magnetic
(WMDM) moments of $Q$.
(Notice that in the literature the WMDM is often
associated with  $F_{2,Q}^Z(s=m_Z^2)$.) 
Numerical values for $t$ and $b$ quarks are given in Table~\ref{tab_anomag}.
\begin{table}[t]
\begin{tabular}{cccc}
& $t$ $(\mu=m_t)$ & $b$ $(\mu=m_b)$ & $b$
$(\mu=m_Z)$\\ \hline
$(g-2)^{\gamma,(1l)}_Q/2$ & $1.53\cdot 10^{-2}$ & $-1.52\cdot10^{-2}$ & $-8.4\cdot10^{-3}$ \\
 $(g-2)^{\gamma,(2l)}_Q/2$ & $4.7\cdot 10^{-3}$  & $-1.00\cdot10^{-2}$
 & $-6.6\cdot10^{-3}$ \\ \hline
 $(g-2)^{\gamma}_Q/2$      & $2.00\cdot 10^{-2}$ & $-2.52\cdot10^{-2}$
 & $-1.50\cdot10^{-2}$ \\ \hline
 $(g-2)^{Z,(1l)}_Q/2$      & $5.2\cdot 10^{-3}$  & $-1.87\cdot10^{-2}$ & $-1.03\cdot10^{-2}$ \\
 $(g-2)^{Z,(2l)}_Q/2$      & $1.6\cdot 10^{-3}$  & $-1.24\cdot10^{-2}$ & $-8.1\cdot10^{-3}$ \\ \hline
 $(g-2)^{Z}_Q/2$           & $6.8\cdot 10^{-3}$  & $-3.11\cdot10^{-2}$ & $-1.85\cdot10^{-2}$ \\
\end{tabular}
\caption{
One- and two-loop QCD contributions, and their sums, 
to the anomalous magnetic and weak magnetic moments of the top and bottom
quark,
for different values of the renormalization scale $\mu$.
 \label{tab_anomag}}
\end{table}

 For the $b$ quark
an upper bound  on its magnetic moment 
 was derived in \cite{Escribano:1993xr} from an analysis of LEP1
 data, which, in our
 convention, reads $|\delta (g-2)^{\gamma}_b/2| < 1.5 \times
 10^{-2}$ (68 \% C.L.). Comparing it with Table~\ref{tab_anomag}
 we see that the QCD-induced
 contributions to the $b$ quark magnetic moment saturate this bound, which
 implies that there is limited room for new physics contributions having the same sign as the QCD contributions to
this quantity.  At a future linear collider \cite{tesla,Abe:2001nq},
when operated at the $Z$ resonance, the sensitivity to this variable
 could be improved
substantially, either by global fits or by analyzing appropriate
angular distributions in $b {\bar b}$ and $b {\bar b} \gamma$ events. 

As to the static form factors of the  top quark, no such tight
constraints exist so far on possible contributions from new
interactions (see Ref.~\cite{Baur:2004uw} for a review). 
These quantities are particularly
sensitive to the dynamics of electroweak symmetry breaking. For
instance, in various models with a strongly coupled symmetry breaking
sector  one may expect contributions from this sector to the static $t$ quark 
form factors at the  5 - 10\% level~\cite{Berger:2005ht}. The QCD-induced
anomalous magnetic moment and the QCD corrections to the axial charge
of the top quark are of the same order of magnitude. Future colliders
have the potential to reach this level of sensitivity.

The form factors presented here have a number of applications, which will be 
addressed in future work. The vector and axial vector form factors 
contribute to the NNLO corrections to the forward-backward asymmetry for
heavy quarks; they can also be used to compute the differential
top quark pair production  cross section
at the ILC in the continuum, where top quark mass effects are still 
non-negligible. The scalar and pseudoscalar form factors enter the NNLO 
corrections to the decay of a Higgs boson into heavy quarks. These 
would become especially important for a very heavy Higgs boson, decaying 
into top quark pairs, where the form factors 
 could be used for a differential description 
of the decay final state.


\subsection{Electroweak Effects at the \mbox{\boldmath$t\bar{t}$} Threshold \\ \small{{\it A. Hoang}}}
\label{shng}

The proper treatment of electroweak effects plays an important role in
high precision measurements at the $t\bar t$ threshold. In the calculation of 
the total cross section, one can categorize electroweak
effects into three classes.

\begin{itemize}
\item[(a)] "Hard" electroweak effects: This class includes electroweak
  effects related to the $t\bar t$ production mechanism itself or
  factorizable corrections to various matching conditions of the
  effective theory that is used to describe the nonrelativistic QCD
  dynamics of the top pair. In general these corrections are
  modifications to the hard QCD matching conditions of the effective
  field theory operators. They can be determined by standard methods
  and are real numbers.
\item[(b)] Electromagnetic effects: Electromagnetic effects are
  relevant for the luminosity spectrum of an $e^+e^-$ initial state (beam
  energy spread, beamstrahlung, initial state radiation) and the
  $t\bar b$ final state (modification to the Coulomb attraction,
  contributions to hard electroweak effects).
\item[(c)] Effects related to the finite top quark lifetime: Apart
  from the top decay (into $Wb$ for the Standard Model) this class
  also includes interference contributions with processes having the
  same final state ($W^+W^-b\bar b$) but only one or even no top quark
  at intermediate stages. This class also accounts for
  interactions involving the top decay products ("non-factorizable"
  effects). 
\end{itemize}

For processes involving highly energetic top quarks, where the
momentum transfer typically involves scales much larger than
$\Gamma_t$, the class (c) effects are small corrections. On the other
hand, for the top threshold dynamics all three classes have to be
accounted for at leading order, because the typical kinetic energy of
the nonrelativistic top quarks is of the same order as the top quark
width,
\begin{equation}
E_{\rm kin} \sim m_t v^2 \sim m_t \alpha_s^2 \sim \Gamma_t \sim m_t \alpha_{\rm em}
\,.
\label{ewcounting}
\end{equation}
The relation in Eq.~(\ref{ewcounting}) gives the power-counting that
has to be employed to systematically account for electroweak effects
within the nonrelativistic expansion: 
\begin{equation}
v^2\sim \alpha_s^2 \sim \alpha_{\rm em}
\,.
\end{equation}
Several analyses concerning electroweak effects belonging to the
various classes have been carried out in the past. However, no coherent and comprehensive treatment that
systematically accounts for all these electroweak effects beyond
leading order currently exists. In the following a brief status report is given
concerning top threshold production in $e^+e^-$ annihilation.

The leading order (LL) electroweak effects belonging to class (a) describe
the production mechanisms of the nonrelativistic top quark pair in
$e^+e^-$ annihilation in the various possible
spin and angular momentum states. Due to the power-counting, the
one-loop hard electroweak corrections already contribute at
next-to-next-to-leading order (NNLL) and involve the standard (real parts of
the) one-loop electroweak corrections to the $e^+e^-\to t\bar t$  
process~\cite{Grzadkowski:1986pm,Guth:1991ab}. There are
also hard electroweak corrections that modify the QCD 
potentials~\cite{Kummer:1995id}, but due to gauge cancellations these
corrections do not contribute at NNLL order. The dominant corrections
caused by an exchanged Higgs boson through a Yukawa potential can
also be considered as a class (a) contribution. 

The most important contribution belonging to the electromagnetic
effects in class (b) is the
luminosity spectrum that affects the c.m.\,energy available in the
$e^+e^-$ collision. While beamstrahlung and the beam energy spread are
machine-dependent and will have to be measured experimentally, the
initial state radiation component is calculable. The luminosity
spectrum leads to sizable smearing of the cross section  and contributes at
LL order. Electromagnetic effects also modify the QCD potential through
photon exchange between the top pair and can lead to hard electroweak
corrections belonging to class (a). In present analyses QED effects
are only accounted for through the luminosity spectrum. In particular
there is no coherent treatment of the calculable QED effects that
systematically accounts for the effects of initial state radiation,
the Coulomb corrections and the hard QED effects.

\begin{figure}[t] 
\includegraphics[width=8cm]{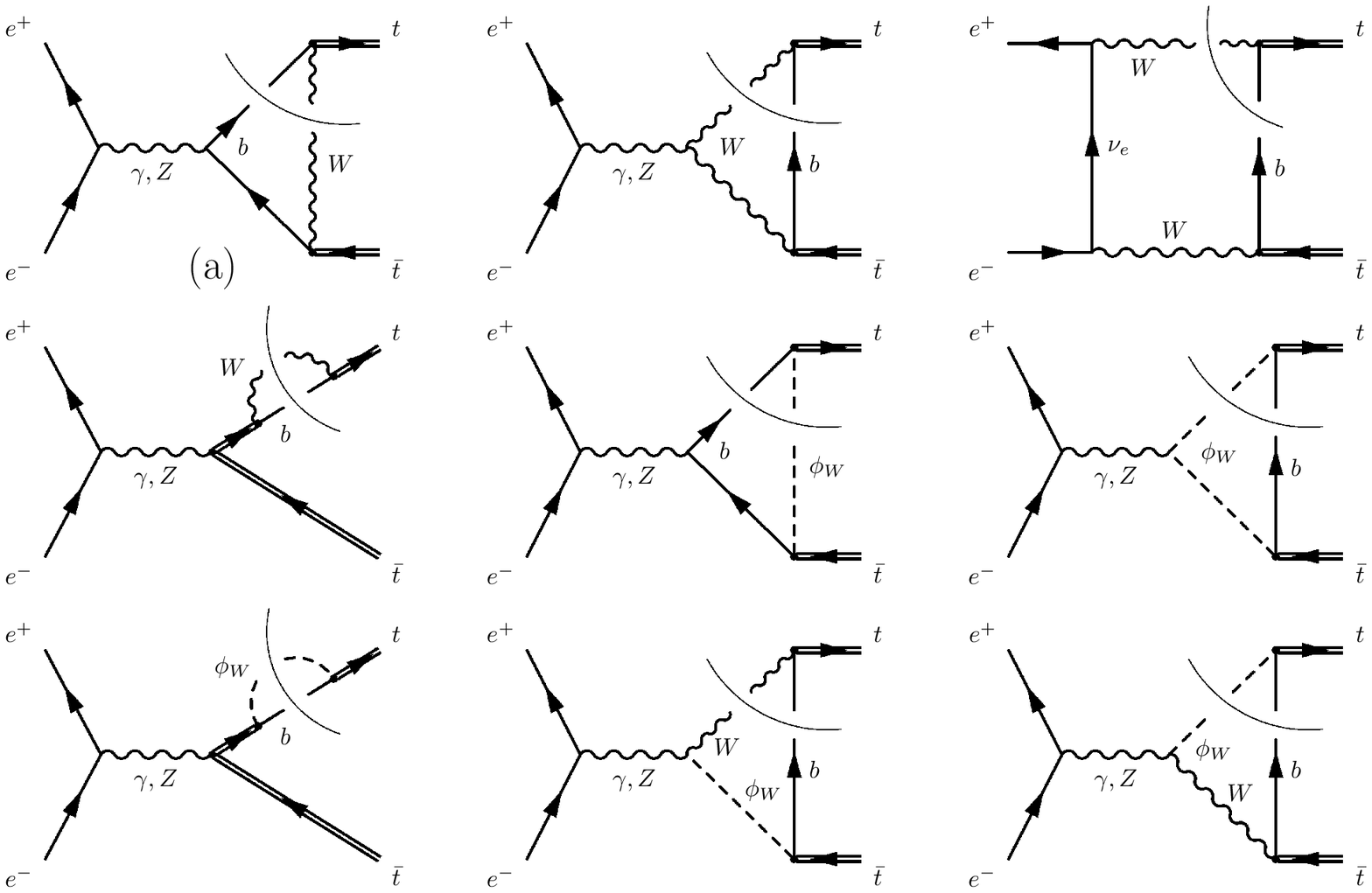}
\qquad
\includegraphics[width=8cm]{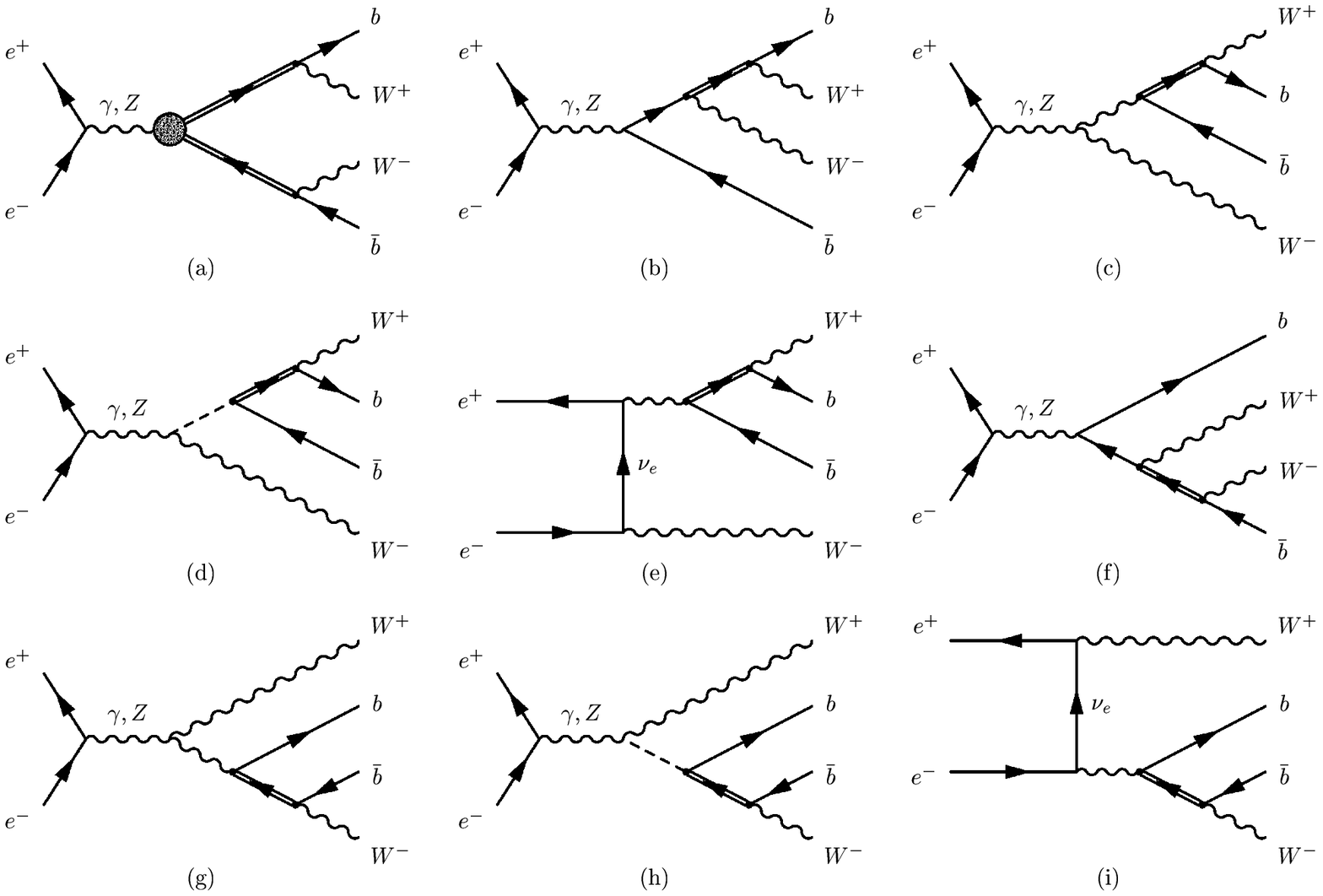}
 \caption{
(a) Cuts in Standard Model diagrams contributing to absorptive parts
   in the  matching conditions for the $t\bar t$ effective theory
   currents. Within the effective theory the absorptive parts describe
   the interference of the diagrams shown in (b). 
 \label{figunstable} }
\end{figure}

The leading order effect belonging to class (c) is the top decay
width, which makes the perturbative treatment to the strong $t\bar t$
dynamics at all possible~\cite{Kuhn:1980gw,Fadin:1987wz}. Within the
nonrelativistic effective theory used to describe QCD effects the top
quark width can be implemented through an imaginary mass shift in the
top quark propagator
\begin{equation}
\frac{i}{k^0-\frac{\bmp^2}{m_t}+\delta m_t+\frac{i}{2}\Gamma_t}
\,,
\end{equation}
where $\delta m_t$ is related to the top quark mass definition that is
used. This effect can be incorporated into the forward $e^+e^-\to
e^+e^-$ scattering amplitude results for stable top quarks by simply
shifting the c.m.\,energy into the complex plane: $\sqrt{s}\to
\sqrt{s}+i\Gamma_t$. Using the optical theorem one can obtain the
total cross section. Beyond leading order the electroweak effects
belonging to class (c) can be systematically incorporated into the
effective theory by accounting for absorptive parts related to the top
decay final states in the matching conditions~\cite{Hoang:2004tg} of
the effective theory.  
This leads to well known effects such as the time dilatation
correction~\cite{Kummer:1995id}, but can also account for interference
contributions with processes where no top pairs are produced, but
which have the same final state~\cite{Hoang:2004tg} (see 
Fig.~\ref{figunstable}).
%
%
%
The gauge cancellation already mentioned
for class (a) also applies here and leads to the cancellation of
interactions among the top quarks and their decay products caused by 
ultrasoft gluons for the total cross
section at NLL order~\cite{Melnikov:1993np,Fadin:1993dz} and even at
NNLL order~\cite{Hoang:2004tg}.  
In Ref.~\cite{Hoang:2004tg} it was also
shown that within the nonrelativistic effective theory the imaginary
matching conditions can lead to ultraviolet phase space divergences 
that require additional renormalization. A complete treatment of all the
class (c) electroweak effects at NLL has not yet been achieved. 

A crucial prediction of the Higgs mechanism is that the Higgs Yukawa coupling to
quarks $\lambda_q$ is related to the quark masses by   $m_q=\lambda_q
V$. At the $e^+e^-$ Linear Collider the top quark Yukawa coupling
can be measured from top quark pair production associated with a
Higgs boson, $e^+e^-\to t\bar t H$. This
process is particularly suited for a light Higgs boson since the cross
section can then reach the $1$-$2$~fb level. Assuming an experimental
precision at the percent level, QCD and electroweak radiative
corrections need to be accounted for in the theoretical predictions.  
The Born cross section was already determined some time ago in
Ref.~\cite{Borneetth}. For the ${\cal O}(\alpha_s)$ QCD one-loop
corrections a number of references in various approximations exist
\cite{Dawson1,Dawson2,Dittmaier1}. On the other hand, the 
full set of one-loop electroweak corrections was obtained in
Refs.~\cite{Belanger1,Denner1} and also in Ref.~\cite{You1}. In
Ref.~\cite{Denner1} a detailed analysis of various differential 
distributions of the cross section $\sigma(e^+e^-\to t\bar t H)$ can
be found. For high energies these fixed order predictions are
sufficient to reach the required theoretical precision.

\begin{figure}[t] 
  \begin{center}
    \includegraphics[width=6cm]{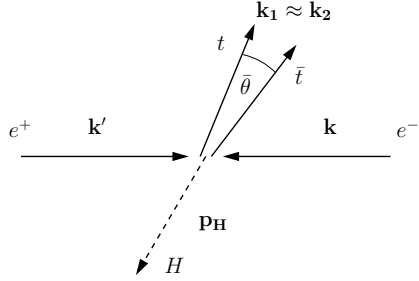}
    \vskip  0.0cm
    \caption{
      Typical constellation of momenta for the process $e^+e^-\to t\bar t H$ 
      in the large Higgs energy endpoint region. 
      \label{figtth} }
  \end{center}
\end{figure}

\begin{figure}[t] 
  \begin{center}
    \includegraphics[width=8cm]{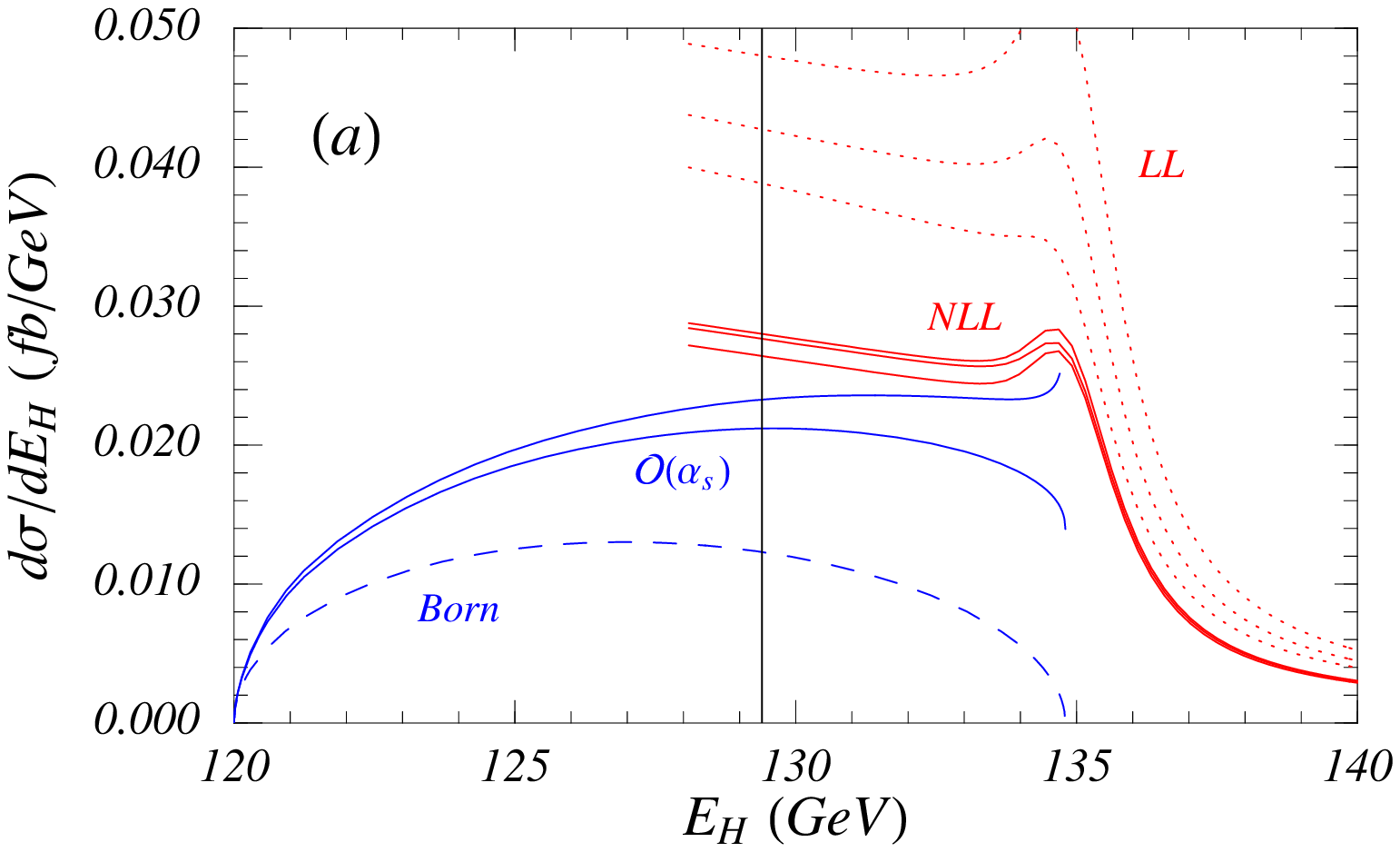}
    \includegraphics[width=8cm]{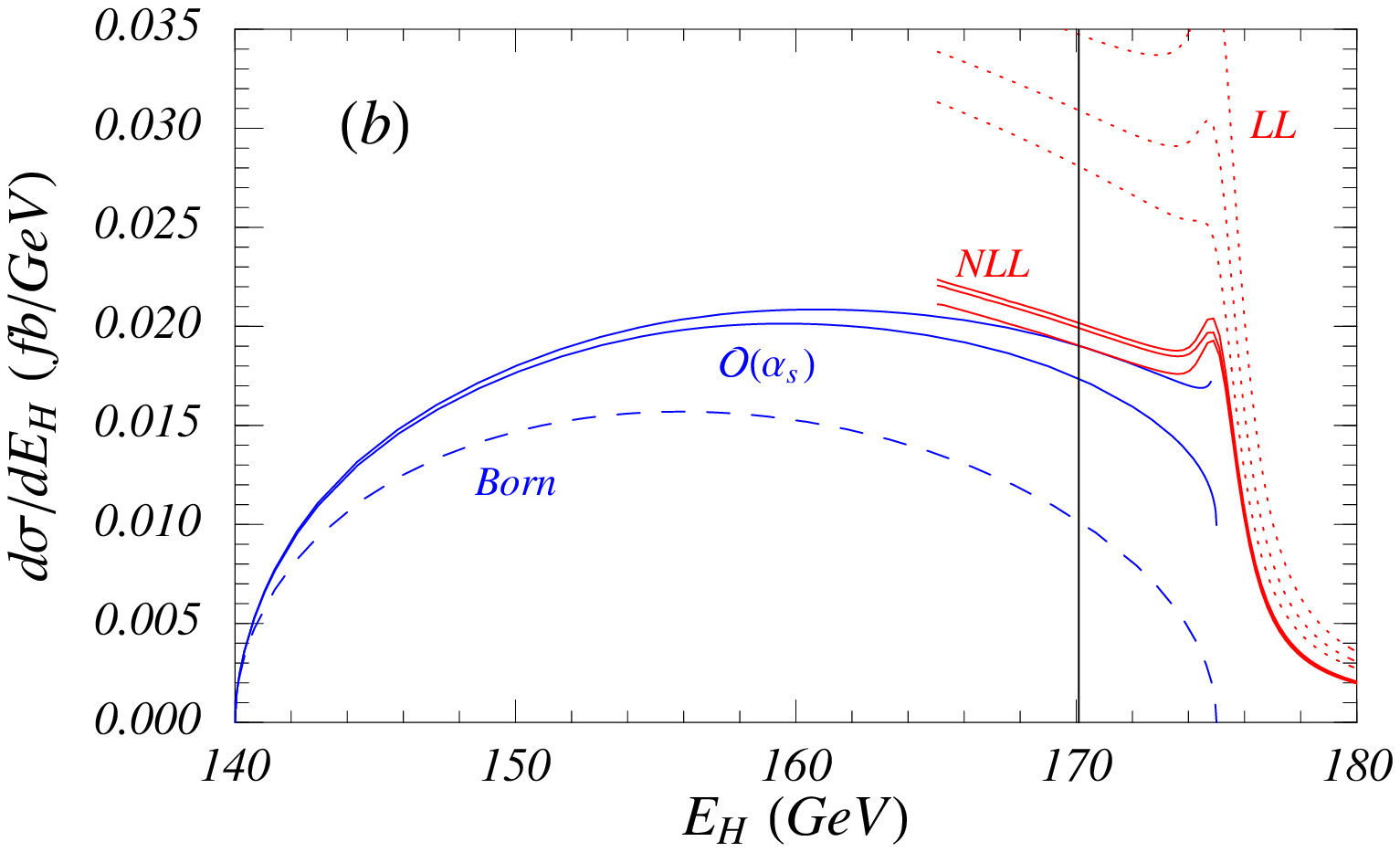}
\vskip -0.3cm
    \caption{
      Higgs energy spectrum at LL (dotted lines) and 
      NLL (solid lines) order in vNRQCD for the soft
      renormalization scales $\mu_S=(0.1,0.2,0.4)m_t$
      and at the Born level and at ${\cal O}(\alpha_s)$ for
      $\mu=\sqrt{s}, \sqrt s |v|$ for the parameters
      (a) $\sqrt{s}=500$~GeV, $m_H=120$~GeV and
      (b) $\sqrt{s}=550$~GeV, $m_H=140$~GeV.
      The top 1S mass has been set to $m_t^{\rm 1S}=180$~GeV and the
      other parameters are $\Gamma_t=1.55~\mbox{GeV}$, 
      $M_Z=91.1876$~GeV, $M_W=80.423$~GeV, $c_w=M_W/M_Z$ and
      $\alpha^{-1}=137.034$.
      \label{figenH} }
  \end{center}
\end{figure}

In Ref.~\cite{Farrell:2005fk} the phase space region where the
Higgs energy is large and the $t\bar t$ pair becomes collinear to
balance the large Higgs momentum was analyzed (see Fig.~\ref{figtth}).
%
%
It was found that since in this kinematic region the $t\bar t$
invariant mass approaches $2m_t$, it is governed by nonrelativistic
dynamics in analogy to the physics relevant for the $t\bar t$
threshold. Thus, the usual fixed-order treatment breaks down and
the nonrelativistic effective theory description known from the $t\bar
t$ threshold has to be applied. For high c.m.\,energies above about
$700$~GeV the large Higgs energy endpoint region is very small, and 
a usual fixed-order treatment is sufficient for the
theoretical determination of the total cross section. However, for
smaller c.m.\,energies (as available  at the first phase of the
ILC running), or for larger Higgs masses, the endpoint region
increases with respect to the full available phase space and an
effective theory treatment for the endpoint region becomes
mandatory. In 
Ref.~\cite{Farrell:2005fk} a QCD endpoint analysis in the framework
of the effective theory vNRQCD~\cite{Luke:1999kz} was carried out.
The results sum singular terms in the endpoint region $\propto
(\alpha_s/v)^n$ and $(\alpha_s\ln v)^n$ at NLL order, where $v$ is the
c.m.~velocity of the top quarks in the $t\bar t$ c.m.~frame. Typical
results for the Higgs energy spectrum for small c.m.\,energies showing
the fixed-order QCD predictions and the LL and NLL order
nonrelativistic effective theory predictions in the large Higgs energy
region are shown in Fig.~\ref{figenH}. The vertical lines mark the
Higgs energy where the top velocity $v=0.2$, which roughly
divides the regions where fixed-order and effective theory
computations are valid. 
In Table~\ref{tab3} the impact of the summation of the
Coulomb singularities and the logarithms of the top quark velocity in
the endpoint region is
analyzed numerically for various choices of the c.m.\,energy and the 
Higgs mass. For all cases the top quark mass $m_t^{\rm 1S}=180$~GeV
is used and the other parameters are fixed as in Fig.~\ref{figenH}.
Here, $\sigma(\mbox{Born})$  
refers to the Born cross section and  $\sigma(\alpha_s)$ to the 
${\cal O}(\alpha_s)$ cross section in fixed-order perturbation theory
using $\mu=\sqrt{s}$ as the renormalization scale. The term
$\sigma(\mbox{NLL})$ refers to the sum of the ${\cal O}(\alpha_s)$
fixed-order cross section for $v>0.2$ using $\mu=\sqrt{s}v$ and the
NLL nonrelativistic cross section for $|v|<0.2$ with the soft vNRQCD
renormalization parameter $\mu_S=0.2 m_t$. The summation of endpoint
singularities is particularly important for smaller c.m.\,energies,
since it leads to an additional significant enhancement of the total
cross section in a region where the cross section is rather small.
This enhancement could improve the prospects of top-Yukawa coupling
measurements at the first phase of the ILC (see section 2.5).
%
%

\begin{table}
\begin{center}
\begin{tabular}{|c|c||c|c||l||c|c|}
\hline
$\sqrt{s}$ [GeV] & 
$m_H$ [GeV] & 
$\mbox{}\;\sigma(\mbox{Born})$ [fb] & 
$\mbox{}\;\sigma(\alpha_s)$ [fb] &
$\mbox{}\;\sigma(\mbox{NLL})$ [fb] & 
$\mbox{}\quad\frac{\sigma(\mbox{\tiny NLL})}{\sigma(\mbox{\tiny Born})}\quad\mbox{}$ &
$\mbox{}\quad\frac{\sigma(\mbox{\tiny
    NLL})}{\sigma(\alpha_s)}\quad\mbox{}$ 
\\ \hline\hline
$500$ & $120$ & $ 0.151$ & $ 0.263$ & $\quad 0.357(20)$ & $ 2.362$ & $
1.359$ \\\hline
$550$ & $120$ & $ 0.984$ & $ 1.251$ & $\quad 1.342(37)$ & $ 1.364$ & 
$ 1.073$ \\\hline
$550$ & $160$ & $ 0.134$ & $ 0.207$ & $\quad 0.254(12)$ & $ 1.902$ & 
$ 1.226$ \\\hline
$600$ & $120$ & $ 1.691$ & $ 1.939$ & $\quad 2.005(30)$ & $ 1.185$ & 
$ 1.034$ \\\hline
$600$ & $160$ & $ 0.565$ & $ 0.700$ & $\quad 0.745(18)$ & $ 1.319$ & 
$ 1.065$ \\\hline
$700$ & $120$ & $ 2.348$ & $ 2.454$ & $\quad 2.485(13)$ & $ 1.058$ & 
$ 1.012$ \\\hline
$700$ & $160$ & $ 1.210$ & $ 1.303$ & $\quad 1.328(11)$ & $ 1.098$ & 
$ 1.020$ \\\hline
\hline
\end{tabular}
\end{center}
{\tighten \caption{
Cross sections and K factors for $\sigma_{\rm tot}(e^+e^-\to t\bar t H)$
for various c.m.\,energies
and Higgs masses and top quark mass $m_t^{\rm 1S}=180$~GeV.}
\label{tab3} }
\end{table}
%


\subsection{Towards a Precise Measurement of the Top Quark Yukawa Coupling at the ILC~\cite{Juste:2005vs} \\ \small{{\it A. Juste}}}
\label{sju}

The top quark Yukawa coupling ($\lambda_t$) is the largest coupling of the Higgs boson to
fermions. A precise measurement of it is very important since it may help unravel the secrets of 
the Electroweak Symmetry Breaking (EWSB) mechanism, in which the top quark could possibly
play a key role. For $m_h<2m_t$, a direct measurement of $\lambda_t$ is 
possible via associated $t\bar{t}h$ production, both at the LHC and a future ILC. 
At the LHC, the expected accuracy~\cite{Weiglein:2004hn} is $\delta\lambda_t/\lambda_t\sim 12-15\%$ for 
$m_h\sim 120-200$ GeV, assuming an integrated luminosity of 300 fb$^{-1}$. Existing feasibility studies at the ILC~\cite{juste1} 
predict an accuracy of $\delta\lambda_t/\lambda_t\sim 6-10\%$ for $m_h\sim 120-190$ GeV, assuming $\sqrt{s}=800$ GeV and 1000 fb$^{-1}$.
However, currently the baseline design for the ILC only contemplates a maximum center-of-mass energy of
500 GeV. Therefore, it is very relevant to explore the prospects of this measurement
during the first phase of the ILC, especially in view of the limited accuracy expected at the LHC; for a number of years, 
the combination of results from the LHC and ILC will yield the most precise determination of $\lambda_t$.

A preliminary feasibility study at $\sqrt{s}=500$ GeV was performed in Ref.~\cite{juste2}, which we briefly overview here.
Indeed, the measurement of $\lambda_t$ at $\sqrt{s}=500$ GeV is more challenging than at $\sqrt{s}=800$ GeV.
On the one hand, the reduced phase-space leads to a large reduction in $\sigma_{t\bar{t}h}$
(e.g. $\sigma^{Born}_{t\bar{t}h}\simeq 0.16(2.5)$ fb at $\sqrt{s}=500(800)$ GeV, for $m_h=120$ GeV). On the other hand, 
the cross section for many background processes is significantly increased. This analysis assumed 
$m_h=120$ GeV and focused on the $t\bar{t}h\rightarrow (\ell\nu b)(jjb)(b\bar{b})$ decay channel ($BR\sim 30\%$).
The dominant background is $t\bar{t}jj$, followed by $t\bar{t}Z$, although other non-interfering
backgrounds (e.g. $W^+W^-$) were also considered. Signal and backgrounds were processed through a fast detector simulation.
After basic preselection cuts, the signal efficiency was found to be $\simeq 50\%$ and the $S:B\simeq 1:100$.
In order to increase the sensitivity, a multivariate analysis using a Neural Network (NN) with 23 variables was performed.
The final selection consisted of an optimized cut on the NN distribution. Assuming an integrated luminosity of
1000 fb$^{-1}$, the expected total numbers of signal and background events were 11 and 51, respectively, resulting
in $(\delta\lambda_t/\lambda_t)_{stat}\simeq 33\%$. Based on previous experience~\cite{juste1}, the addition of the 
fully hadronic decay channel was expected to ultimately lead to $(\delta\lambda_t/\lambda_t)_{stat}\simeq 23\%$. 
While this analysis is already rather sophisticated,
significant improvements are expected from the usage of a more efficient $b$-tagging algorithm or a more optimal 
treatment of the kinematic information. In the next sections we discuss additional sources of improvement which are 
currently under investigation.

The precise measurement of $\lambda_t$ requires accurate theoretical predictions for $\sigma_{t\bar{t}h}$. Currently, one-loop
QCD and electroweak corrections are available. However, at $\sqrt{s}=500$ GeV and for $m_h\geq 120$ GeV,
the kinematic region where the $t\bar{t}$ system is non-relativistic dominates. As discussed in Ref.~\cite{Farrell:2005fk} and Section 2.4, 
in this regime 
Coulomb singularities are important and need to be resummed within the framework of the vNRQCD effective theory, leading to
large enhancements factors in the cross section relative to the Born level. At the ILC, because of ISR and beamstrahlung (BS),
the event-by-event center-of-mass energy ($\sqrt{\hat{s}}$) will be lower than the nominal one, 
thus bringing the $t\bar{t}$ system deeper into the non-relativistic regime. In order to compute the expected $\sigma_{t\bar{t}h}$ including these effects,
the 11-fold Born differential cross section for $e^+e^- \rightarrow t\bar{t}h \rightarrow W^+bW^-\bar{b}h$ was multiplied 
by a K-factor defined as $K(E_h,\sqrt{\hat{s}})=(d\sigma_{t\bar{t}h}^{NLL}/dE_h)/(d\sigma_{t\bar{t}h}^{Born}/dE_h)$,
where $E_h$ stands for the Higgs boson energy in the $e^+e^-$ rest-frame, and then folded with ISR and BS structure functions.
The NLL differential cross section was kindly provided by the authors of Ref.~\cite{Farrell:2005fk}.
Fig.~9 (left and center) compares the Born (for off-shell top quarks) and NLL differential cross sections 
for different values of $\sqrt{\hat{s}}$, assuming $m_t^{1S}=180$ GeV and 
$m_h=120$ GeV. The ratio of these two curves defines $K(E_h,\sqrt{\hat{s}})$ and can be significantly
larger than 1, especially for low values of $\sqrt{\hat{s}}$. Since the NLL prediction is only valid for 
$E_h\leq E^{max}_h$ (where $E^{max}_h$  effectively corresponds to a cut on the top quark velocity in the $t\bar{t}$ rest-frame of $\beta_t<0.2$), 
we currently set $K(E_h,\sqrt{\hat{s}})=1$ for $E_h>E^{max}_h$, although in practice, it should be possible to use
$K(E_h,\sqrt{\hat{s}})=(d\sigma_{t\bar{t}h}^{O(\alpha_s)}/dE_h)/(d\sigma_{t\bar{t}h}^{Born}/dE_h)$.
Table~IV compares the predicted Born and ``NLL-improved'' $\sigma_{t\bar{t}h}$
for different scenarios, illustrating the large impact of radiative effects in the initial state.
This underscores the importance of being able to predict these effects to the percent level. 
While the impact of ISR cannot be reduced, it might be possible to find an optimal operating point of the accelerator, as far as this measurement
is concerned, in terms of BS and total integrated luminosity. 
Finally, it is found that, for $m_h=120$ GeV, resummation effects can increase $\sigma_{t\bar{t}h}$ by a
factor of $\sim 2.4$ with respect to the Born cross section  used in the previous feasibility study.

\begin{table}[t]
\begin{center}
\caption{Comparison of the Born and NLL $\sigma_{t\bar{t}h}$ for different scenarios regarding radiative effects in the initial state.}
\begin{tabular}{|c|c|c|c|}
\hline \textbf{(ISR,BS)}  & \textbf{\boldmath$\sigma_{t\bar{t}h}$ (fb) (Born)} & \textbf{\boldmath$\sigma_{t\bar{t}h}$ (fb) (``NLL-improved'')} 
& \textbf{Enhancement factor} \\
\hline 
(off,off) & 0.157(1) & 0.357(2) & 2.27 \\
\hline 
(off,on) & 0.106(1) & 0.252(3) & 2.38 \\
\hline 
(on,on) & 0.0735(8) & 0.179(2) & 2.44 \\
\hline
\end{tabular}
\label{sigmatth}
\end{center}
\end{table}

\begin{figure*}[t]
\centering
\includegraphics[width=5.5cm]{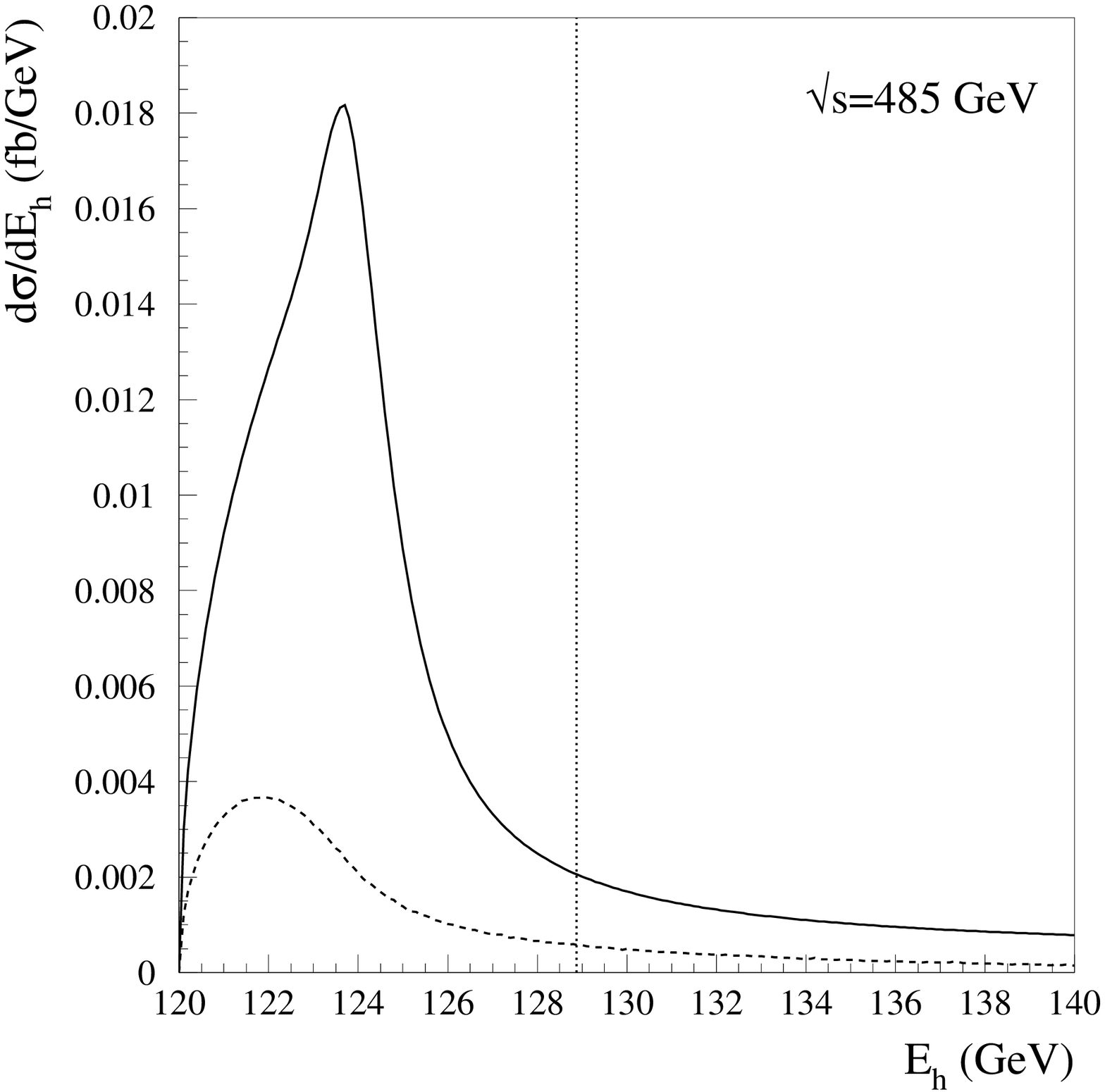}
\includegraphics[width=5.5cm]{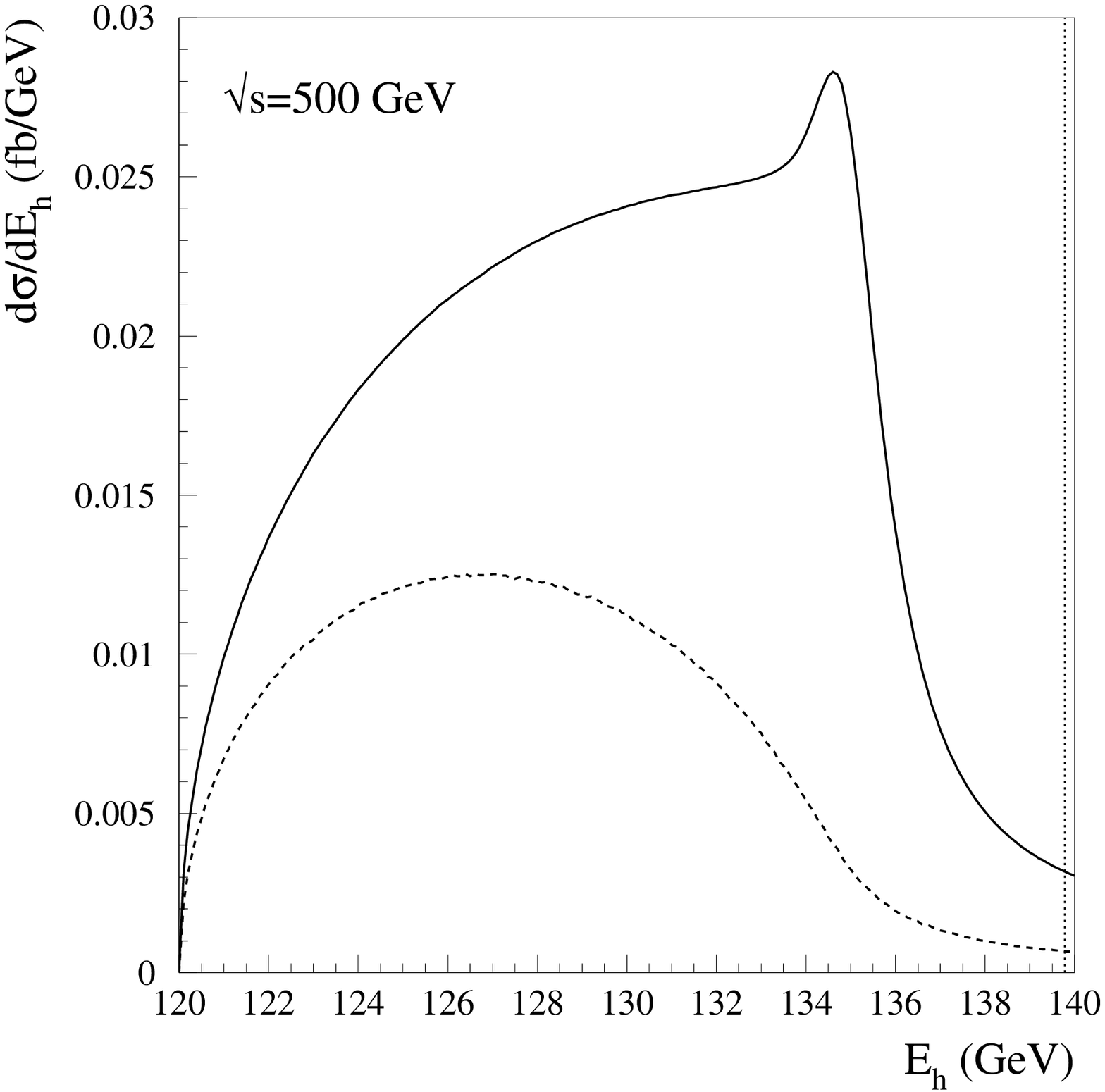}
\includegraphics[width=5.5cm]{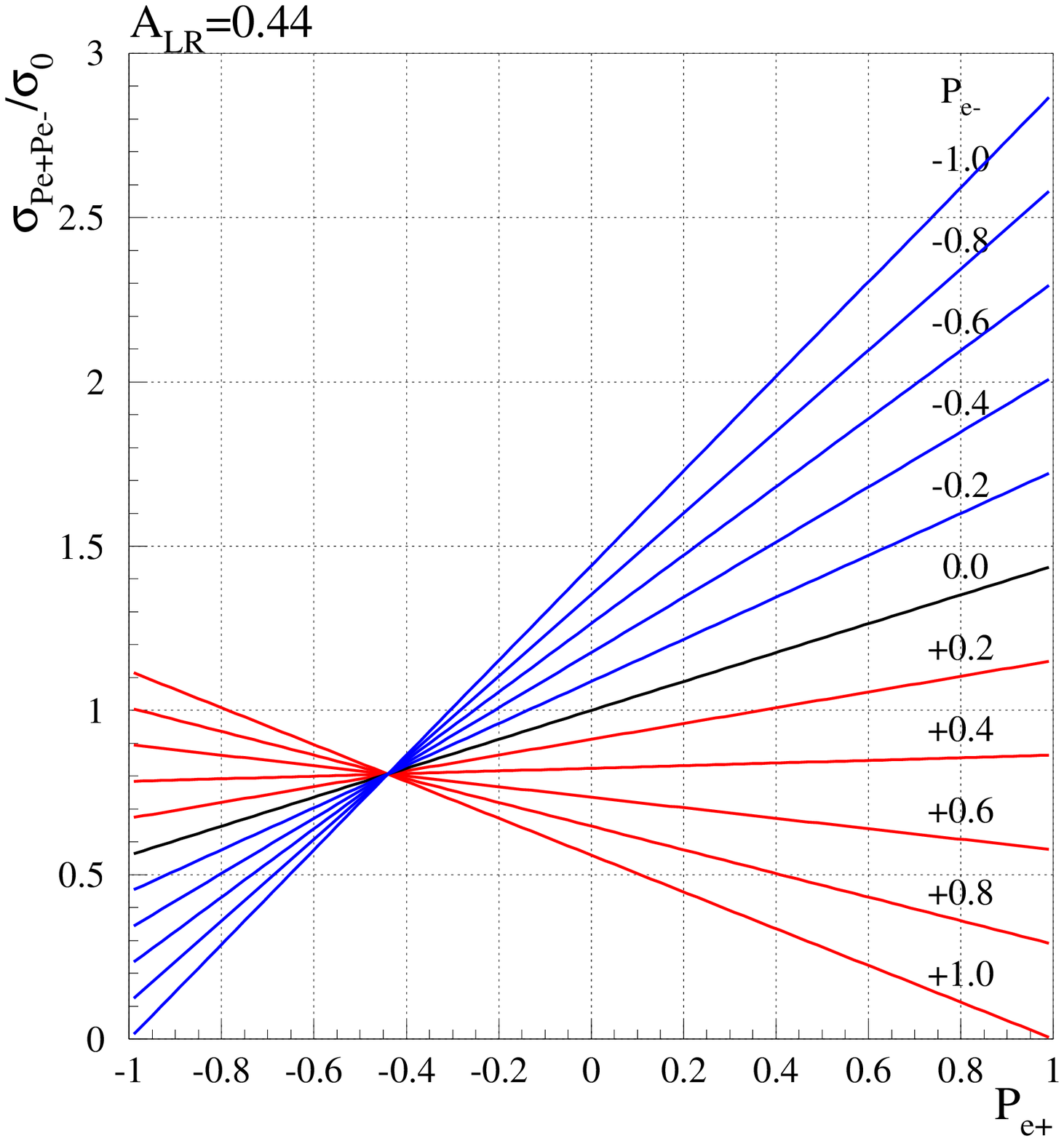}
\caption{Left and center: comparison of the Born (dashed) and NLL (solid) $d\sigma_{t\bar{t}h}/dE_h$ for different values of $\sqrt{s}$, assuming
$m_t^{1S}=180$ GeV and $m_h=120$ GeV. The dotted line indicates the value of $E^{max}_h$.
Right: ratio of polarized to unpolarized cross section for different values of ($P_{e^-},P_{e^+}$).}
\label{combo}
\end{figure*}

So far, all feasibility studies of this measurement have assumed unpolarized beams. Currently, the baseline design for the ILC only includes
longitudinal electron beam polarization ($|P_{e^-}|\simeq 0.8$). Positron beam polarization ($|P_{e^+}|\simeq 0.6$) is considered as an option.
The ratio of the polarized cross section (for arbitrary longitudinal beam polarization) 
($\sigma_{P_{e^-}P_{e^+}}$) to the unpolarized cross section ($\sigma_0$) is given by
$\sigma_{P_{e^-}P_{e^+}}/\sigma_0 = (1-P_{e^-}P_{e^+})(1-P_{eff}A_{LR})$, where $P_{eff}=(P_{e^-}-P_{e^+})/(1-P_{e^-}P_{e^+})$ denotes the
``effective polarization'' and $A_{LR}$ is the ``left-right asymmetry'' of the process of interest~\cite{gudi}.
Therefore, two potential enhancement factors can in principle be exploited: the first one requires having both beams polarized, the second
one requires $A_{LR}\neq 0$ and a judicious choice of the signs of $P_{e^-}$ and $P_{e^+}$ in order to have $P_{eff}A_{LR}<0$.
In the case of SM $t\bar{t}h$ production, $A_{LR} \simeq 0.44$, essentially independent of $\sqrt{s}$ 
in the range $\sim 0.5-1.0$ TeV. Assuming $(A_{LR})_{SM}$, Fig.~9 (right) shows the cross section enhancement factor as a function 
of $P_{e^+}$, for different values of $P_{e^-}$. The optimal (realistic) operating point would be $(P_{e^-},P_{e^+})=(-0.8,+0.6)$, achieving an
increase in $\sigma_{t\bar{t}h}$ by a factor of $\simeq 2.1$ with respect to the unpolarized case. Unfortunately, this choice does not
help reduce the dominant background, which is increased by a similar factor. Nevertheless, the net result is still an improvement
in the statistical precision on $\lambda_t$ by $\sim 45\%$, which would be an argument in favor of including positron polarization 
in the baseline design. For $(P_{e^-},P_{e^+})=(-0.8,0)$, only a modest increase in $\sigma_{t\bar{t}h}$ by a factor of $\sim 1.3$ would be achieved.
It is important to realize that, in order to choose the sign of $P_{e^-}$ and $P_{e^+}$, it is necessary to know the sign of $A_{LR}$.
Anomalous couplings in the $tt\gamma$ and $ttZ$ vertices could possibly lead to
deviations in $A_{LR}$ from the SM prediction. Unfortunately, due to the limited precision
in the measurement of the $ttZ$ couplings~\cite{Baur:2004uw,new}, the LHC is not expected to provide any useful constraints on the sign of $A_{LR}$.
Therefore, at the ILC the first step should be to perform measurements of the polarized $\sigma_{t\bar{t}}$ in order to determine
the sign of $A_{LR}$, and thus fix the signs of $P_{e^-}$ and $P_{e^+}$ (the magnitudes should be the largest possible). On the other hand, the 
measurement of $\lambda_t$ requires a percent-level and model-independent determination of the $t\bar{t}\gamma$ and $t\bar{t}Z$ couplings, which
typically benefits from changing the beam polarization. Therefore, it would be desirable to optimize the running strategy to maintain the largest
possible $\sigma_{t\bar{t}h}$, needed for a precise measurement of $\lambda_t$, while meeting the precision goals for measurements of top quark couplings.

We have studied the prospects of a precise measurement of the top quark Yukawa coupling during the first phase of the ILC.
Taking into consideration an existing feasibility study, and the additional enhancement factors to $\sigma_{t\bar{t}h}$ discussed 
here, we anticipate a precision of $(\delta\lambda_t/\lambda_t)_{stat}\sim 10\%$ for $m_h=120$ GeV, assuming $\sqrt{s}=500$ GeV and 1000 fb$^{-1}$.

\newpage


\subsection{The \mbox{\boldmath$t\bar{t}$} Threshold at an \mbox{\boldmath$e^+e^-$} Collider \\ \small{{\it Y. Kiyo}}}
\label{skiyo}

It is well known that QCD corrections to $t\bar{t}$ production
near threshold develop a Coulomb singularity.  The structure of the
cross section is
\begin{eqnarray}
\sigma_{tot}(e^+e^-\rightarrow t\bar{t})=\sigma_{Born}
      \bigg[1+ c^{(1)}\,\bigg(\frac{\alpha_s}{v}\bigg)
             + c^{(2)}\,\bigg(\frac{\alpha_s}{v}\bigg)^2
             + c^{(3)}\,\bigg(\frac{\alpha_s}{v}\bigg)^3
             + \cdots
             + c^{(n)}\,\bigg(\frac{\alpha_s}{v}\bigg)^n+\cdots
        \bigg],
\label{eq1}
\end{eqnarray}
where $\sigma_{Born}$ is the Born cross section and
$v=\sqrt{1-4m_t^2/s}$ is the velocity of the top quarks.  In Eq. 7, we have pulled out the Coulomb singularity
$(\alpha_s/v)^n$ explicitly so that the $n$th coefficient starts with
${\cal O}(v^0)$, 
\begin{eqnarray}
c^{(n)}=c^{(n, 0)}+c^{(n, 1)}\,v+c^{(n, 2)}\, v^2+\cdots. \label{eq2}
\end{eqnarray}
Near threshold the kinematics of the top quarks is non-relativistic
and $v \sim \alpha_s$ holds.  To get a meaningful cross section we thus
have to sum up the Coulomb singularities $\sim \alpha_s/v \sim 1$.
The leading order (LO) cross section then contains the Coulomb singularities 
$\sim(\alpha_s/v)^n$ to all orders of the coupling expansion,
\begin{eqnarray}
\sigma_{tot}^{LO}=\sigma_{Born}
      \sum_{n=0} c^{\,(n, 0)}\,\bigg(\frac{\alpha_s}{v}\bigg)^n.
\label{eq3}
\end{eqnarray}
The $c^{n, 1}$ terms are suppressed by $v$ compared with the LO and
result in next-to-leading order (NLO) corrections.  These terms are
known since long.  Next-to-next-to leading order (NNLO) calculations
have been completed some time ago by several groups \cite{TWGR}.  They
sum up all the corrections of $c^{(n,2)}\, v^2\,
(\alpha_s/v)^n$ to all orders $n$.

\begin{figure*}[htb]
\centering
\includegraphics[width=40mm]{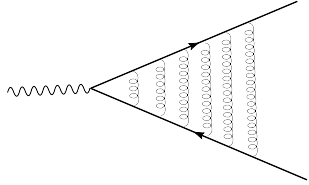}
\caption{Typical QCD loop diagram which yields the Coulomb singularity
  $\sim (\alpha_s/v)^n$, $n=$ the number of gluon exchanges.} \label{tt01}
\end{figure*}

The Coulomb singularity originates from potential gluons, which have
a typical momentum $k^\mu \sim m_t (v^2,\vec{v})$, exchanged
between almost on-shell top and anti-top quarks in Fig.~10. 
For the kinematics of potential gluons, the corresponding propagator
reduces to the Coulomb potential
\begin{eqnarray}
\widetilde{V_{C}}=-\frac{4\pi C_F \alpha_s}{{\bf q}^2}, \label{eq4}
\end{eqnarray}
where $C_F=4/3$.  In the gluon propagator, the energy component of $q$
was set to zero because of the potential gluon's kinematics.  Using
this gluon propagator in place of normal gluon propagators in the 
figure, one may reproduce the $c^{(i, 0)}$ coefficients to all 
orders in Eq. 8.  This procedure can be systematically 
extended to higher orders using effective field theory (EFT)
techniques.  The constructed effective field theories are versions of
non-relativistic QCD called pNRQCD/vNRQCD \cite{Luke:1999kz}.  We do not go
into details of the EFT but summarize several features.
\begin{itemize}
\item  The $t\bar{t}$ pair at threshold is created by the production
  current $\vec{J}=C_J\,[t\vec{\sigma}\bar{t}]+\cdots$ in the EFT,
  where $C_J$ is the Wilson coefficient of the corresponding current
  in the EFT, and the dots denote subleading operators.
\item The produced top quark pair forms bound state resonances by
  exchanging potential gluons; the binding Coulomb potential is 
$\displaystyle \widetilde{V}_0={\cal V}_0 \widetilde{V}_C ({\bf q})$,
  where ${\cal V}_0=1+{\cal O}(\alpha_s)$ includes QCD corrections to
  the leading order Coulomb potential from loop diagrams.
\item There are subdominant potentials, e.g.
$\displaystyle \widetilde{V}_1 = {\cal V}_1 g_s^2/|{\bf q}|$,
  which have to be taken into account if one wants to go beyond the
  NLO cross section.
\item A yet new type of corrections is known at
  next-to-next-to-next-to leading order (NNNLO), which is referred to
  as ultra-soft corrections because of the typical gluon momentum of 
  $k^0\sim|\vec{k}|\sim mv^2$.
\end{itemize}
An important point here is that the EFT scheme is a systematic way to sum up
higher order corrections, and it makes practical calculations easier
compared to those of full QCD.  Currently complete NNLO total cross
sections are known in both (semi-) analytical and numerical ways.  The
results are summarized in Ref.~\cite{TWGR}. 

In the following we briefly survey recent attempts in going beyond
NNLO and in resumming potentially large logarithms for the threshold
cross section in the EFT framework.\\

In the EFT calculation of the $t\bar{t}$ cross section, higher order
QCD corrections enter through the higher order coefficients ${\cal
  V}_i$ of potentials and through subdominant potentials,
e.g. $\widetilde{V}_{1}({\bf q})$.  Furthermore, there are corrections 
from ultra-soft gluons, starting to contribute at NNNLO.  They cannot
be written in form of potentials, as they include a noninstantaneous,
dynamical propagation in time, while the potentials are all instantaneous. 
The first correction to ${\cal V}_{0}$ is referred to as a Coulomb correction,
\begin{eqnarray}
{\cal V}^{(n)}_0=1+\frac{\alpha_s}{4\pi}\big(\beta_0 L(q)+\frac{43}{9}\big)+
\cdots +\big(\frac{\alpha_s}{4\pi}\big)^n \cdot \big(\mbox{group
  factors, $L(q)$, etc.}\big),
\end{eqnarray}
where $\beta_0=11-(2/3)n_f$ is the coefficient of the QCD $\beta-$function,
$L(q)=\ln(\mu^2/{\bf q}^2)$, and $\mu$ is the QCD renormalization
scale.  The QCD corrections to all the coefficients can be found in
Ref.~\cite{KPSS} up to ${\cal O}(\alpha_s^3)$, except the third order
constant term, so-called $a_3$ in the literature.  Using
$\widetilde{V}_0^{(3)}={\cal V}_0^{(3)}\widetilde{V}_C$, the Coulomb
correction to the total cross section was computed in Ref.~\cite{BKS}.
The other corrections to the $t\bar{t}$ cross section at NNNLO due to 
subleading potentials and ultra-soft gluons are not known yet. 
\begin{figure*}[htb]
\begin{center}
\includegraphics[bb=180 505 430 670,width=8cm]{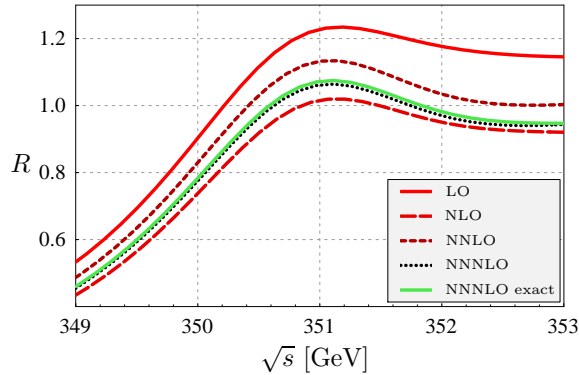}
\caption{Normalized cross section $R \equiv \sigma(e^+ e^- \to t\bar
  t)/\sigma(e^+ e^- \to \mu^+ \mu^-)$ in the threshold regime
  including the resummation of Coulomb corrections.  Figure from Ref.~\cite{BKS}.}
\label{tt02}
\end{center}
\end{figure*}
In Fig.~11 we show the $t\bar{t}$ threshold total cross
section including the Coulomb corrections as calculated in Ref.~\cite{BKS}.
The curves are obtained using analytical results of successive Coulomb
corrections from LO to NNNLO, and the line denoted by "NNNLO exact" is
obtained by numerically solving the Schr\"odinger equation using
${\cal V}^{(3)}_0$ at NNNLO.  The difference between NNNLO and "NNNLO
exact" is yet of higher order as the numerical method contains rigid
NNNLO plus additional contributions which come from iterations of the
potential $\widetilde{V}_0$. Studying the renormalization scale 
dependence of these predictions, the theoretical uncertainty for the 
NNNLO Coulomb corrections was estimated to be about 5\% of the cross 
section. Although the full corrections at NNNLO are not known, the $1S$
energy of the $t\bar{t}$ resonance, which corresponds to the peak position of
the cross section, is known analytically at NNNLO from
Refs.~\cite{KPSS,ThirdOrder}.  The result shows that the uncertainty 
in the choice of the scale is small and the perturbative 
series for the $1S$ energy is stabilized (see Ref.~\cite{KS}).

In higher order calculations there appear potentially large logarithms of
ratios of scales.  In the threshold regime we have large logarithms of
$v$ in the coefficients $c^{(n, i)}$, arising from ratios of the
largely different scales of the problem, namely the energy of the top
quarks, $E_t\sim m v^2$, the momentum $p\sim m_t v$ of almost on-shell
top quarks, and the c.m. energy $\sqrt{s} \sim 2m_t$.  The
origin of these logarithms is related to UV divergences in the EFT.
They can be resummed using renormalization group (RG) arguments
\cite{Luke:1999kz,RGI}, resulting in RG improved cross section predictions,
$\sigma_{LL}, \sigma_{NLL}$, etc.  The RG improvement was extensively
studied recently, and the next-next-leading-log (NNLL) cross section
was calculated in Ref.~\cite{HMST} (apart from the NNLL running of a
current for which the anomalous dimension is only fully known at NLL,
see also Ref.~\cite{Hoang,pinedanew}).  The RG improved cross section then contains
all terms of order $\sum_{n,m} (\alpha_s/v)^n (\alpha_s \ln
v)^m$.  The resummation of the logarithms leads to a reduction of the
scale dependence of the normalization from 20\% to about 6\% (at
NNLL), especially at energies around the 1$S$ peak. This is illustrated in
Fig.~12, where the scale dependence of the fixed order
(left panel) and RG improved (right panel) cross sections are shown
for different orders. 

The different studies discussed above strengthen our confidence that the 
estimate of the uncertainty of the top quark mass determination from 
Ref.~\cite{TWGR}, $\Delta m_t < 100$ MeV, is realistic and will not 
be spoiled by uncalculated higher order corrections.  There is even hope 
to believe that, if different approaches turn out to converge, the 
estimated error may shrink. For determinations of the top quark total
width, the top Yukawa coupling $y_t$ and the strong coupling constant 
from the threshold scan theoretical uncertainties are comparable to 
the expected experimental errors.  An ultimate theoretical
normalization error of at most $3\%$ is desirable.

\begin{figure*}[htbp]
\begin{center}
\includegraphics[bb=90 450 525 725, clip, width=8cm]{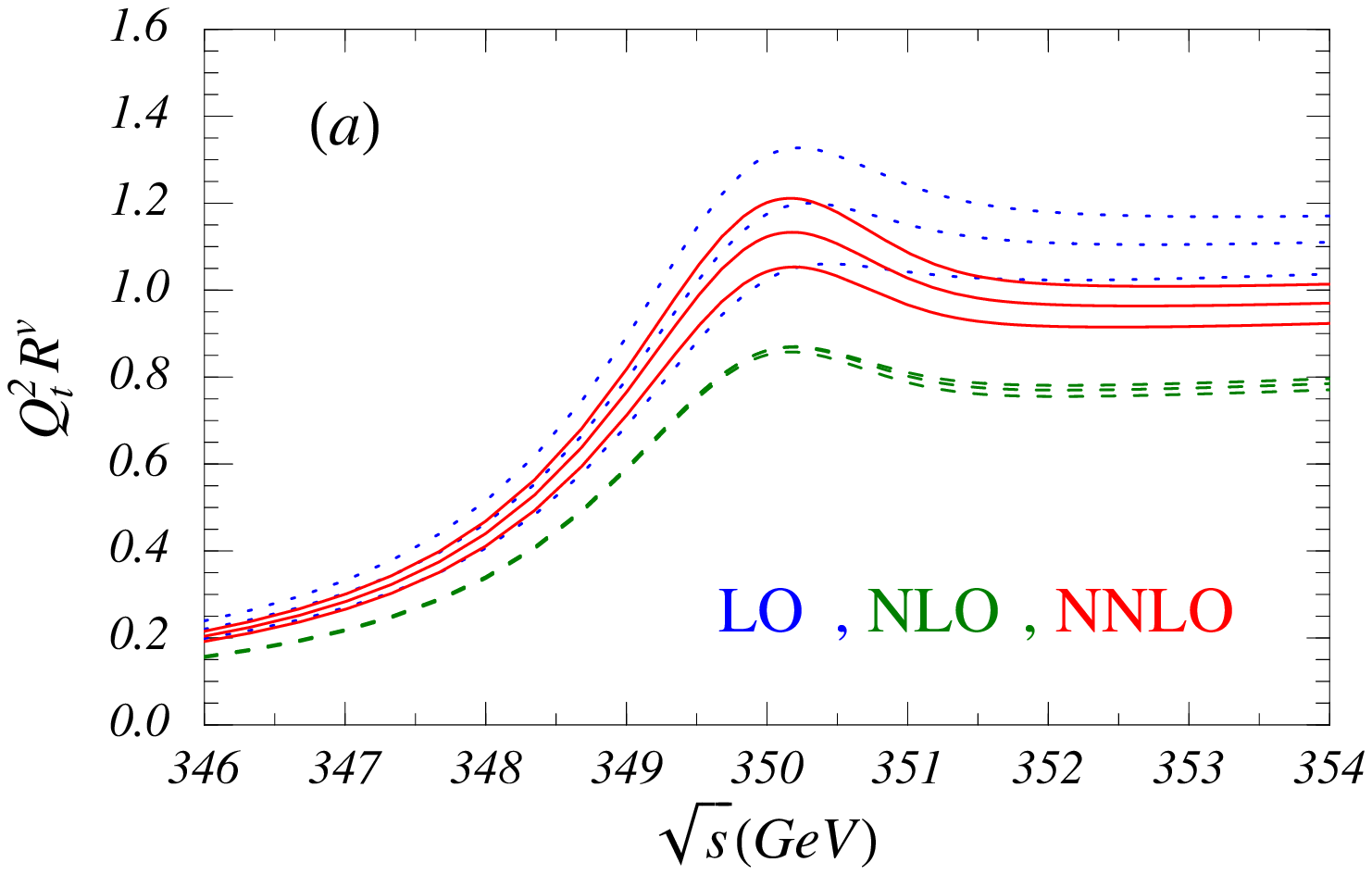}
\includegraphics[bb=90 450 525 725, clip,width=8cm]{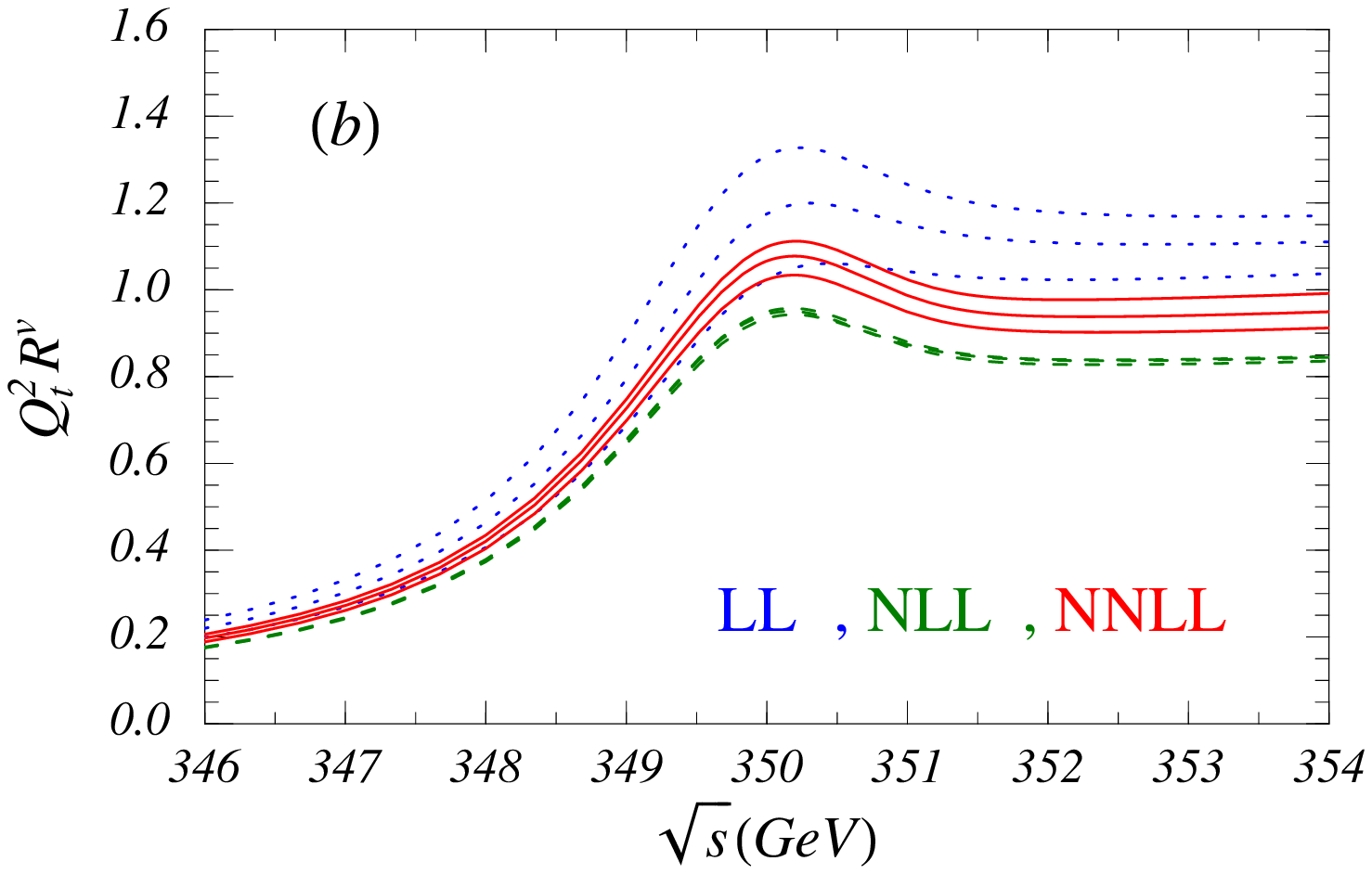}
\caption[]{Scale dependence of the fixed order cross section (a) and the
  RG improved cross section predictions (b) in different orders.  Each
  band of curves is obtained by varying the vNRQCD renormalization
  scale, choosing $\nu = 0.15, 0.2, 0.3$.
  Figures from Ref.~\cite{Hoang}.} 
\label{figfixedRGI}
\end{center}
\end{figure*}


\subsection{Differential Distributions and Experimental Aspects of the \mbox{\boldmath$t\bar t$} Threshold Scan \\ \small{{\it T. Teubner}}}
\label{steub}

The most precise determination of the top quark mass will come from
a dedicated threshold scan at $\sqrt{s} \sim 2m_t$.  This is mainly
a consequence of the fact that by performing a counting
experiment of color singlet $t\bar t$ pairs one can avoid most of the
systematic uncertainties inherent in a kinematic reconstruction of the
decay products of the colored quarks.  These uncertainties, which are
closely related to the problem of the mass definition, will ultimately
limit the determination of $m_t$ as done at the Tevatron and soon at the
LHC.  Detailed studies have shown that via a threshold scan,
measurements with very small statistical and systematic errors will be 
possible at the ILC~\cite{Martinez:2002st}.  A multi-parameter fit for
the top quark mass $m_t$, width $\Gamma_t$, the strong coupling
$\alpha_s$ and the top Yukawa coupling resulted in experimental errors
of about $\Delta m_t
\sim 20$ MeV, $\Delta \Gamma_t \sim 30$ MeV, $\Delta \alpha_s \sim
0.0012$, depending on details of the fit.  (At threshold, the top
Yukawa coupling can only be measured for a light Higgs and even then
with less than $30\%$ accuracy.)  Such a precision will only be achieved if
the accuracy of the theoretical predictions can match the experimental
one, and recently a lot of effort has been invested in further
improvements of the theory (see sections 2.4 and 2.6). Although most of
the information in the fit will come from the precise measurement of the total
$t\bar t$ cross section, differential distributions are needed for
several reasons. 
\begin{itemize}
\item Experimentally, cuts are needed to discriminate the signal
  form backgrounds, so the measured cross section can never be the
  fully inclusive total cross section.
\item Distributions are required to build realistic (higher order)
  Monte Carlo generators for the signal process.
\item Using additional observables beyond $\sigma_{\rm tot}$ adds
  information, helping to disentangle correlations among the parameters
  determined from a threshold scan, and increases the sensitivity to
  possible New Physics in top production and decay.
\end{itemize}
Differential distributions used so far are the top quark momentum
distribution ${\rm d}\sigma/{\rm d}p_t$ and the forward-backward
asymmetry $A_{\rm FB}$ of the cross section, but theoretical studies
exist also for the polarization of the top quarks.  In the following
we will briefly discuss these observables and their role in the
threshold scan, and comment also on the issue of rescattering corrections.\\ 

\begin{figure*}[htb]
\begin{center}
\includegraphics[bb=105 300 475 540,width=8cm]{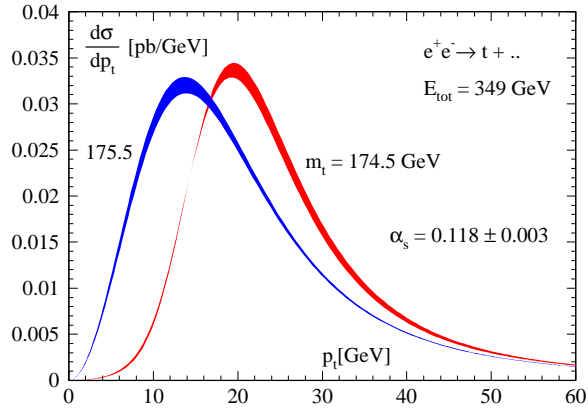}
\caption{Top quark momentum distribution in the threshold regime.}
\label{fig:dsdp}
\end{center}
\end{figure*}
The top quark momentum distribution ${\rm d}\sigma/{\rm d}p_t$ is
available in the framework of the next-to-next-to leading order (NNLO)
calculations~\cite{Hoang:1999zc} (see also Ref.~\cite{Nagano:1999nw}).
Close to threshold its form resembles a $1S$ Coulomb-like wave
function, becoming more symmetric well above threshold.
Fig.~13 shows ${\rm d}\sigma/{\rm d}p_t$ at a fixed c.m. energy of $349$ GeV 
for two different masses $m_t = 174.5, 175.5$
GeV, where the width of the band is obtained by varying $\alpha_s$
between $0.115$ and $0.121$.  The peak position of ${\rm d}\sigma/{\rm
  d}p_t$ is not much dependent on $\alpha_s$ but is very sensitive to
$m_t$, and a change of $\alpha_s$ mainly affects the normalization of
the distribution.  This is in contrast to $\sigma_{\rm tot}$, where
$m_t$ and $\alpha_s$ are strongly correlated parameters when fitting
theoretical predictions to (simulated) data. 

\begin{figure*}[htb]
\begin{center}
\includegraphics[bb=30 335 550 500,width=17cm]{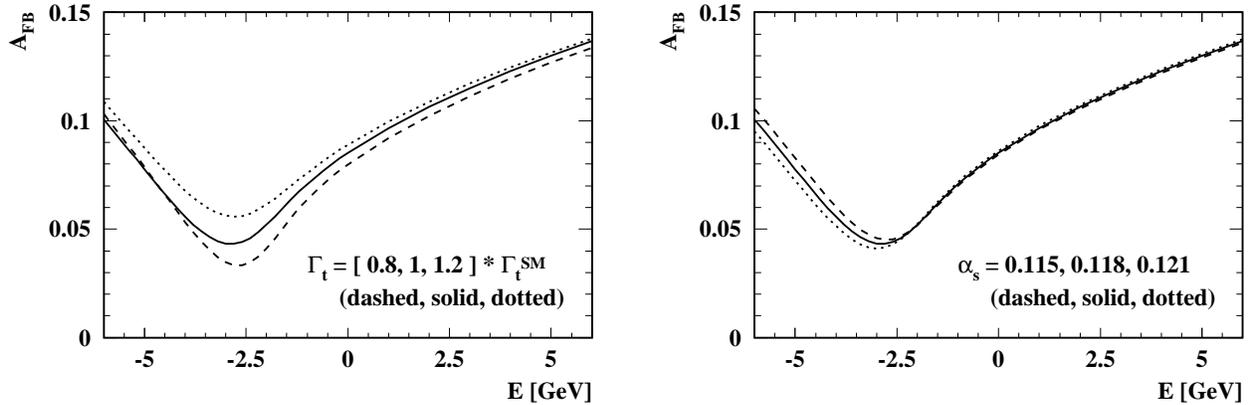}
\caption{Dependence of the forward-backward asymmetry $A_{\rm FB}$ on
  the top quark width (left) and the strong coupling $\alpha_s$ (right
  panel).} 
\label{fig:afb}
\end{center}
\end{figure*}
Interference of the leading vector current (through $\gamma$ and $Z$
exchange) with the suppressed axial vector contribution (from $Z$
only) leads to a forward-backward asymmetry $A_{\rm FB}$.  The size of
the asymmetry depends on how much the corresponding $S$ and $P$ wave
resonance contributions overlap and hence on the top quark width
$\Gamma_t$, but less on $\alpha_s$ (and $m_t$), see Fig.~14.

As already demonstrated in experimental studies~\cite{Martinez:2002st}, the two
observables ${\rm d}\sigma/{\rm d}p_t$ and $A_{\rm FB}$ can be used
together with $\sigma_{\rm tot}$ to make best use of the
available information and to disentangle the correlations between the
parameters $m_t, \Gamma_t, \alpha_s (, y_t)$ in a multi-paramter fit.  
However, the latest theoretical developments such as NNNLO corrections,
renormalization group improvement by summing large logarithms at NNLL
order (see section 2.6) or the effect of EW corrections (see
section 2.4) are dealing with the total cross section, whereas
distributions are available at fixed NNLO only.  This situation is quite
common, as higher order corrections are increasingly difficult for
differential cross sections, and not many distributions are known
beyond NLO.  Nevertheless, from a pragmatic point of view and
for use in Monte Carlos, it is legitimate to use the best available
prediction for the total cross section together with distributions
available at lower order.  By rescaling the distributions using
the total cross section, their scale dependence will decrease and a
consistent normalization will be ensured. 

\begin{figure*}[htb]
\begin{center}
\includegraphics[bb=20 575 210 650,width=6cm]{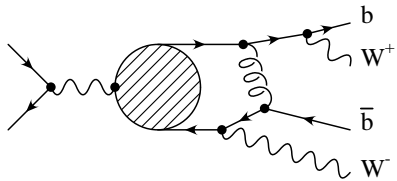}
\caption{Feynman diagram responsible for rescattering corrections
  between the top and the $\bar b$ from the $\bar t$ decay.}
\label{fig:rescat}
\end{center}
\end{figure*}
All recent NNLO, NNNLO as well as the renormalization group improved
calculations are not taking into account the so-called ``rescattering
corrections'', i.e. interactions between the $t, \bar t$ quarks and the
$b, \bar b$ quarks from the top decays, or between $b$ and $\bar b$ 
(see Fig.~16 for a typical Feynman diagram).  However,
such corrections were calculated numerically in
Refs.~\cite{Harlander:1996vg,Peter:1997rk} to NLO accuracy.  They are strongly
suppressed (vanishing at NLO) in the inclusive $\sigma_{\rm tot}$ but
generally important for distributions and typically of order $10\%$.
They therefore should be included in a realistic Monte Carlo description. 

In addition to differential distributions, also the top quark
polarization has been studied in the threshold
regime~\cite{Harlander:1996vg,Peter:1997rk,RefsPolar}.  Predictions
are available at NLO accuracy for all three polarization components,
including rescattering corrections.  Even for unpolarized $e^+, e^-$
beams the top quarks are (longitudinally) polarized to $40\%$, and for
the foreseen beam polarization the top quarks will be highly polarized.
Due to the short life-time the top polarization will be transmitted
undisturbed to the decay $b$'s and $W$'s and allow very interesting
studies.  Making use of the polarization, observables can be defined
which are sensitive to the top's electric dipole moment (probing
CP-violation beyond the Standard Model) or to anomalous top couplings
like $V+A$ admixtures to the weak interaction.  However, more realistic
experimental studies are needed in order to fully explore the
potential of polarization in the threshold regime.\\

\newpage

\noindent
{\em Experimental aspects of the threshold scan}\footnote{This section
    can only provide a brief review and the reader interested in more
    details is referred to the presentations of Stewart Boogert at this
    workshop.}

Precision measurements from a $t\bar t$ threshold scan require a very
precise knowledge of both the average c.m. energy $\langle \sqrt{s}
\rangle$ and of the luminosity spectrum ${\rm d}L/{\rm
  d}\sqrt{s}$~\cite{Boogert:2002jr}.  
This is not an easy task, as the beam dynamics at a linear collider is
not as constrained as it was e.g. at LEP, and only a single
measurement of bunches before collision will be possible.  The average
c.m. energy can be determined by energy (up- and downstream)
spectrometers in dedicated
beam line inserts, discussed in Working Group 4 of the Snowmass
Workshop.  A possible problem here could be a bias between the
spectrometer measurements and the collision c.m. energy.  Another
possibility to measure the beam energy is the use of physics
processes like $Z$ pair production, or radiative return (through
photon radiation) calibrated to the $Z$ peak~\cite{Hinze:2005kh}. 

The luminosity spectrum is determined by (a) the beam
spread, (b) beamstrahlung, and (c) initial state radiation (ISR).  All
three effects will lead to a smearing of the $t\bar t$ threshold cross
section, resulting in a significant reduction of the effective
luminosity and hence the observed cross section,
\begin{equation}
\sigma^{\rm obs}(\sqrt{s}) = \frac{1}{L_0} \int_0^1 L(x)\,
\sigma(x\sqrt{s})\,{\rm d}x\,.
\end{equation}
The influence of the three effects is demonstrated in Fig.~16. 
The beam spread will typically be $\sim 0.1\%$ at the ILC and will cause
comparably little smearing (though additional beam diagnostics may be
required to measure and monitor the beam spread), but beamstrahlung
and ISR are very important.  The luminosity spectrum ${\rm d}L/{\rm
  d}\sqrt{s}$ will lead to a systematic shift in the extracted top
mass which must be well understood; otherwise it could become the
dominant systematic error.  In the past, $t\bar t$ threshold studies
were carried out under the assumption that the spectrum is basically
known.  However it has turned out that the precise determination of
${\rm d}L/{\rm d}\sqrt{s}$ is a challenging task.  The proposed method
is to analyse the acollinearity of (large angle) Bhabha scattering
events, which is sensitive to a momentum mismatch between the beams
but insensitive to the absolute energy scale~\cite{Monig:2000bm}. For
this, the envisioned high resolution of the forward tracker will be
very important to achieve the required accuracy.\footnote{An
  additional complication is, that in Bhabha scattering ISR is not
  completely factorizable from final state radiation.} 
\begin{figure*}[htb]
\begin{center}
{\includegraphics[bb=0 10 567 350,width=8cm]{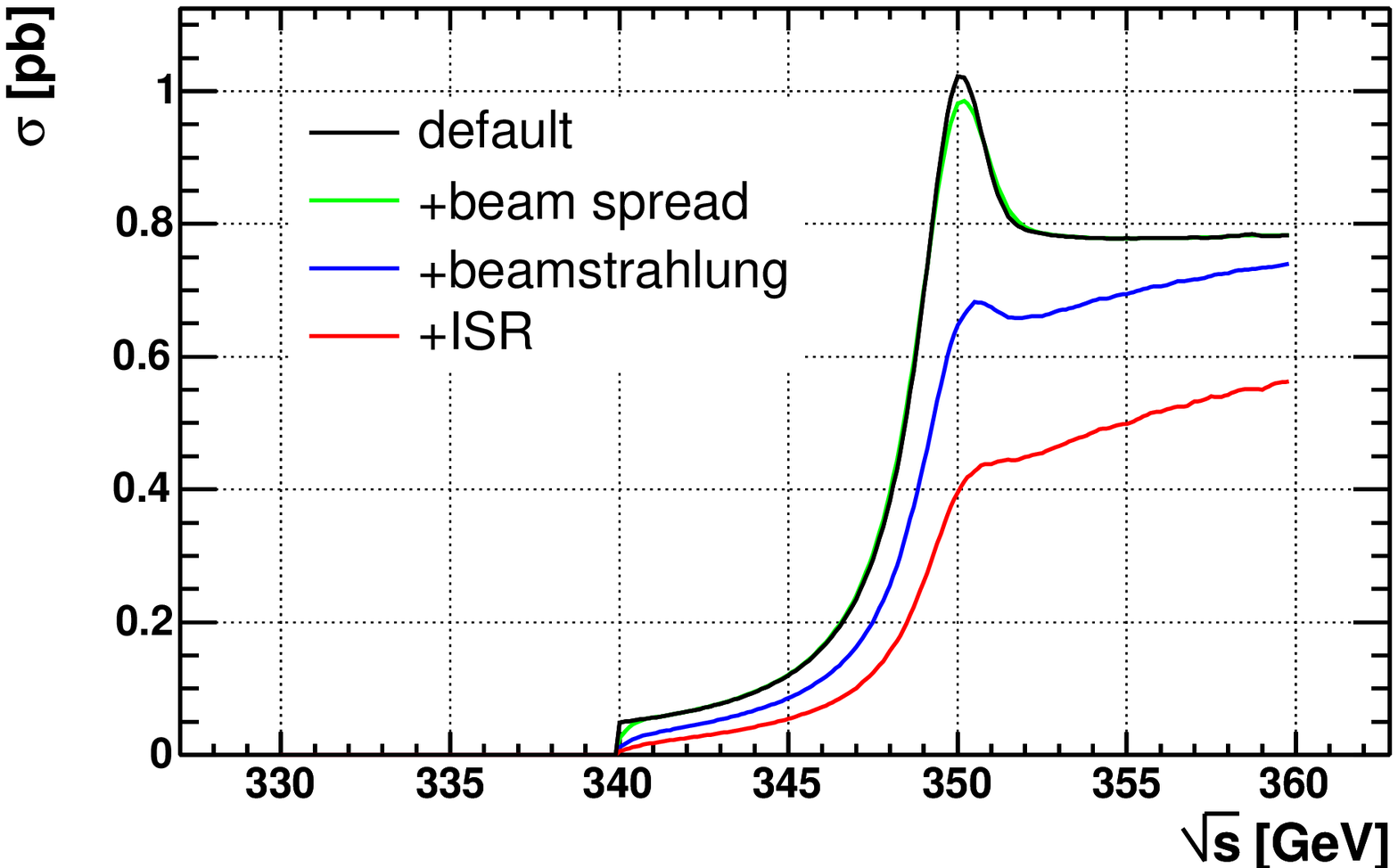}
\includegraphics[bb=0 10 567 350,width=8cm]{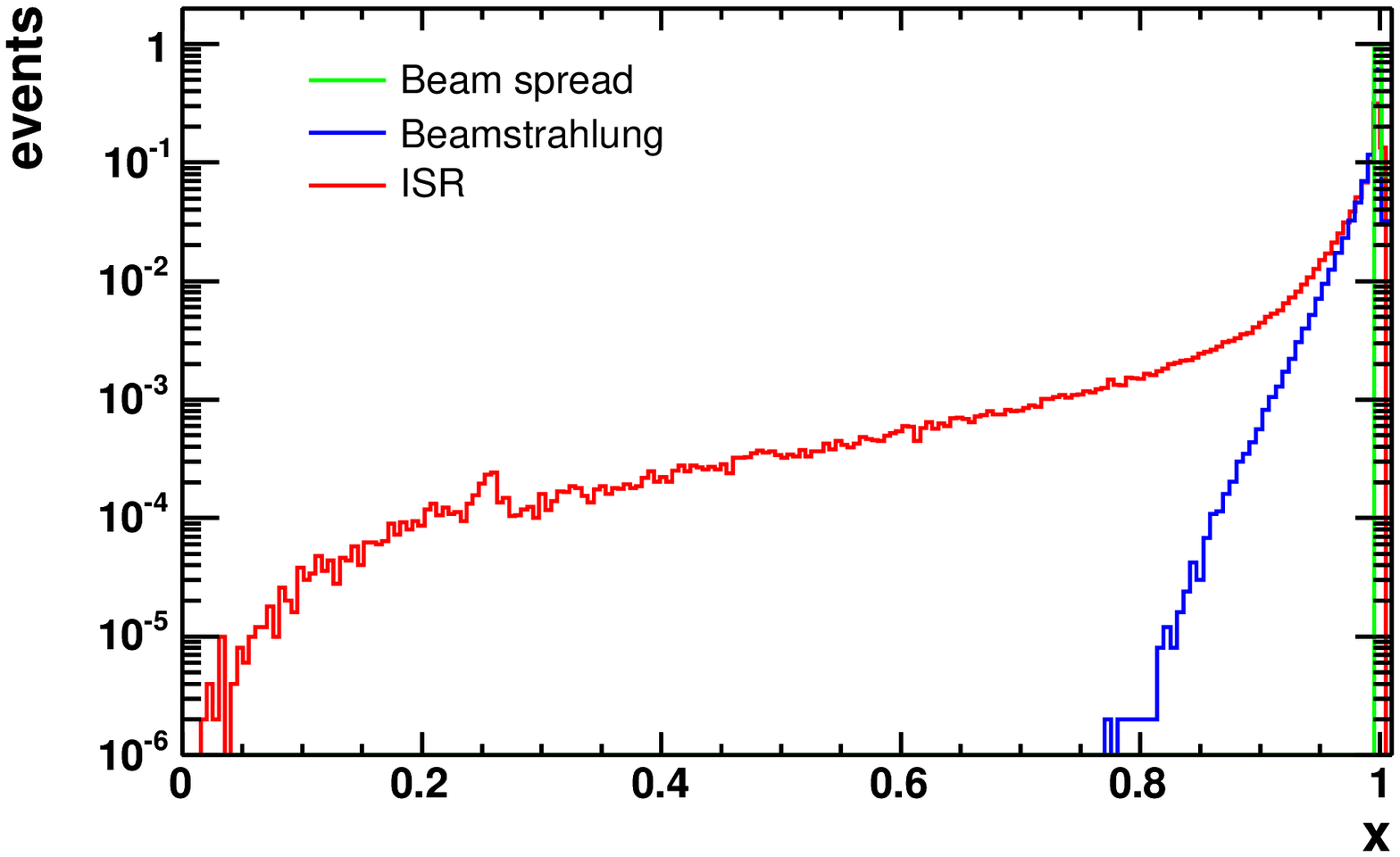}}
\caption{Left: Smearing of the theoretical $t\bar t$ cross section
  (`default') by beam effects and initial state radiation.  Right
  panel: Simulation of beam spread, beamstrahlung and ISR as
  distributions of $x=\sqrt{s}/\sqrt{s_0}$ (where $\sqrt{s_0}$ is the nominal
  c.m. energy of the machine).}
\label{fig:sigmaexp}
\end{center}
\end{figure*}

In the simulation of the luminosity spectrum beamstrahlung is the main
unknown.  Integrated ILC accelerator simulations for these machine
dependent effects are becoming available, including effects from the
linac beam spread, the beam diagnostics and delivery system and the
simulation of the collision dynamics using the package Guinea-pig.  At
this workshop Stewart Boogert presented first results of spectrum
parameterizations (using the package CIRCE), based on new simulations
from G. White for $\sqrt{s}=350$ GeV.  The new parameterizations show 
variations from earlier results which were based on beam simulations at
higher energies only.  The parameterizations will be used as input in the
simulation of the Bhabha scattering (based on the package BHWIDE) for
the acollinearity analysis.  These involved simulations are required to
extract ${\rm d}L/{\rm d}\sqrt{s}$ and to understand the uncertainties
of the reconstructed spectrum.  Only then can one quantify in physics
analyses like the $t\bar t$ threshold scan, to which extent the
uncertainties in the luminosity spectrum affect the accuracy of the
intended measurements like the top quark mass, width, etc.

The effects due to ISR are in principle machine independent and
calculable to high precision in QED.  Nevertheless, ISR effects
complicate the measurement of beamstrahlung and accelerator energy
spread, and one has to ensure that the precision of the theoretical
formulae used in the Monte Carlo codes is sufficient.  Similarly, the
accuracy of differential distributions from programs used for the
simulation of the wide angle Bhabha events must be assessed, and
recent theoretical results~\cite{Bhabha} may have to be included.

The recent developments discussed here make it clear that much work
remains to be done before the sophisticated analysis techniques for
the $t\bar t$ threshold scan are understood at the required level of
accuracy.  For a complete study, which should take into account the total cross
section, distributions, full beam effects including asymmetric
boosts and detector effects, a full Monte Carlo generator will be needed,
and a project for this has been started at Snowmass.  With a more detailed 
experimental study it will also be possible to optimize the scan
strategy, i.e. how the available luminosity should be distributed
among different energy scan points.

The analysis of beamstrahlung and its impact on the physics program at
different energies will also be relevant for the optimization of the
machine design.  Concerning detectors, the demands of the top pair counting
experiment are most probably not problematic for the existing detector
designs.  However, the requirements for Bhabha scattering and
radiative $Z$ return might set the requirements of the low angle
tracking system and the electromagnetic calorimeter.  Finally, it
should also be noted, that the understanding of other physics
processes like the $WW$ or SUSY thresholds will largely benefit from
the top case, and $t\bar t$ at threshold should be regarded as the benchmark.

\section{QCD EFFECTS IN ILC PHYSICS}
\label{sec_pQCD}


\subsection{Precision QCD at the ILC: \mbox{\boldmath$e^+e^- \rightarrow 3$} Jets \\ \small{{\it A. Gehrmann-De Ridder, 
      T. Gehrmann, E.W.N. Glover}}}
\label{sg1}

The production of light quark-antiquark pairs in electron-positron 
annihilation gives rise to final states containing QCD jets. Depending on the 
amount of additional hard QCD radiation, one obtains final states with 
a certain number of jets: if only quark and antiquark are hard, two-jet final
states are produced, one additional hard gluon yields a three-jet final state,
two extra gluons or a secondary quark-antiquark pair can give rise to four-jet 
final states and so on. Studying these multi-jet final states,
one can probe many aspects of perturbative QCD. Three-jet final 
states and related event shape observables were studied extensively at LEP in 
order to determine the strong coupling constant $\alpha_s$, which 
controls the probability of radiating a hard gluon in these events. 
The measurement of four-jet-type observables at LEP established the 
gauge group structure of QCD. Five-jet and higher multiplicities were 
sometimes considered in new physics searches, where QCD-induced 
processes form a theoretically calculable background. 

The LEP measurements of three-jet observables are of a very high 
statistical precision. The extraction of $\alpha_s$ from these 
data sets relies on a comparison of the data with theoretical predictions. 
Comparing the different sources of error in this extraction~\cite{bethke}, 
one finds that the purely experimental error is negligible compared to 
the theoretical uncertainty. There are two sources of theoretical 
uncertainty: the theoretical description of the parton-to-hadron 
transition (hadronization uncertainty)
and the theoretical calculation of parton-level jet production 
(perturbative or scale uncertainty). Although the precise 
size of the hadronization uncertainty is debatable and perhaps often 
underestimated, it is certainly appropriate to consider the scale 
uncertainty as the dominant source of theoretical error on the precise 
determination of  $\alpha_s$ from three-jet observables. This scale uncertainty
arises from truncating the perturbative QCD expansion of jet observables 
at the next-to-leading order (NLO) and can be improved considerably by 
computing next-to-next-to-leading order (NNLO) corrections. 

Given the planned luminosity of the ILC, one expects that this collider 
will deliver jet-production data of a statistical quality similar to LEP. 
An attractive perspective 
of such a measurement at the ILC would be to determine the evolution of 
$\alpha_s$ over a wide energy range, which is potentially sensitive to
new physics thresholds. Concerning uncertainties on such a determination, it
is worthwhile to note that the hadronization corrections become less important 
at higher energies, thus leaving the scale uncertainty as dominant source 
of theoretical error.
Experimental aspects of such measurements, especially issues related to 
the beam energy profile (which were irrelevant at LEP) were not studied up 
to now, and certainly deserve further attention.

In the recent past, many steps towards the 
NNLO calculation of 
$e^+e^- \to 3$~jets have been accomplished
(see Ref.~\cite{ourant} and references therein).
Foremost, the relevant two-loop $1 \to 3$ matrix elements 
are now available.
One-loop corrections to $1\to 4$ matrix elements 
have been known for longer and form part 
of NLO calculations of $e^+e^- \to 4$~jets.
These NLO matrix elements naturally
contribute to $e^+e^- \to 3$~jets
 at NNLO if one of the 
partons involved becomes unresolved (soft or collinear). 
In this case, the infrared 
singular parts of the matrix elements need to be extracted and integrated 
over the phase space appropriate to the unresolved configuration 
to make the infrared pole structure explicit. 
As a final ingredient, the tree level $1\to 5$ processes also
contribute to  $e^+e^- \to 3$~jets
 at NNLO.
These contain double real radiation singularities corresponding to two 
partons becoming simultaneously soft and/or 
collinear. To compute the contributions from single unresolved radiation 
at one-loop and double real radiation at tree-level,  one has to 
find subtraction terms which coincide with the full matrix elements 
in the unresolved limits 
and are still sufficiently simple to be integrated analytically in order 
to cancel  their  infrared pole structure with the two-loop virtual 
 contributions. In the following, we present a new method, named 
antenna subtraction, to carry out NNLO calculations of jet observables and 
discuss its application to $e^+e^- \to 3$~jets.

In electron-positron annihilation, 
an $m$-jet cross section at NLO is obtained by summing contributions from 
$(m+1)$-parton tree level and $m$-parton one-loop processes:
\begin{displaymath}
{\rm d}\sigma_{NLO}=\int_{{\rm d}\Phi_{m+1}}\left({\rm d}\sigma^{R}_{NLO}
-{\rm d}\sigma^{S}_{NLO}\right) +\left [\int_{{\rm d}\Phi_{m+1}}
{\rm d}\sigma^{S}_{NLO}+\int_{{\rm d}\Phi_{m}}{\rm d}\sigma^{V}_{NLO}\right].
\end{displaymath}
The cross section ${\rm d}\sigma^{R}_{NLO}$ is the $(m+1)$-parton tree-level
cross section, 
while 
${\rm d}\sigma^{V}_{NLO}$ is the one-loop virtual correction to the 
$m$-parton Born cross section ${\rm d}\sigma^{B}$. Both contain infrared 
singularities, which are explicit poles in $1/\e$ in ${\rm d}\sigma^{V}_{NLO}$,
while becoming explicit in ${\rm d}\sigma^{R}_{NLO}$ only after integration 
over the phase space.  In general, this integration  involves the
(often iterative) definition of the jet observable, such that  an analytic
integration is not feasible (and also not appropriate). Instead,   one would
like to have a flexible method that can be easily adapted to  different jet
observables or jet definitions. Therefore, the infrared singularities  
of the real radiation
contributions should be extracted using  infrared subtraction  terms.
One introduces ${\rm d}\sigma^{S}_{NLO}$, which is a counter-term for  
 ${\rm d}\sigma^{R}_{NLO}$, having the same unintegrated
singular behavior as ${\rm d}\sigma^{R}_{NLO}$ in all appropriate limits.
Their difference is free of divergences 
and can be integrated over the $(m+1)$-parton phase space numerically.
The subtraction term  ${\rm d}\sigma^{S}_{NLO}$ has 
to be integrated analytically over all singular regions of the 
$(m+1)$-parton phase space. 
The resulting cross section added to the virtual contribution 
yields an infrared finite result. 
Several methods for  constructing
 NLO subtraction terms systematically  were proposed in the 
literature~\cite{cs,singleun,cullen,ant}. For some of these methods, 
extension to NNLO was discussed~\cite{nnlosub} and partly worked out. 
We focus on the  antenna subtraction method~\cite{cullen,ant}, 
which we extend to NNLO. 

The basic idea of the antenna subtraction approach at NLO is to construct 
the subtraction term  
${\rm d}\sigma^{S}_{NLO}$
from antenna functions. Each antenna function encapsulates 
all singular limits due to the 
 emission of one unresolved parton between two color-connected hard
partons (tree-level three-parton antenna function).
This construction exploits the universal factorization of 
phase space and squared matrix elements in all unresolved limits,
depicted in Fig.~17.
The individual antenna functions are obtained by normalizing 
three-parton tree-level matrix elements to the corresponding two-parton 
tree-level matrix elements. 
\begin{figure}[t] 
\epsfig{file=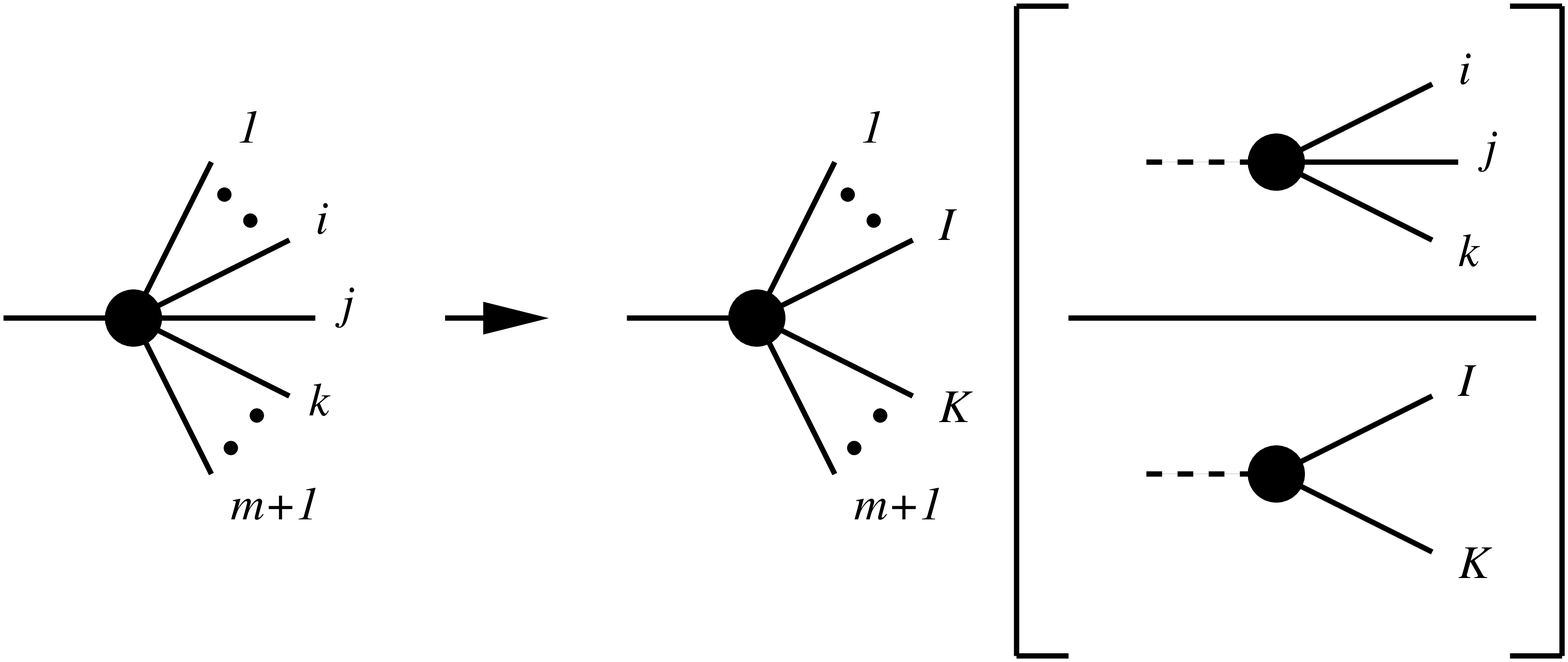,height=3.6cm}
\caption{Illustration of NLO antenna factorization representing the
factorization of both the squared matrix elements and the 
$(m+1)$-particle phase
space. The term in square brackets
represents both the antenna function and the antenna phase space.
\label{fig:nloant}
}\vspace{3mm}
\epsfig{file=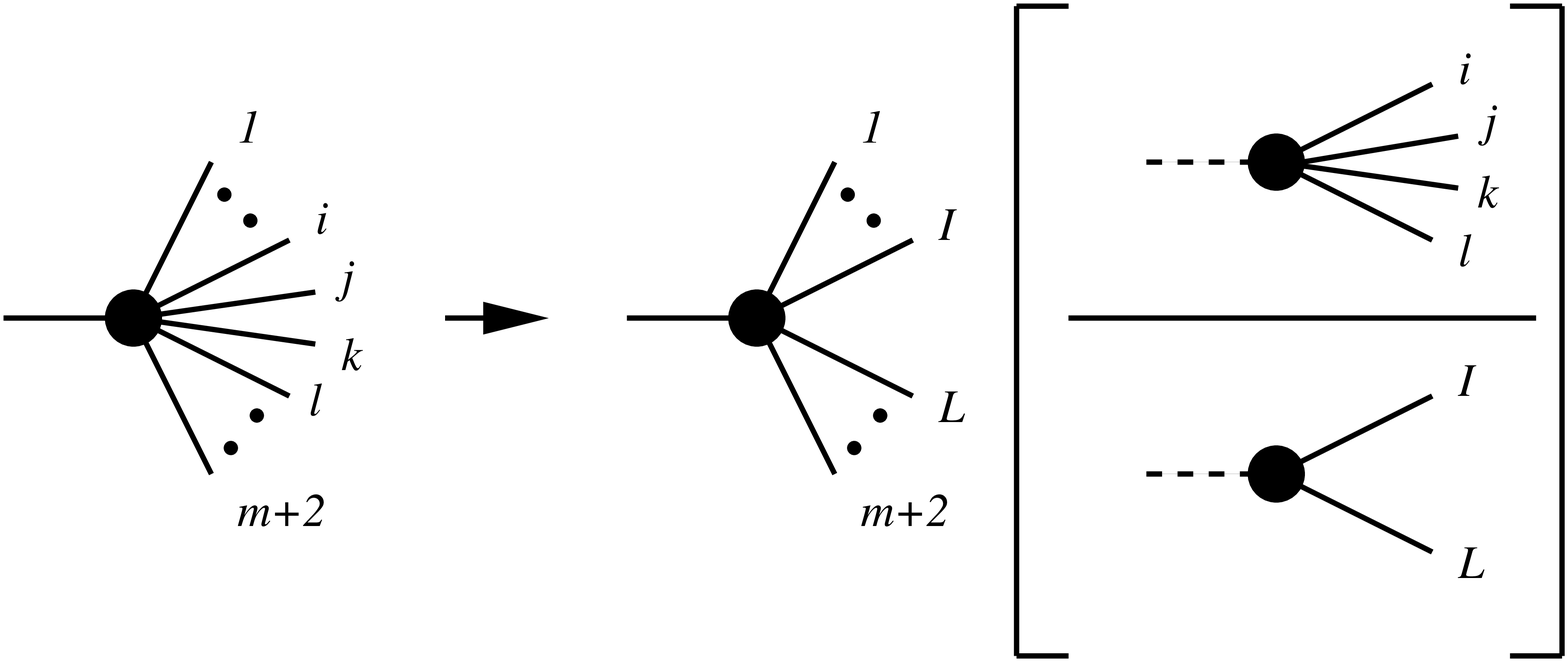,,height=3.6cm}
\caption{\label{fig:sub2a} Illustration 
of NNLO antenna factorization representing the
factorization of both the squared matrix elements and the $(m+2)$-particle 
phase
space when the unresolved particles are color connected. 
The term in square brackets
represents both the antenna function and the antenna phase space.}
\end{figure}

At NNLO, the $m$-jet production is induced by final states containing up to
$(m+2)$ partons, including the one-loop virtual corrections to $(m+1)$-parton final 
states. As at NLO, one has to introduce subtraction terms for the 
$(m+1)$- and $(m+2)$-parton contributions. 
Schematically the NNLO $m$-jet cross section reads,
\begin{eqnarray*}
{\rm d}\sigma_{NNLO}&=&\int_{{\rm d}\Phi_{m+2}}\left({\rm d}\sigma^{R}_{NNLO}
-{\rm d}\sigma^{S}_{NNLO}\right) + \int_{{\rm d}\Phi_{m+2}}
{\rm d}\sigma^{S}_{NNLO}\nonumber \\ 
&&+\int_{{\rm d}\Phi_{m+1}}\left({\rm d}\sigma^{V,1}_{NNLO}
-{\rm d}\sigma^{VS,1}_{NNLO}\right)
+\int_{{\rm d}\Phi_{m+1}}{\rm d}\sigma^{VS,1}_{NNLO}  
\nonumber \\&&
+ \int_{{\rm d}\Phi_{m}}{\rm d}\sigma^{V,2}_{NNLO}\;,
\end{eqnarray*}
where $\d \sigma^{S}_{NNLO}$ denotes the real radiation subtraction term 
coinciding with the $(m+2)$-parton tree level cross section 
 $\d \sigma^{R}_{NNLO}$ in all singular limits~\cite{doubleun}. 
Likewise, $\d \sigma^{VS,1}_{NNLO}$
is the one-loop virtual subtraction term 
coinciding with the one-loop $(m+1)$-parton cross section 
 $\d \sigma^{V,1}_{NNLO}$ in all singular limits~\cite{onelstr}. 
Finally, the two-loop correction 
to the $m$-parton cross section is denoted by ${\rm d}\sigma^{V,2}_{NNLO}$.

Both types of NNLO subtraction terms can be constructed from antenna 
functions. In ${\rm d}\sigma^{S}_{NNLO}$, we have to distinguish four
different types of unresolved configurations:
(a) One unresolved parton but the experimental observable selects only
$m$ jets;
(b) Two color-connected unresolved partons (color-connected);
(c) Two unresolved partons that are not color connected but share a common
radiator (almost color-unconnected);
(d) Two unresolved partons that are well separated from each other 
in the color 
chain (color-unconnected). Among those, configuration (a) is properly 
accounted for by a single tree-level three-parton antenna function 
like used already at NLO. Configuration (b) requires a 
tree-level four-parton antenna function (two unresolved partons emitted 
between a pair of hard partons) 
as shown in Fig.~18, while (c) and (d) are accounted for by 
products of two tree-level three-parton antenna functions. 
\begin{figure}
\epsfig{file=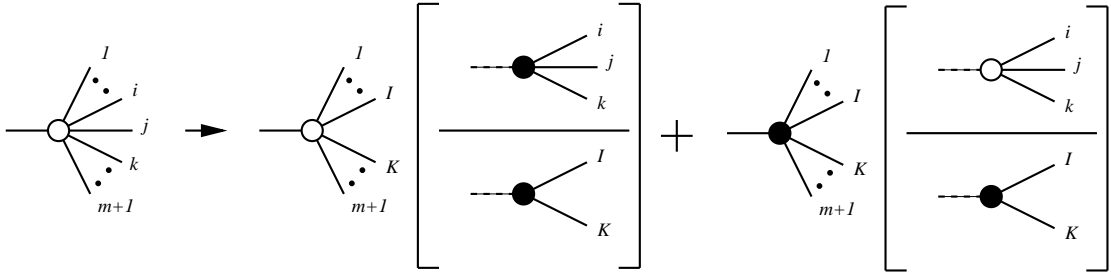,height=3.6cm}
\caption{Illustration of NNLO antenna factorization representing the
factorization of both the one-loop
``squared" matrix elements (represented by the white blob)
and the $(m+1)$-particle phase
space when the unresolved particles are color connected. 
\label{fig:subv}}
\end{figure}

In single unresolved limits, the one-loop cross section 
$\d \sigma^{V,1}_{NNLO}$ is described by the sum of two terms~\cite{onelstr}: 
a tree-level splitting function times a one-loop cross section 
and a one-loop splitting function times a tree-level cross section. 
Consequently, the 
one-loop single unresolved subtraction term $\d \sigma^{VS,1}_{NNLO}$
is constructed from tree-level and one-loop three-parton antenna functions,
as sketched in Fig.~19. Several other terms in  
 $\d \sigma^{VS,1}_{NNLO}$ cancel with the results
from the integration of terms in 
the double real radiation subtraction term  $\d \sigma^{S}_{NNLO}$
over the phase space appropriate to one of the unresolved partons, thus 
ensuring the cancellation of all explicit infrared poles in the difference 
$\d \sigma^{V,1}_{NNLO}-\d \sigma^{VS,1}_{NNLO}$.

Finally, all remaining terms in 
$\d \sigma^{S}_{NNLO}$ and $\d \sigma^{VS,1}_{NNLO}$ have to be integrated 
over the four-parton and three-parton antenna phase spaces. After   
integration, the infrared poles are rendered explicit and
cancel with the 
infrared pole terms in the two-loop squared matrix element 
$\d \sigma^{V,2}_{NNLO}$. 

The  subtraction terms $\d \sigma^{S}_{NLO}$,
$\d \sigma^{S}_{NNLO}$ 
and $\d \sigma^{VS,1}_{NNLO}$ require three different types of 
antenna functions corresponding to the different pairs of hard partons 
forming the antenna: quark-antiquark, quark-gluon and gluon-gluon antenna 
functions. In the past~\cite{cullen,ant}, NLO antenna functions were 
constructed by imposing definite properties in 
all single unresolved limits (two collinear limits
and one soft limit for each 
antenna). 
This procedure turns out to be impractical at NNLO, where each antenna 
function must have definite behaviors in a large number of single and 
double unresolved limits. Instead, we derive these antenna functions in 
a systematic manner from physical matrix elements known to possess the 
correct limits. The quark-antiquark antenna functions can be obtained 
directly from 
the $e^+e^- \to 2j$ real radiation corrections at NLO and NNLO~\cite{our2j}. 
For quark-gluon and gluon-gluon antenna functions, effective Lagrangians 
are used to obtain tree-level processes yielding a quark-gluon or 
gluon-gluon final state. The antenna functions are then obtained from 
the real radiation corrections to these processes. 
Quark-gluon antenna functions 
were derived~\cite{chi} from the purely QCD 
(i.e.\ non-supersymmetric) NLO and NNLO corrections to the decay of 
a heavy neutralino into a massless gluino plus partons~\cite{hw}, while 
gluon-gluon antenna functions~\cite{h} result from the QCD corrections 
to Higgs boson decay into partons~\cite{hgg}. 

All tree-level three-parton and four-parton antenna functions 
and  three-parton one-loop antenna functions are listed in Ref.~\cite{ourant}, 
where we also provide their integrated forms, obtained using the 
phase space integration techniques described in Ref.~\cite{ggh}.

In Refs.~\cite{ourant,CFsquare} we derived
the $1/N^2$-contribution to 
the NNLO corrections to  $e^+e^- \to 3$~jets. This color factor receives
contributions from 
$\gamma^*\to q\bar q ggg$ and $\gamma^*\to q\bar q q\bar qg$
at tree-level~\cite{tree5p},  
$\gamma^*\to q\bar q gg$ and $\gamma^*\to q\bar q q\bar q$  at
one-loop~\cite{onel4p} 
and $\gamma^*\to q\bar q g$ at two-loops~\cite{twol3p}. 
The four-parton and 
five-parton final states contain infrared singularities, which need to
be extracted using the antenna subtraction formalism.

In this contribution, all gluons are effectively photon-like, and couple 
only to the quarks, but not to each other. Consequently, only quark-antiquark
antenna functions appear in the construction of the subtraction terms. 

Starting from the program {\tt EERAD2}~\cite{cullen}, which computes 
the four-jet
production at NLO, we implemented the NNLO antenna subtraction method 
for the $1/N^2$ color factor contribution to $e^+e^-\to 3j$. {\tt EERAD2}
already  contains the five-parton and four-parton 
matrix elements relevant here, as well as the NLO-type subtraction terms. 

The implementation contains three channels, classified 
by their partonic multiplicity: 
(a) in the five-parton channel, we
integrate ${\rm d}\sigma_{NNLO}^{R} - {\rm d}\sigma_{NNLO}^{S}$;
(b) in the four-parton channel, we integrate
${\rm d}\sigma_{NNLO}^{V,1} - {\rm d}\sigma_{NNLO}^{VS,1}$;
(c) in the three-parton channel, we integrate
${\rm d}\sigma_{NNLO}^{V,2} +{\rm d}\sigma_{NNLO}^{S}
+ {\rm d}\sigma_{NNLO}^{VS,1}$.
The numerical integration over these channels is carried out by Monte Carlo 
methods. 

By construction, the integrands in the four-parton and 
three-parton channel are free of explicit infrared poles. In the 
five-parton and four-parton channel, we tested the proper implementation of 
the subtraction by generating trajectories of phase space points approaching 
a given single or double unresolved limit. 
Along these trajectories, we observe that the 
antenna subtraction terms converge locally towards the physical matrix 
elements, and that the cancellations among individual 
contributions to the subtraction terms take place as expected. 
Moreover, we checked the correctness of the 
subtraction by introducing a 
lower cut (slicing parameter) on the phase space variables, and observing 
that our results are independent of this cut (provided it is 
chosen small enough). This behavior indicates that the 
subtraction terms ensure that the contribution of potentially singular 
regions of the final state phase space does not contribute to the numerical 
integrals, but is accounted for analytically. 

As a final point, we noted in Ref.~\cite{ourant} that the 
infrared poles of the two-loop (including one-loop times one-loop) correction
to $\gamma^*\to q\bar qg$ are canceled in all color factors by a 
combination of integrated three-parton and four-parton
antenna functions.
This highly non-trivial cancellation 
clearly illustrates that the antenna functions derived 
here 
correctly approximate QCD matrix elements in all infrared singular limits at 
NNLO. They also outline the structure of infrared 
cancellations in $e^+e^-\to 3j$ at NNLO, and indicate the structure of the 
subtraction terms in all color factors.

In this talk, we discussed the theoretical prerequisites for 
performing precision QCD studies on existing LEP data and at the ILC. 
In particular, the precise extraction of the strong coupling constant 
$\alpha_s$ requires improved theoretical predictions to reduce the 
scale error inherent to calculations in perturbative QCD. At present, 
this extraction relies on the calculation of $e^+e^- \to 3$~jets at NLO 
accuracy, and we reported on progress towards the NNLO calculation. 

This calculation requires a new method 
for the subtraction of infrared 
singularities 
which we call antenna subtraction. We introduced 
subtraction terms for double real radiation at tree level and 
single real radiation at one loop based on 
antenna functions. These antenna 
functions describe the color-ordered radiation of unresolved 
partons between a 
pair of hard (radiator) partons. All antenna functions at NLO and NNLO 
can be derived systematically from physical matrix elements. 

Using this method, we implemented 
the NNLO corrections to the subleading color contribution to 
$e^+e^- \to 3$~jets into a flexible parton level event generator program, 
 and are currently proceeding with the implementation~\cite{new3j}
 of the full set 
of color factors relevant to 
the  NNLO corrections to $e^+e^- \to 3$~jets. 


\subsection{Study of \mbox{\boldmath$V_L V_L \rightarrow t\bar{t}$} at the ILC Including \mbox{\boldmath${\cal O}(\alpha_s^2)$} 
Corrections~\cite{Godfrey:2005gp}
\\ \small{{\it S. Godfrey}}}
\label{sgod}

Understanding the mechanism of electroweak symmetry breaking (EWSB) is a 
primary goal of the LHC and ILC \cite{Weiglein:2004hn}.  
While much effort has been devoted to the weakly interacting weak 
sector scenario the strongly interacting weak sector (SIWS) remains 
a possibility.  
Because the $t$-quark mass is the same order of magnitude as the scale of 
EWSB it has long been suspected that $t$-quark properties may provide 
hints about the nature of EWSB and the subprocess $V_LV_L\to t\bar{t}$ 
has been suggested as a probe.  
While $V_LV_L\to t\bar{t}$ can be studied at both hadron colliders and
$e^+e^-$ colliders, the overwhelming QCD backgrounds will 
likely make it impossible to study 
the $V_LV_L\to t\bar{t}$ subprocess at the LHC \cite{Han:2003pu}.  
In contrast, the ILC offers a much cleaner environment.
But to be able to attach meaning to precision 
measurements it is necessary to understand radiative corrections, both 
electroweak and QCD.  
Here we show 
 ${\cal O} (\alpha_s) $ corrections to the tree level electroweak 
$V_L V_L \rightarrow t \bar t$ process in the SM at the ILC.
Due to space limitations we point the interested reader to 
Ref.~\cite{Godfrey:2004tj} for a more detailed account and a more 
complete set of references.

We are interested in the subprocesses $VV\to t\bar{t}$ which occur in 
the processes
$e^+e^- \to  \ell_1 \ell_2 +  V V \to \ell_1 \ell_2 + t\bar{t}$
where $\ell_1 \ell_2$ is $\nu\bar{\nu}$ for the $W^+W^-\to t\bar{t}$ 
subprocess and $e^+e^-$ for the $ZZ \to t\bar{t}$ subprocess. The 
vector bosons are treated as partons inside the $e^+$ and $e^-$ using 
the effective boson approximation \cite{eva,Dawson:1986tc}.  
The total cross section is then
obtained by integrating the $W$ (or $Z$) 
luminosities with the subprocess cross section \cite{Kauffman:1989aq}.

The ${\cal O}(\alpha_s)$ corrections for 
the processes $W^+W^-\to t\bar{t}$ $ZZ\to t\bar{t}$ are calculated 
using the
FeynArts, FormCalc and LoopTools packages \cite{fclt}.  
The QCD corrections to $W^+W^- \to t\bar{t}$ are shown in 
Fig.~20 (left side).
The infrared singularity in the vertex 
corrections are canceled by the soft contributions from the process 
$W^+W^-\to t\bar{t}g$ which are shown in Fig.~20 (right side). 
We regulate the IR-singularity by introducing a gluon mass which 
is equivalent to standard dimensional regularization
for processes with no triple gluon vertex present.
This approach has the additional benefit that
varying the value of the gluon mass acts as a check of 
the numerical cancellations between the different contributions.

\begin{figure}[t]
\includegraphics[width=70mm]{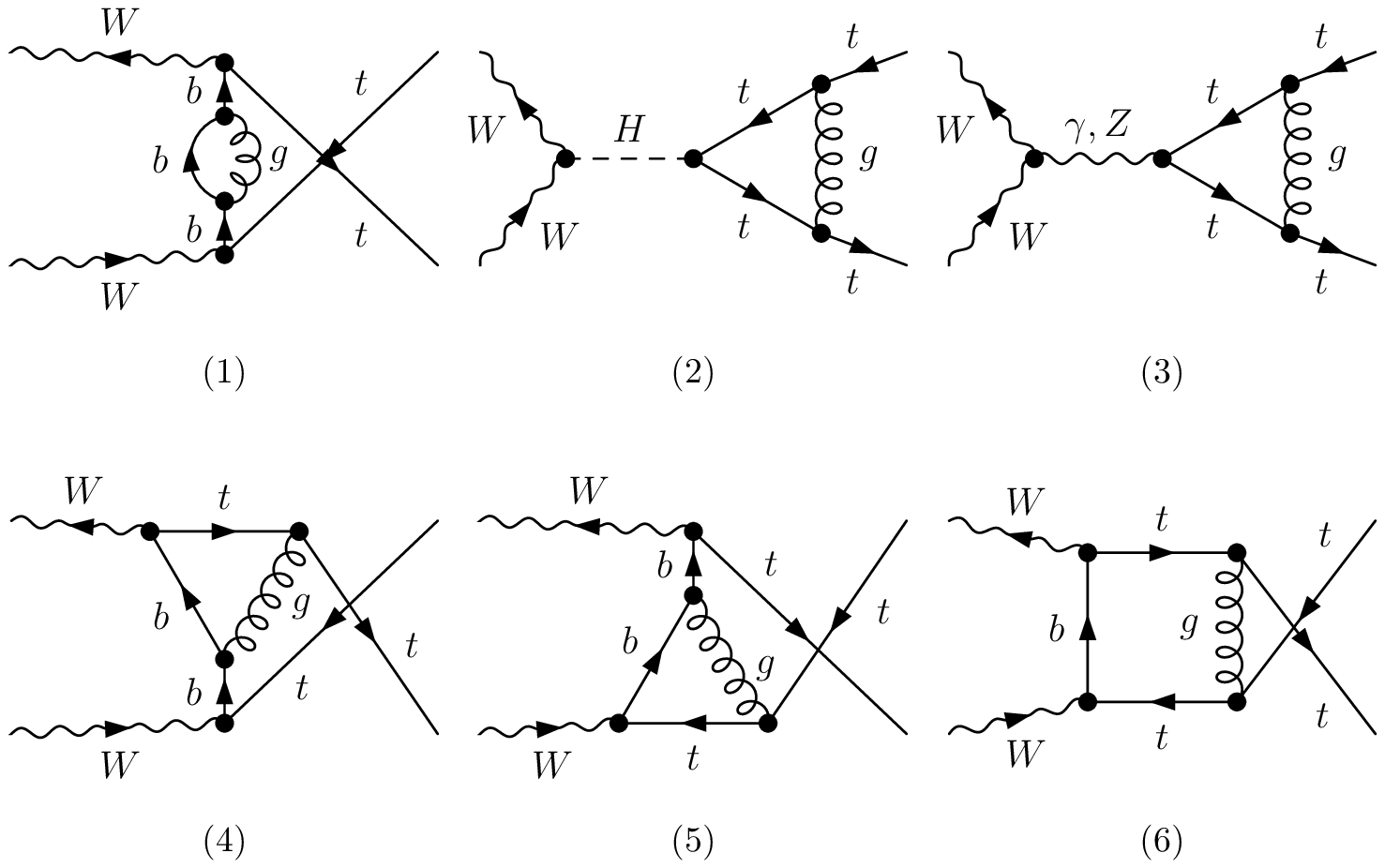}
\qquad\qquad
\includegraphics[width=70mm]{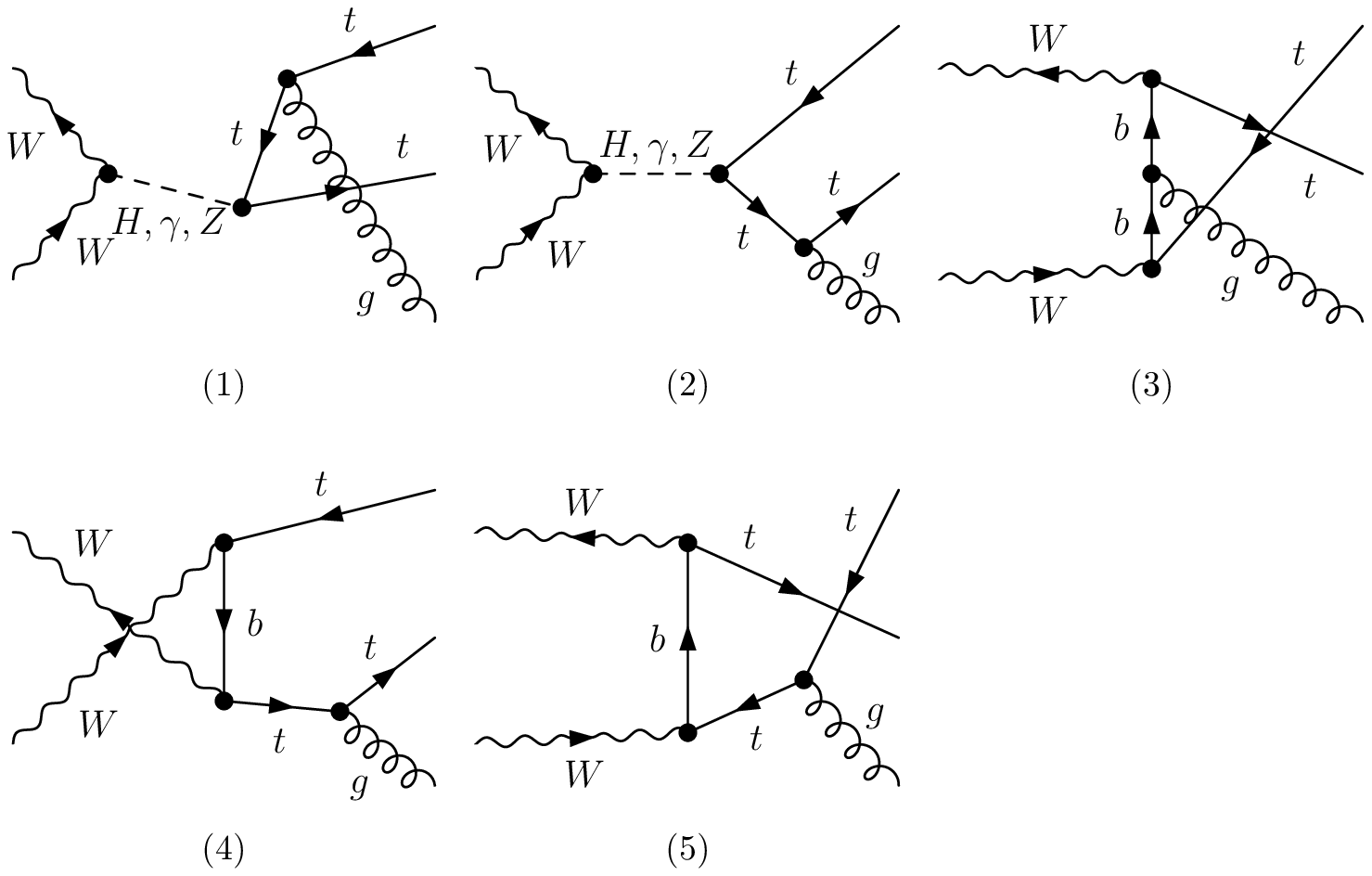}
\caption{${\cal O} (\alpha_s) $ QCD corrections to $W^+W^-\to t\bar{t}$.
(a) Virtual QCD contributions to $W^+W^-\to t\bar{t}$.
(b) Feynman diagrams for $W^+W^-\to t\bar{t}+g$. } \label{diagrams}
\end{figure}

We include in our results the kinematic cuts
$m_{t\bar{t}}>400$~GeV and $p_T^{t,\bar{t}}>10$~GeV.  
Since the longitudinal scattering cross section is much larger 
than the $TT$ and $TL$ cases and it is the longitudinal gauge 
boson processes which corresponds to the Goldstone bosons of the 
theory we will henceforth only include results for $V_L V_L$ scattering.

\begin{figure}[t]
\includegraphics[width=70mm]{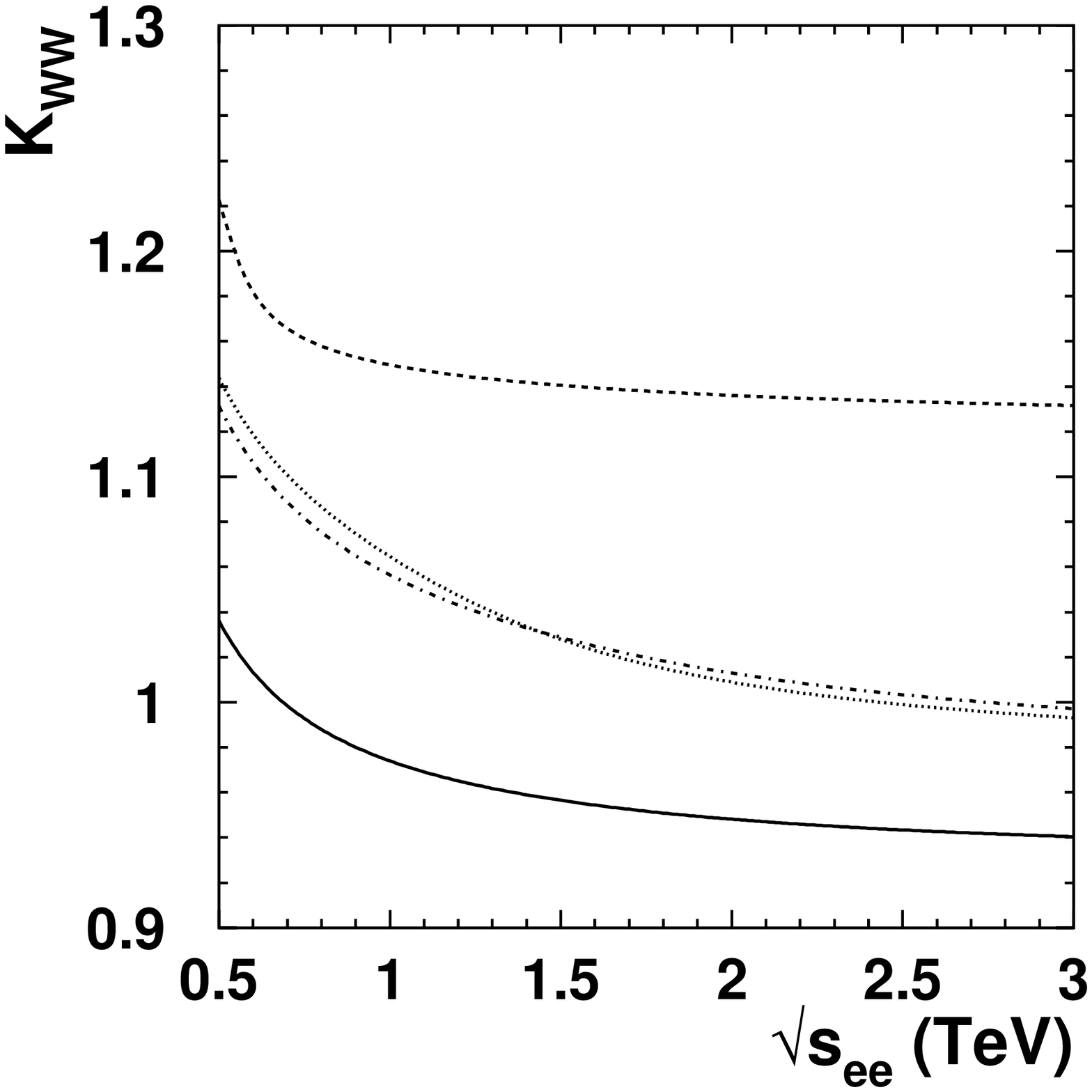}
\includegraphics[width=70mm]{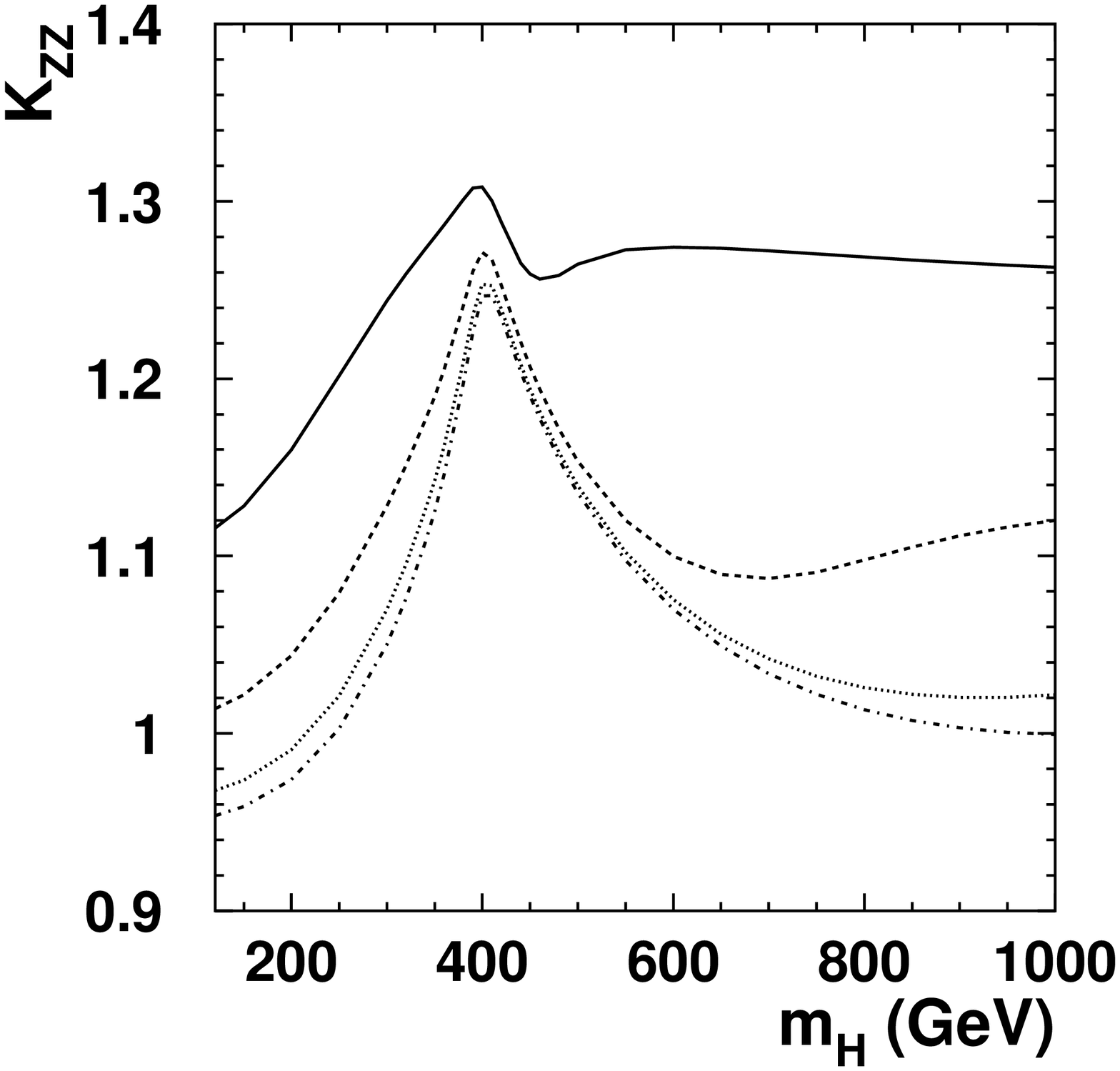}
\caption[]{(left side)
The K-factors as a function of  $\sqrt{s}_{e^+e^-}$
for $e^+e^- \to \nu\bar{\nu} t \bar t$ (via $W^+_LW^-_L$ fusion).
The solid line is for $M_H=120$~GeV, the dashed line for
$M_H=500$~GeV, the dotted line for $M_H=1$~TeV, and the dot-dashed 
line for $M_H=\infty$ (LET). 
(right side)
The K-factor as a function of  $M_H$
for $e^+e^- \to \nu\bar{\nu} t \bar t$ (via $W^+_LW^-_L$ fusion).
The solid line is for $\sqrt{s}_{e^+e^-}=500$~GeV,
the dashed line for $\sqrt{s}_{e^+e^-}=1$~TeV,
the dotted line for $\sqrt{s}_{e^+e^-}=2$~TeV, and
the dot-dashed line for $\sqrt{s}_{e^+e^-}=3$~TeV.
See text for an explanation of the K-factor.
} \label{kfactor}
\end{figure}

The QCD corrections to longitudinal scattering 
are often presented 
as a K-factor, normally defined as the ratio of the NLO to LO cross sections. 
Because the ${\cal O} (\alpha_s) $ 
QCD corrections we calculated are LO 
corrections to a tree level electroweak result we take the 
K-factor to be the ratio of the cross section with the 
${\cal O} (\alpha_s) $ QCD 
corrections and the tree level electroweak cross sections. 
The K-factors for 
$\sigma(e^+e^- \to \nu\bar{\nu} t \bar t)$ which goes via
$W^+_LW^-_L$ fusion and for 
$\sigma(e^+e^-\to e^+e^-  t \bar t)$ is
shown in Fig.~21 (left side) 
as a function of $\sqrt{s}_{e^+e^-}$.
The ${\cal O} (\alpha_s) $ 
QCD corrections are largest for $M_H=500$~GeV with K-factors  
ranging from over 1.2 for $\sqrt{s}_{e^+e^-}=500$~GeV to 1.15 for 
$\sqrt{s}_{e^+e^-}=1$~TeV. The corrections decrease
as $\sqrt{s}_{e^+e^-}$ increases.
The variation of the K-factor with $M_H$ is shown in 
Fig.~21 (right side).  
The fact that the K-factor is largest for 
$M_H=500$~GeV in Fig.~21 (left side) 
and that it peaks at $M_H\simeq 400$~GeV in Fig.~21 (right side)
is a threshold effect which
is an artifact of the kinematic cut we imposed on the $t\bar{t}$ 
invariant mass.  The important point 
is that the QCD corrections are not insignificant 
compared to the effects we might wish to study such as top Yukawa 
couplings or anomalous $VVt\bar{t}$ couplings.


\subsection{QCD at a Photon Collider~\cite{Sullivan:2005pb} \\ \small{{\it Z. Sullivan}}}
\label{ssull}

A terascale photon collider will provide a unique opportunity to understand
the resolved hadronic structure of light beyond a few GeV.  This structure
completely dominates the QCD cross section if the invariant mass of hadronic
final state particles is less than $\sim W_{\textrm{max}}/3$.  This is clear
from Fig.~22(a), where we see the cross section for $bb$
production as a function of invariant mass $M_{bb}$.
The first calculation of the uncertainty in this cross section is presented in
detail in the Proceedings of this workshop \cite{Sullivan:2005pb}.  The
result, the dashed band surrounding the upper solid curve in 
Fig.~22(b), is that the cross section cannot be predicted to better
than a factor of 5.  There is no way to improve this prediction without
measuring the gluon, charm, and bottom parton distributions for the photon
\textit{in situ} at this collider.

\begin{figure*}[tb]
\centering
\includegraphics[width=3.25in]{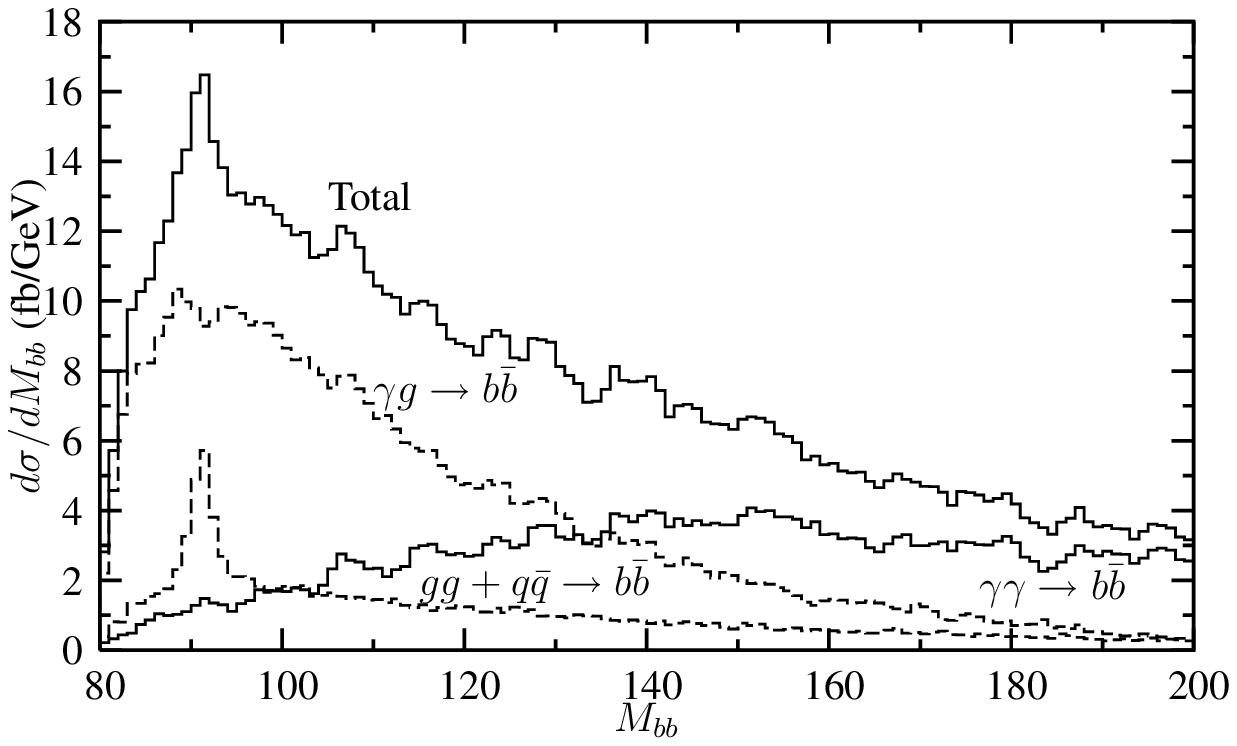}\hspace*{0.25in}\includegraphics[width=3.25in]{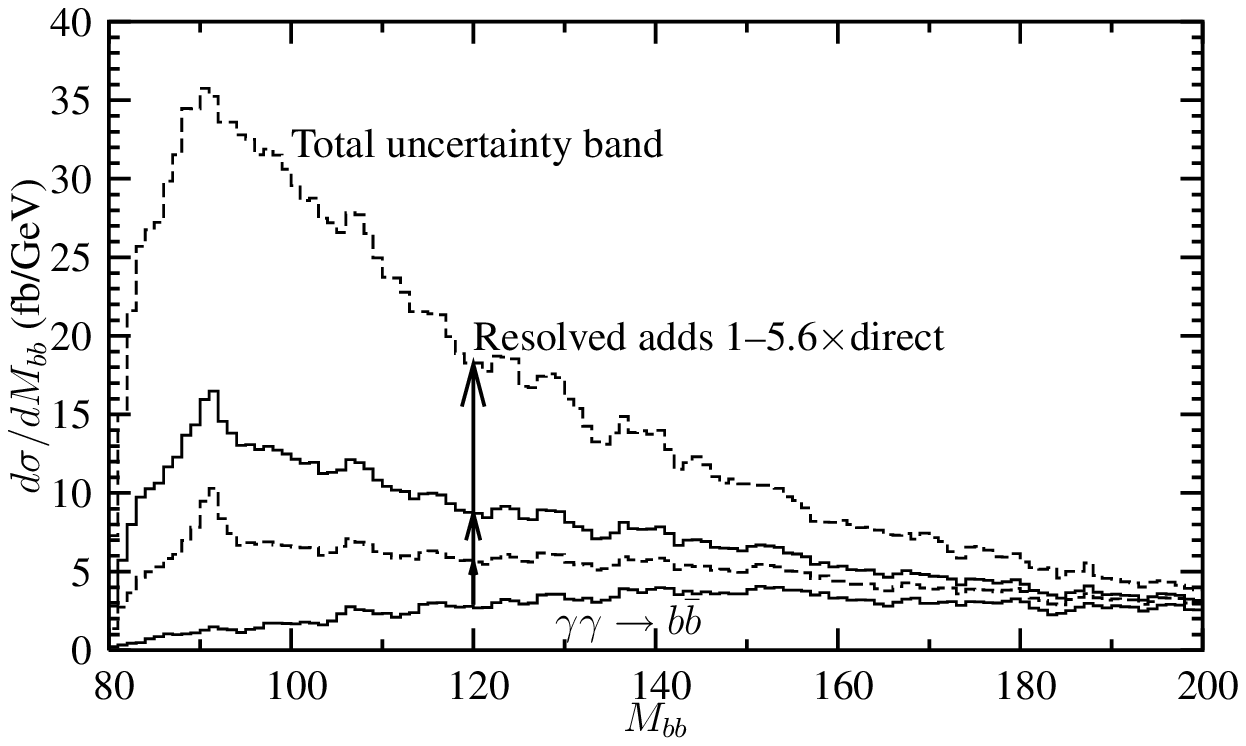}\\
{ \hspace*{0.25in}(a)\hspace*{3.35in}(b) }
\caption{(a) Breakdown of contributions to the $bb$-dijet cross section
vs.\ dijet invariant mass $M_{bb}$.
(b) Range of theoretical predictions for $bb$-dijet cross section
vs.\ dijet invariant mass $M_{bb}$.}\label{fig:zstotbb}
\end{figure*}

Loosening the cuts used for Fig.~22(a) will provide an
extremely clean sample of events with a gluon in the initial state.  Hence,
the gluon structure should be quickly well-measured.  The peak near 90 GeV is
from a $Z$ resonance, for which $40\%$ of the events come from a $c\bar c$
initial state.  Therefore, a clean extraction of the charm structure can be
made by reconstructing the $Z$ peak in multiple channels (particularly
$Z\to\mu^+\mu^-$).  Finally, the long resolved-resolved tail above the $Z$
peak is almost entirely due to $bb+\bar b\bar b+b\bar b$ collisions.
Therefore, given enough data, even the bottom structure may be accessible.

The study \cite{Sullivan:2005pb} focused on a photon-photon collider, but
another option being considered is a photon-electron collider.  In general, it
is more difficult to cleanly extract the gluon PDF in a $\gamma$--$e$
collision than a $\gamma$--$\gamma$ collision, because the cross section at
high invariant mass ($>20$ GeV) is smaller, the decay products tend to be
boosted more forward into less-well instrumented regions of the detector, and
an additional deconvolution must be performed to remove the effect of
extracting an almost-real photon from the electron.  Nevertheless, this option
should be examined in more detail as it may be simpler to construct a
$\gamma$--$e$ collider.

\newpage

\section{NEW PHYSICS EFFECTS ON TOP QUARK PROPERTIES}
\label{sec_np}


\subsection{Top Compositeness at Colliders \\ \small{{\it K. Agashe}}} 
\label{sagh}

Consider the Randall-Sundrum (RS1) model 
\cite{Randall:1999ee}
which is
a compact slice of AdS$_5$,    
\begin{eqnarray} 
ds^2 & = & e^{-2k |\theta| r_c} \eta^{\mu \nu} dx_{\mu} dx_{\nu} + r_c^2 d 
\theta^2, \; - \pi \leq \theta \leq \pi,
\label{metric}
\end{eqnarray} 
where $k$ is
the curvature scale
and the extra-dimensional interval is realized as an orbifolded circle of 
radius $r_c$. The two orbifold fixed points, $\theta = 0, \pi$, correspond 
to the ``UV'' (or ``Planck) and ``IR'' (or ``TeV'') branes respectively. In 
warped spacetimes the relationship between
5D mass scales and 4D mass 
scales (in an effective 4D description) depends on 
location in the extra dimension through the 
warp factor, $e^{-k |\theta| r_c}$. This allows large 4D mass hierarchies to 
naturally arise without large hierarchies in the defining 5D theory, whose
mass parameters are taken to be of order the observed 
Planck scale, $M_{ Pl } \sim 10^{18}$ GeV.
For example, the 4D massless graviton 
mode is localized near the UV brane  
while the Higgs sector is taken to be 
localized on the IR brane. In the 4D effective theory one then finds
\begin{equation}
{\rm Weak ~Scale} \sim M_{ Pl } e^{-k \pi r_c} .  
\end{equation}
A modestly large radius, i.e.,
$k \pi r_c \sim \log \left( M_{ Pl } / \hbox{TeV} \right)
\sim 30$, can then accommodate a TeV-size weak scale.  
Kaluza-Klein (KK) graviton resonances have masses
$m \sim k e^{ - k \pi r_c }$.  These masses are at the TeV-scale, since their wave functions 
are also localized near the IR brane.


In the original RS1 model, it was assumed that the entire SM was localized on the TeV brane, and 
that only gravity propagated in the full 5D space. Thus, the effective UV cut-off
for gauge and fermion fields 
and hence the scale suppressing higher-dimensional operators is
at a TeV, the same scale which sets the Higgs sector. However,
bounds from electroweak precision tests (EWPT) 
on this cut-off are approximately 5-10 TeV, while those from flavor changing neutral 
currents
(FCNCs)
such as $K - \bar{K}$ mixing
are around 1000 TeV. Stabilizing the electroweak scale thus requires
fine-tuning; even though RS1 explains the big hierarchy between
the Planck and electroweak scales, it has a ``little'' hierarchy problem
between the weak scale and the TeV scale cut-off.



An attractive solution to this problem is to allow the SM gauge \cite{Davoudiasl:1999tf}
and fermion \cite{Grossman:1999ra, Gherghetta:2000qt} 
fields to propagate in the extra dimensional bulk.
We first explain how bulk fermions enable us to evade flavor constraints.
The localization of the wavefunction of the massless chiral mode
is controlled by the 5d mass term for each fermion, which in units of
$k$ is denoted by the $c$-parameter.
In the warped scenario, for
$c>1/2$ ($c<1/2$) the zero mode is localized near the Planck (TeV) brane,
whereas for $c = 1/2$, the wave function is flat. 
We therefore choose $c > 1/2$ for light fermions so that the
effective UV cut-off at the location
of the light fermions is much greater than a TeV, suppressing dangerous FCNCs.  
This naturally results in a small $4D$ Yukawa coupling
to the Higgs on the TeV brane without any hierarchies in the
fundamental $5D$ Yukawa couplings 
\cite{Grossman:1999ra, Gherghetta:2000qt, Huber:2000ie}. 
Similarly, we choose
$c \ll 1/2$ for the top quark to obtain an $O(1)$ Yukawa coupling.  We can also show that
in this scenario with bulk gauge fields 
high-scale unification of gauge couplings
can be accommodated.


Since gauge fields are also in the bulk, their excited KK modes induce 
additional effects in flavor physics and in EWPT, which are calculable in 
the 5D effective field theory. 
For example, the couplings of KK modes to light fermions are 
flavor-dependent, giving FCNCs. 
However, this flavor dependence is small. KK modes, just like the 
Higgs, are localized near the TeV brane, whereas the light fermions are 
near the Planck brane. The FCNCs are therefore proportional to the 
Yukawa couplings, resulting in a suppression of flavor violation. 
This scenario has an 
analog of the
Glashow-Iliopoulos-Maiani
(GIM) mechanism of the SM, resulting in the suppression of the 
calculable FCNCs \cite{Gherghetta:2000qt, Huber:2000ie}.


Early studies showed that the corrections 
to EWPT from gauge KK modes are too large, unless 
the KK mass is greater than 10 TeV \cite{Huber:2000fh}. Such a large KK scale
results in a little hierarchy problem 
between the weak and KK scales. The localization of the
light fermions near the 
Planck brane reduces the contribution of gauge KK modes
to two observables -- the $S$ parameter 
and $4$-fermion operators, but the observable 
called the $T$ parameter still gives stringent constraints.


In Ref.~\cite{Agashe:2003zs}, it was shown that this 
problem can be avoided by extending the electroweak gauge group in 
the bulk to $SU(2)_L \times
SU(2)_R \times U(1)_{ B - L }$. Such an extension provides a custodial isospin 
symmetry to protect the $T$ parameter from large corrections. Thus, KK masses 
as low as 3 TeV are allowed by oblique EW data, significantly ameliorating the little 
hierarchy problem. 


We now consider the
top and bottom quarks. It is clear that we prefer $c \ll 1/2$ for $t_L$ 
to obtain a top Yukawa coupling of $O(1)$ without too large
a 5D Yukawa coupling, but this implies a large shift in coupling of $b_L$
to the $Z$-boson
unless the KK scale is larger than a TeV. This shift occurs due to the large
coupling of $b_L$ to the KK $Z$ modes, which mix with 
the zero-mode $Z$ via
the Higgs vev \cite{Agashe:2003zs}:
\begin{eqnarray}
\frac{ \delta \left( g^{ t_R }_Z \right) }{ g^{ t_R}_Z } 
\Big|_{ \hbox{ KK gauge } }
& \approx & \frac{ m_Z^2 } { \left( 0.41 m_{ KK } \right) ^2 } 
\frac{ 1 - 2 c_R}{ 3 - 2 c_R } 
\left( - \frac{ k \pi r_c }{2} + \frac{ 5 - 2 c_R}{ 4 ( 3 - 2 c_R ) } \right).
\end{eqnarray}
Here, $m_{ KK } \approx 
2.45 \; k e^{ - k \pi r_c }$ 
is defined to be the mass of the lightest gauge KK mode.
Thus, 
there is a tension between obtaining the top Yukawa
and not shifting
the coupling of $b_L$ to $Z$. 
As a compromise, 
KK masses $\sim 5$ TeV are consistent with a shift in 
$g_Z^{ b_L } $ of approximately $0.25\%$ for $c_L \sim 0.4$.
The $t_R$ must therefore be localized
near the TeV brane:
$c_R \stackrel{<}{\sim} 0$. Such a profile for
$t_R$ leads to sizable shift in its coupling to the $Z$
via exchange of KK $Z$-bosons, and also via KK $t_L$ modes.
The additional contribution from KK $t_L$ exchange
is due to
zero-mode $t_R$ mixing with KK $t_L$ via a Higgs vev which then couples 
to the $Z$, giving the shift
\begin{eqnarray}
\frac{ \delta \left( g_Z^{ t_R } \right) }{ g^{ t_R }_Z } 
\Big|_{ \hbox{ KK fermion } } & \approx & 
\sum_{ n } \frac{ 1/2 }{ - 2 / 3 \sin^2 \theta_W } 
\left( \frac{ m_t \sqrt{ 1 / 2 - c_L } }{ m_{ t^{ ( n ) }_L } } \right)^2.
\end{eqnarray}
Here $m_t \sqrt{ 1 / 2 - c_l }$ is
the mass term coupling zero-mode and KK top quarks, and the KK $t_L$ masses are given by $m_{ t^{ ( n ) }_L } 
\approx  \pi \; k e^{ - k \pi r_c } ( n - c_L /2 ) 
\approx 0.78 \; m_{ KK } \; ( n - c_L /2 )$.

\begin{figure}
\centering
\includegraphics[width=75mm]{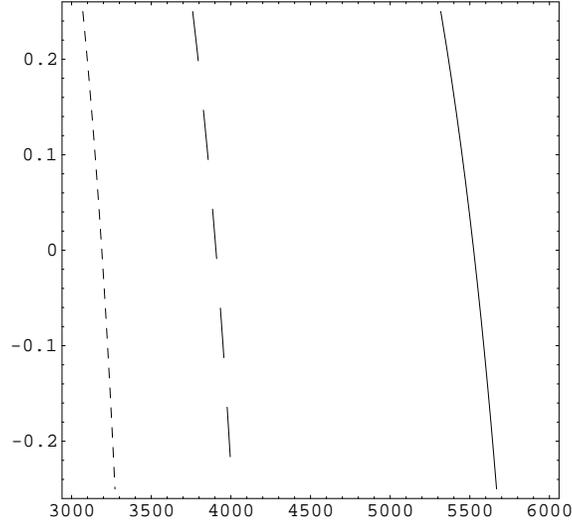} 
\caption{Shift in the coupling of $t_R$ 
to $Z$ as a function of KK mass in GeV on horizontal axis and 
$c_R$ on vertical axis for the choice 
$c_L = 0.4$. The solid, long-dashed and short-dashed 
lines correspond to shifts of $5 \%$, $10 \%$ and $15 \%$, respectively.}
\label{shift}
\end{figure}

The total shift in the coupling
of $t_R$ is plotted in Fig.~23.
As seen in the figure, we obtain an $10\%$ shift for KK masses of a few TeV.  Smaller KK masses are not 
allowed by EWPT, and larger masses lead to large fine-tuning.
Observability of an effect of this size might 
be difficult at the LHC since
the sensitivity of the LHC is only at the
$20 \%$ level for a shift in the axial coupling of the
top quark to the $Z$.  It should be possible at the ILC,
which has sensitivity at the few percent level 
for shifts in
both axial and vector couplings of the top quark to $Z$ \cite{Abe:2001nq}.
There are similar shifts in couplings of Higgs to $W/Z$ due to its profile being 
localized near
TeV brane just like for $t_R$.

Finally,
an intriguing aspect is that via the AdS/CFT correspondence 
\cite{Maldacena:1997re}, such a scenario is 
conjectured to be dual to a purely $4D$
theory with a composite Higgs boson \cite{Arkani-Hamed:2000ds}, with
the light fermions being elementary and $t_R$ being
composite. This provides an intuitive understanding for the 
large shifts in the couplings of $t_R$ and Higgs
to $Z$.
Hence, these signals might be valid for general
composite Higgs models as well.


\subsection{Top quark properties in Little Higgs Models~\cite{Berger:2005ht} \\ 
\small{{\it C.F. Berger, M. Perelstein, F. Petriello}}} 
\label{sberg}

In this section, we study the corrections to
the top quark properties in ``Little Higgs'' models of electroweak symmetry
breaking~\cite{review}, and compare the expected deviations from the SM
 predictions with expected sensitivities of experiments at the LHC and
the ILC.  In the Little Higgs models, electroweak symmetry is driven by the
radiative effects from the top sector, including the SM-like top and its
heavy counterpart, a TeV-scale ``heavy top'' $T$. Probing this structure 
experimentally is quite difficult. While the LHC should be able to 
discover the $T$ quark, its potential for studying its couplings is 
limited~\cite{PPP,ATLAS}. Direct production of the $T$ will 
likely be beyond the kinematic reach of the ILC. However, we will show below
that the corrections to the gauge couplings of the SM top, induced by its
mixing with the $T$, will be observable at the ILC throughout the parameter 
range consistent with naturalness. Measuring these corrections will provide 
a unique window on the top sector of the Little Higgs.

Little Higgs models contain a light Higgs boson which is a
{\it composite} of more fundamental 
degrees of freedom.
A generic composite Higgs model must become strongly coupled at an energy
scale around 1 TeV, leading to unacceptably large corrections to precision
electroweak observables. In contrast, Little Higgs models remain perturbative
until
a higher energy scale, around 10 TeV. The hierarchy between the Higgs mass
and the strong coupling scale is natural and stable with respect to radiative
corrections.  Because of the special symmetry structure of the theory,
the Higgs mass vanishes at tree level, as do one-loop
quadratically divergent diagrams.
The mass term is dominated by the
logarithmically divergent one-loop contribution from the top quark, which 
triggers electroweak symmetry breaking.

Many Little Higgs models have been proposed in the literature. We will
consider two examples in this study, the ``Littlest Higgs''
model~\cite{littlest}, and its variation incorporating T parity~\cite{LHT}.  We will 
study the effects on the $t\bar{t}Z$ vertex in these models; for a more detailed study, 
see Ref.~\cite{Berger:2005ht}.

\begin{figure*}[t]
\centering
\includegraphics[width=60mm,angle=90]{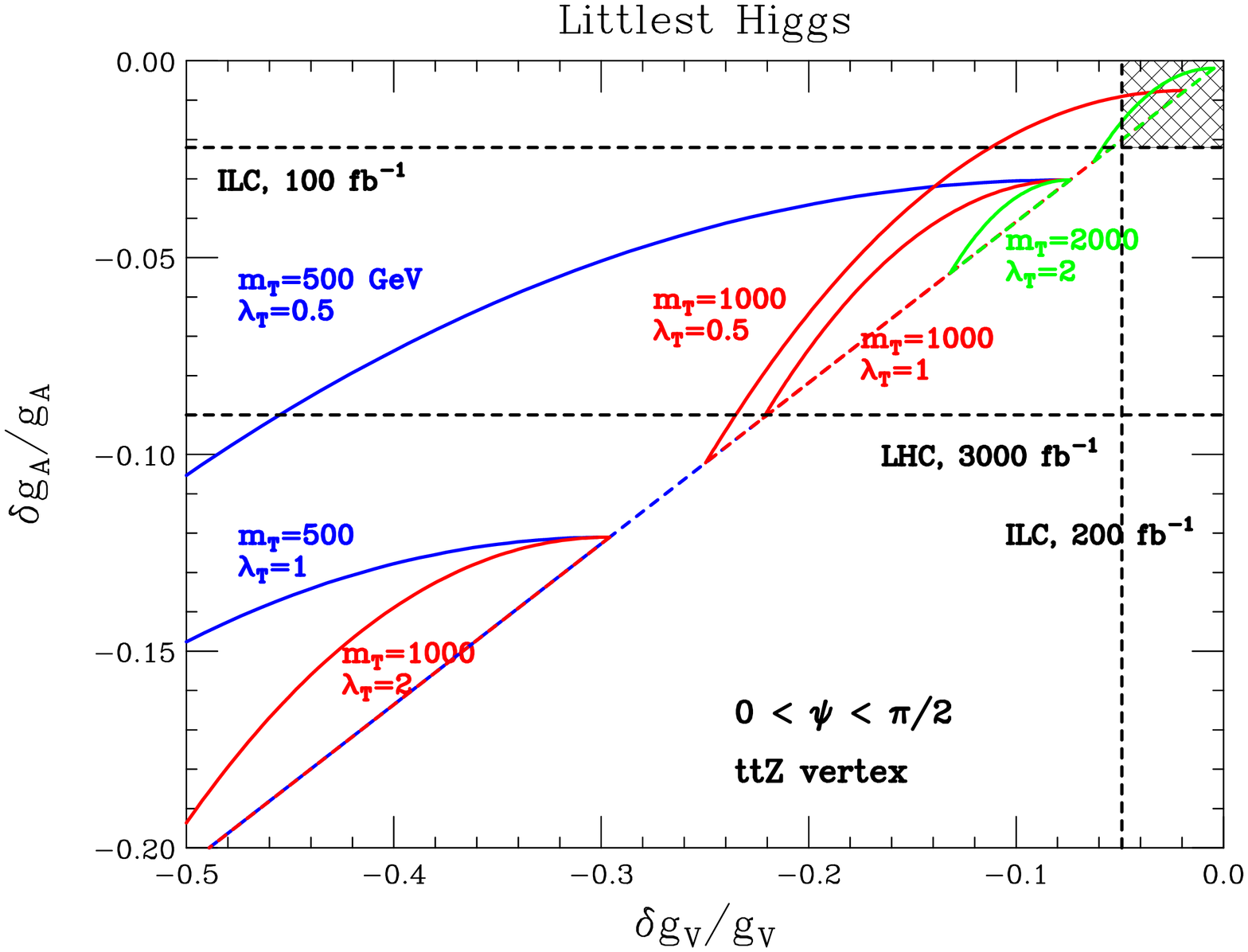}
\hskip1cm
\includegraphics[width=60mm,angle=90]{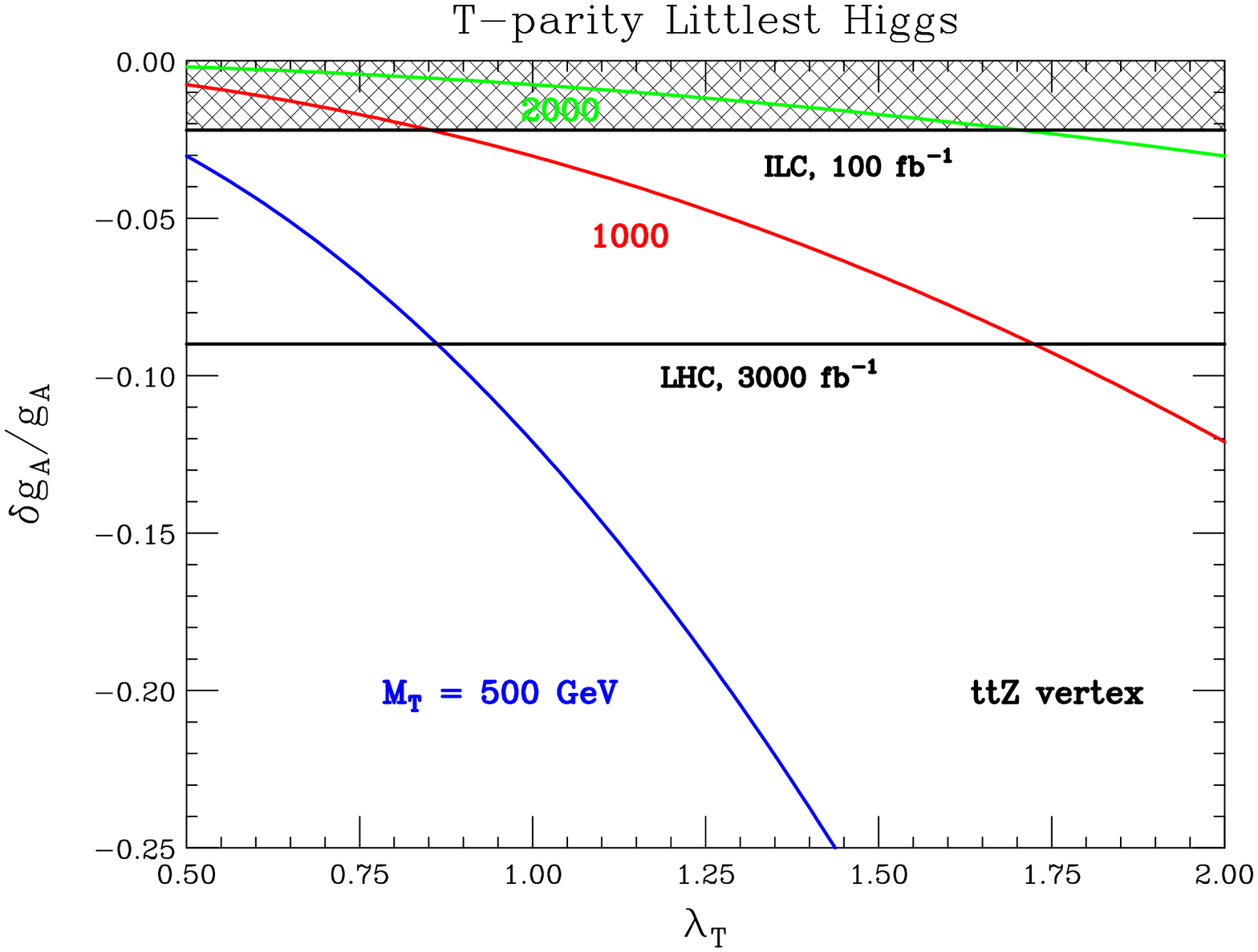}
\caption{The corrections to the $t\bar{t}Z$ axial and vector couplings in the original 
Littlest Higgs model (left panel) and its extension with T-parity (right panel). 
The regions in which the ILC would observe no 
deviation from the SM are shaded.}
\label{fig:LH}
\end{figure*}

Corrections to the gauge couplings of the top quark in Little Higgs model arise
from two
sources: the mixing of the (left-handed) top with the heavy top $T$, and
the mixing of the SM gauge bosons $W^\pm, Z^0$ with their heavy counterparts,
$W^\pm_H$ and $W_H^3$. Using the superscripts ``t'' and ``g'' to denote the
contributions from these two sources, the corrections to the $t\bar{t}Z$
coupling can be written as
\beqa
\delta g^{Z{\rm t}}_R  = 0, ~~& &~~
\delta g^{Z{\rm g}}_R = \frac{v^2}{4 f^2}\frac{c_{\psi}^2 s_{\psi}^2}
{c_{W}^2-s_{W}^2}g_R^Z, \nonumber \\
\delta g^{Z{\rm t}}_L = \frac{\lambda_T^2 v^2 g_A^Z}{m_T^2}, ~~& &~~
\delta g^{Z{\rm g}}_L = \frac{v^2}{4 f^2}\left[2g_{A}^Z s_{\psi}^4+g_R^Z
\frac{c_{\psi}^2 s_{\psi}^2}{c_{W}^2-s_{W}^2}\right].
\eeqa{delta_ttZ}
Here, $g_{L,R}^Z$ are the SM left- and right-handed  $t\bar{t}Z$ couplings,
$g_V^Z=(g_R^Z+g_L^Z)/2$ and $g_A^Z=(g_R^Z-g_L^Z)/2$ are their vector and
axial combinations, $c_W,s_W$ are
respectively the cosine and sine of the weak mixing angle, and $s_\psi\equiv
\sin\psi$, $c_\psi\equiv\cos\psi$.  In the original Littlest Higgs model~\cite{littlest}, 
both the gauge sector shift and the top sector shift occur; in the T-parity 
model~\cite{LHT}, only the top sector shift is present.  The predicted shifts in the $t\bar{t}Z$ axial and vector couplings
for $m_T=0.5, 1.0,$ and 2.0 TeV, and $\lambda_T=0.5, 1, 2$, are
plotted in Fig.~24 (left panel), along with the experimental
sensitivities expected at the LHC~\cite{Baur:2004uw} and the ILC~\cite{Abe:2001nq}.
The mixing angle $\psi$ is varied between $0$ and $\pi/2$.
Note that the shifts have a definite sign. While only a rather small part of
the parameter space is accessible at the LHC even with 3000 fb$^{-1}$
integrated luminosity, the ILC experiments will be able to easily observe the
shifts in most of the parameter space preferred by naturalness considerations (however, the prospects for observation at the 
LHC improve when additional final states such as $b\bar{b}+4\,j$ are included; see Ref.~\cite{newbaur}).  Similarly, shifts 
in the $Wtb$ coupling can be probed via deviations in the top quark width at the ILC~\cite{Berger:2005ht}.


\subsection{Testing CPT Symmetry with Top Quark Physics~\cite{CRT}  \\ \small{{\it J.A.R. Cembranos}}}
\label{sjarc}

The viability to observe evidence of CPT violation in the top sector has been analyzed through 
the measurement of a mass difference between top and anti-top \cite{CRT}. This study has been focused on the 
CPT violating ratio of the top quark, $R_{CPT}(t)\equiv 2(m_t-m_{\bar t})/(m_t+m_{\bar t})$. The present 
constraints from the Tevatron are approximately 10\%, and they could be reduced by one order of magnitude at the LHC or ILC. 
The most promising studied channel is the lepton plus jets channel for top anti-top production.  However, 
other techniques to reconstruct the top mass could also be very interesting, such as the analysis of the 
J/psi from $b$ decay at the LHC, which improves the systematic uncertainties.  Single top production could also 
be studied, since a combination of different measurements would be 
necessary in order to consider CPT violation as the explanation of any exotic data.

\underline{Di-lepton channel}: Di-lepton events originating predominantly from  
 $t\bar t \rightarrow W^+ 
(\rightarrow \ell^+\nu)\,b\,W^-(\rightarrow \ell^-\bar \nu)\,\bar b $,
 with $\ell = e$ or $ \mu$, have been used in Tevatron to 
measure the top quark pole mass supposing an identical mass for the top and anti-top quarks. The same data can be used to 
study CPT violation through a double peak in the reconstructed invariant mass  associated to the lepton and b quark 
coming from the single decay of the top or anti-top \cite{CRT}. 
\begin{figure}[ht]
 \centerline{ 
    \includegraphics[width=0.4\textwidth,clip]{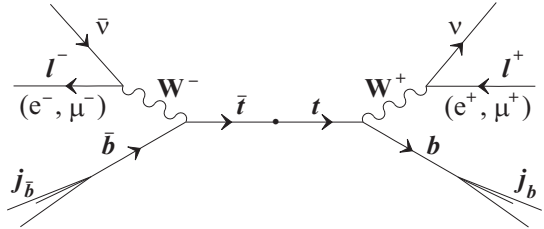}} 
  \caption{Schematics of the top and anti-top decays in the dilepton channel.} 
  \label{dileptonSch} 
\end{figure} 
By using the Tevatron data accumulated at Fermilab from 1992 through 1995 \cite{CDFdilept1}, it is possible 
to find the bound $|R_{CPT}(t)|<0.13$ at the 95\% C.L. \cite{CRT}. On the other hand, the sensitivity of 
the LHC can be estimated as $R_{CPT}(t)=0.03$ at the 95\% c.l. following an analogous analysis \cite{CRT}.

\underline{Lepton plus jets channel}: A more promising signal is provided when one of the $W$ decays leptonically 
while the other one decays hadronically:  
$t\bar t \rightarrow W^+(\rightarrow \ell^+\nu b)\,b\,W^-(\rightarrow q \bar q')\,\bar b$. In fact, the inclusive lepton plus 
jets channel provides a larger and cleaner 
sample of top quarks, whose mass can be reconstructed directly using the
 hadronic part of the decay. The invariant mass of the 
three jets coming from the top ($m_{jjb}\equiv m_{j_q j_{\bar q} j_{b}}$) or anti-top ($m_{jjb} \equiv m_{j_q j_{\bar q} j_{\bar b}}$)
presents a peak at the top ($m_t$) 
or anti-top ($m_{\bar t}$) mass respectively.  

The estimate combining the CDF \cite{CDFl+jets} and D{\O} data \cite{D0l+jets} gives a more constraining bound of 
$R_{CPT}(t) < 9.2\times 10^{-2}$. The sensitivity of the LHC is also better in this channel since both statistical 
and systematic uncertainties are expected to be improved. Indeed, the LHC will be able to test the CPT violation of 
the top quark to almost one order of magnitude better than the present constraints: $|R_{CPT}(t)|\simeq 0.014$ at the 
95\% C.L. or equivalently, $m_t-m_{\bar t}\simeq 2.4$ GeV \cite{CRT}. 
\begin{figure}[ht]
 \centerline{ 
    \includegraphics[width=0.44\textwidth,clip]{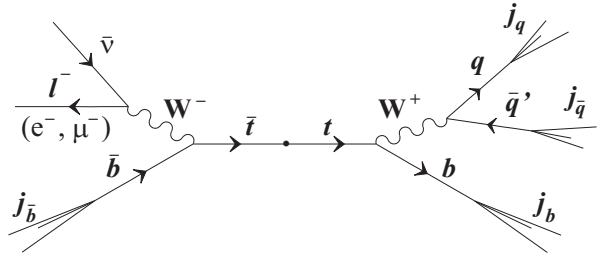}} 
  \caption{Schematic example of the top and anti-top decays in the lepton plus jets channel.} 
  \label{fig:avto1} 
\end{figure} 

The same analyses can be performed with the ILC, where an increase of the statistical 
uncertainties but a decrease of the systematic ones is expected \cite{Biernacik:2003xv}. The importance of these last uncertainties 
leads to a small improvement of the sensitivity in relation to the LHC.

\bigskip\bigskip\bigskip
\noindent
{\Large\bf ACKNOWLEDGMENTS}

\bigskip

\noindent
CFB is supported by the US Department of Energy under contract DE-AC02-76SF00515.  YK is supported by the DFG Sonderforschungsbereich/Transregio 9 
"Computer-gest\"utzte Theoretische Teilchenphysik". 
FP is supported by
the University of Wisconsin Research Committee with funds granted by the
Wisconsin Alumni Research Foundation.

\end{document}